\PassOptionsToPackage{unicode}{hyperref}
\PassOptionsToPackage{hyphens}{url}
\PassOptionsToPackage{dvipsnames,svgnames,x11names}{xcolor}
\documentclass[
  english,
  11pt,
  a4paper,
]{article}
\usepackage{xcolor}
\usepackage[margin=1in]{geometry}
\usepackage{amsmath,amssymb}
\usepackage{iftex}
\ifPDFTeX
  \usepackage[T1]{fontenc}
  \usepackage[utf8]{inputenc}
  \usepackage{textcomp} 
\else 
  \usepackage{unicode-math} 
  \defaultfontfeatures{Scale=MatchLowercase}
  \defaultfontfeatures[\rmfamily]{Ligatures=TeX,Scale=1}
\fi
\usepackage{lmodern}
\ifPDFTeX\else
\fi
\IfFileExists{upquote.sty}{\usepackage{upquote}}{}
\IfFileExists{microtype.sty}{
  \usepackage[]{microtype}
  \UseMicrotypeSet[protrusion]{basicmath} 
}{}
\makeatletter
\@ifundefined{KOMAClassName}{
  \IfFileExists{parskip.sty}{%
    \usepackage{parskip}
  }{
    \setlength{\parindent}{0pt}
    \setlength{\parskip}{6pt plus 2pt minus 1pt}}
}{
  \KOMAoptions{parskip=half}}
\makeatother
\makeatletter
\ifx\paragraph\undefined\else
  \let\oldparagraph\paragraph
  \renewcommand{\paragraph}{
    \@ifstar
      \xxxParagraphStar
      \xxxParagraphNoStar
  }
  \newcommand{\xxxParagraphStar}[1]{\oldparagraph*{#1}\mbox{}}
  \newcommand{\xxxParagraphNoStar}[1]{\oldparagraph{#1}\mbox{}}
\fi
\ifx\subparagraph\undefined\else
  \let\oldsubparagraph\subparagraph
  \renewcommand{\subparagraph}{
    \@ifstar
      \xxxSubParagraphStar
      \xxxSubParagraphNoStar
  }
  \newcommand{\xxxSubParagraphStar}[1]{\oldsubparagraph*{#1}\mbox{}}
  \newcommand{\xxxSubParagraphNoStar}[1]{\oldsubparagraph{#1}\mbox{}}
\fi
\makeatother

\usepackage{longtable,booktabs,array}
\usepackage{calc} 
\usepackage{etoolbox}
\makeatletter
\patchcmd\longtable{\par}{\if@noskipsec\mbox{}\fi\par}{}{}
\makeatother
\IfFileExists{footnotehyper.sty}{\usepackage{footnotehyper}}{\usepackage{footnote}}
\makesavenoteenv{longtable}
\usepackage{graphicx}
\makeatletter
\newsavebox\pandoc@box
\newcommand*\pandocbounded[1]{
  \sbox\pandoc@box{#1}%
  \Gscale@div\@tempa{\textheight}{\dimexpr\ht\pandoc@box+\dp\pandoc@box\relax}%
  \Gscale@div\@tempb{\linewidth}{\wd\pandoc@box}%
  \ifdim\@tempb\p@<\@tempa\p@\let\@tempa\@tempb\fi
  \ifdim\@tempa\p@<\p@\scalebox{\@tempa}{\usebox\pandoc@box}%
  \else\usebox{\pandoc@box}%
  \fi%
}
\def\fps@figure{htbp}
\makeatother

\NewDocumentCommand\citeproctext{}{}

\makeatletter
 \let\@cite@ofmt\@firstofone
 \def\@biblabel#1{}
 \def\@cite#1#2{{#1\if@tempswa , #2\fi}}
\makeatother
\newlength{\cslhangindent}
\newlength{\csllabelwidth}
\newenvironment{CSLReferences}[2] 
 {\begin{list}{}{%
  \setlength{\itemindent}{0pt}
  \setlength{\leftmargin}{0pt}
  \setlength{\parsep}{0pt}
  \ifodd #1
   \setlength{\leftmargin}{\cslhangindent}
   \setlength{\itemindent}{-1\cslhangindent}
  \fi
  \setlength{\itemsep}{#2\baselineskip}}}
 {\end{list}}
\usepackage{calc}

\ifLuaTeX
\usepackage[bidi=basic,shorthands=off]{babel}
\else
\usepackage[bidi=default,shorthands=off]{babel}
\fi
\ifLuaTeX
  \usepackage{selnolig} 
\fi

\providecommand{\tightlist}{%
  \setlength{\itemsep}{0pt}\setlength{\parskip}{0pt}}

\usepackage[T1]{fontenc}
\makeatletter
\@ifpackageloaded{caption}{}{\usepackage{caption}}
\AtBeginDocument{%
\ifdefined\contentsname
  \renewcommand*\contentsname{Table of contents}
\else
  \newcommand\contentsname{Table of contents}
\fi
\ifdefined\listfigurename
  \renewcommand*\listfigurename{List of Figures}
\else
  \newcommand\listfigurename{List of Figures}
\fi
\ifdefined\listtablename
  \renewcommand*\listtablename{List of Tables}
\else
  \newcommand\listtablename{List of Tables}
\fi
\ifdefined\figurename
  \renewcommand*\figurename{Figure}
\else
  \newcommand\figurename{Figure}
\fi
\ifdefined\tablename
  \renewcommand*\tablename{Table}
\else
  \newcommand\tablename{Table}
\fi
}
\@ifpackageloaded{float}{}{\usepackage{float}}
\floatstyle{ruled}
\@ifundefined{c@chapter}{\newfloat{codelisting}{h}{lop}}{\newfloat{codelisting}{h}{lop}[chapter]}
\floatname{codelisting}{Listing}

\makeatother
\makeatletter
\@ifpackageloaded{caption}{}{\usepackage{caption}}
\@ifpackageloaded{subcaption}{}{\usepackage{subcaption}}
\makeatother
\usepackage{bookmark}
\IfFileExists{xurl.sty}{\usepackage{xurl}}{} 
\makeatletter
\@ifundefined{xmpquote}{}{}
\makeatother
\hypersetup{
  pdftitle={Profit Redistribution in a TANK Model with Price and Wage Rigidities: An Analytical Benchmark and an Own-Lag Comparison},
  pdfauthor={Kenji Miyazaki},
  pdflang={en},
  colorlinks=true,
  linkcolor={blue},
  filecolor={Maroon},
  citecolor={Blue},
  urlcolor={Blue},
  pdfcreator={LaTeX via pandoc}}

\title{Profit Redistribution in a TANK Model with Price and Wage
Rigidities: An Analytical Benchmark and an Own-Lag Comparison}
\author{Kenji Miyazaki\footnote{Hosei University, Faculty of Economics.
  Email: \texttt{miya\_ken@hosei.ac.jp}. The author thanks Mitsuhiro
  Okano and Masataka Eguchi for useful comments and suggestions. The
  author acknowledges financial support from JSPS KAKENHI (Grant Numbers
  23K01384 and 26K04923).}}
\date{2026-08-25}
\begin{document}
\maketitle
\begin{abstract}
This paper studies profit redistribution in a Two-Agent New Keynesian
model with price and wage rigidities, type-specific wages, and
type-specific labor supply. An aggregate-lag benchmark yields
closed-form distributional allocations and impact and infinite-horizon
cumulative monetary-policy multipliers. The local effect of the
hand-to-mouth share on impact multipliers requires an additional root
condition; its local effect on infinite-horizon cumulative multipliers
requires determinacy. The own-lag comparison holds primitive cost
coefficients fixed rather than equating economic rigidity across
timings. Across a grid of policy feedback, household wage-subsidy
financing, household shares, and duration targets, expansionary monetary
shocks raise output and inflation and lower profits. The hand-to-mouth
share increases 21-quarter cumulative responses under both timings and
impact responses under own-lag adjustment, whereas the aggregate-lag
impact ordering can reverse slightly. Timing changes inflation and
profit persistence. The aggregate-lag benchmark compresses dispersion
losses into a higher price-inflation weight; own-lag adjustment leaves
inherited relative wages and dynamic dispersion terms.\\
\textbf{Keywords}: TANK, heterogeneity, price rigidity, wage rigidity,
monetary policy\\
\textbf{JEL classification}: E21, E24, E31, E44, E52
\end{abstract}

\newpage

\section{Introduction}\label{introduction}

Household heterogeneity matters for monetary transmission because policy
shocks redistribute income across households with different marginal
propensities to consume. This paper studies that redistribution in a
Two-Agent New Keynesian (TANK) model with both price rigidity and wage
rigidity. It derives a closed-form equilibrium for an economy in which
patient savers own firms and smooth consumption, while hand-to-mouth
households consume current labor income. The solution preserves
type-specific wages and labor supply, rather than imposing common wages
or common labor allocations across households.\footnote{The claim refers
  to analytical TANK models that jointly allow price rigidity, wage
  rigidity, and type-specific wages and labor supply. Existing
  sticky-wage TANK models often obtain tractability by imposing common
  wages or common labor supply across household types.}

The closed-form analysis uses quadratic adjustment costs defined
relative to lagged aggregate price and wage indices. This timing
assumption isolates the profit-redistribution mechanism but differs from
conventional own-lag Rotemberg adjustment. Within this analytical
benchmark, heterogeneity does not automatically amplify monetary policy.
After an expansionary monetary shock, price rigidity raises real wages
and lowers firm profits. Since profits accrue to savers while
hand-to-mouth households rely on labor income, this redistribution can
raise aggregate demand when the hand-to-mouth share is larger. Wage
rigidity acts through two margins rather than weakening the mechanism
unconditionally: analytically, it shifts the allocation of a given
profit change toward the consumption gap, while in the benchmark
duration calibration it dampens nominal and profit responses. The impact
inflation and output multipliers rise locally with the hand-to-mouth
share only under an additional root condition. For the infinite-horizon
cumulative multipliers, Proposition 3 establishes a local increase with
the hand-to-mouth share as long as the benchmark equilibrium remains
determinate.

The conventional own-lag framework developed in Miyazaki (2026) provides
a direct test of whether these conclusions depend on the tractability
assumption. Prices adjust relative to each firm's own lagged price,
wages adjust relative to each household type's own lagged wage, and both
Phillips curves contain expected future inflation. The exercise changes
adjustment-cost timing while holding primitive cost coefficients fixed;
it does not equate reduced-form slopes or economic rigidity across the
two specifications.

Under the baseline policy rule and household-side wage-subsidy-financing
arrangement, all 27 combinations of the hand-to-mouth share and the
price- and wage-duration targets preserve the monetary-shock sign and
heterogeneity orderings reported below. The expanded grid also varies
four output-gap coefficients in the policy rule and two ways of
financing household wage subsidies, while retaining the firm-side
sales-subsidy offset. Expansionary monetary shocks raise output and
inflation and lower profits on impact and in 21-quarter sums throughout
that grid. Own-lag impact and 21-quarter cumulative responses increase
monotonically with the hand-to-mouth share throughout the extended grid.
Under aggregate-lag adjustment, the finite-horizon cumulative responses
retain that ordering throughout, whereas the impact ordering reverses
slightly in some cases with stronger output-gap feedback and a longer
price-duration target. The largest reversal in the impact output effect
is 0.0105 percentage points, or 1.32 percent of the corresponding
representative-agent response.

Replacing type-specific wage-subsidy rebates with a common lump-sum tax
has little effect on macroeconomic responses but increases the offset
from relative wage income and narrows the consumption gap. A matched
common-wage/common-labor closure leaves benchmark own-lag impact output
and inflation close to the type-specific specification, while its wider
consumption gap is consistent with an earnings-reallocation-buffer
interpretation. At the benchmark calibration, the output impact remains
similar across timings, while inflation and profit persistence differ
materially. Thus the sign and finite-horizon cumulative amplification of
the redistribution mechanism are robust within the reported grid, but
the impact ordering and quantitative magnitudes remain conditional on
policy and adjustment timing.

The analytical benchmark gives three closed-form results. First, the
TANK Phillips curve shows how price and wage rigidities jointly
determine the slope of inflation with respect to the output gap. Its
structural coefficients are independent of the hand-to-mouth share, even
though the distributional consequences of inflation are not. Second, the
heterogeneity-adjusted IS curve contains an expected
profit-redistribution term. When price inflation changes expected
profits, the induced consumption gap between savers and hand-to-mouth
households shifts aggregate demand. Third, the impact and
infinite-horizon cumulative monetary-policy multipliers identify local
conditions under which heterogeneity strengthens transmission in the
analytical benchmark.

The welfare calculation also marks the boundary of the closed-form
result. In the aggregate-lag benchmark, all cross-household dispersion
costs can be represented by a higher contemporaneous weight on price
inflation. This compression is specific to the common-lag timing: under
conventional own-lag adjustment, relative wages are inherited state
variables, so consumption, hours, and wage-inflation dispersion remain
separate dynamic welfare terms. For the benchmark monetary shock,
raising the hand-to-mouth share from zero to 0.2 increases the own-lag
loss by 3.4 percent, but direct dispersion accounts for only 0.047
percent of total loss. The coefficient change and realized loss share
are distinct objects from different timing specifications; the loss
ordering survives within the examined calibration grid, whereas the
single-inflation-weight representation and its policy interpretation do
not.

The paper contributes to a growing literature that uses
heterogeneous-agent models to study policy transmission. Heterogeneous
Agent New Keynesian (HANK) models incorporate realistic variation in
income, wealth, and credit access (McKay et al. 2016; Kaplan et al.
2018), but their quantitative richness often requires numerical methods
that obscure the mapping from structural parameters to aggregate
outcomes (Kaplan and Violante 2018). TANK models provide a tractable
alternative by dividing households into savers and hand-to-mouth
consumers (Galí et al. 2007). Prior analytical TANK models show that
hand-to-mouth households can amplify fiscal and monetary shocks (Galí et
al. 2007; Bilbiie 2008), and Debortoli and Galí (2024) show that a
well-designed TANK model can match key aggregate implications of a HANK
model.

This paper brings wage rigidity into that analytical TANK framework
without suppressing the cross-household labor and wage margins that are
central to the redistribution mechanism. Canonical New Keynesian models
show that sticky wages matter for monetary transmission (Erceg et al.
2000; Christiano et al. 2005). In TANK economies, however, wage rigidity
interacts with profits, labor supply, and household heterogeneity, which
makes closed-form analysis harder. Quadratic adjustment costs relative
to lagged aggregate indices make the benchmark tractable. This price
reference is closest to the analytical benchmark in Bilbiie (2025) and
to the timing comparison in Maliar and Naubert (2025), whereas the
common-primitive comparison here adds type-specific sticky wages, profit
incidence, distributional allocations, and welfare accounting. That
comparison then shows which conclusions survive when forward-looking
price and wage setting and inherited relative wages are restored.

The analysis proceeds as follows. Section 2 positions the paper relative
to analytical TANK models, sticky-wage TANK models, and quantitative
HANK work. Section 3 presents the model environment and the benchmark
equilibrium conditions. Section 4 derives the aggregate-lag TANK
Phillips curve, the heterogeneity-adjusted IS curve, the monetary-policy
multipliers, and the benchmark welfare result. Section 5 evaluates the
analytical conditions, compares the benchmark with the common-primitive
own-lag model, and tests policy-rule and household-side
wage-subsidy-financing robustness. Section 6 concludes.

\section{Related Literature and
Contribution}\label{related-literature-and-contribution}

This section places the paper in the closest TANK and HANK literatures
by focusing on the assumptions that matter for the mechanism.

\subsection{Analytical TANK Models with Price
Rigidity}\label{analytical-tank-models-with-price-rigidity}

Analytical TANK models show how far a two-household structure can go in
explaining heterogeneous-agent transmission without the full state space
of HANK models. Bilbiie (2008) show that hand-to-mouth households can
amplify monetary and fiscal shocks in a tractable nominal-rigidity
environment. Debortoli and Galí (2024) show that a carefully constructed
TANK model can approximate important aggregate responses of richer HANK
economies. These papers establish the value of closed-form TANK
analysis. Their main analytical results, however, are built around price
rigidity and do not jointly characterize price rigidity, wage rigidity,
type-specific wages, and type-specific labor supply.

Two recent papers are especially close on analytical method and
price-adjustment timing. Bilbiie (2025) develops a richer analytical
heterogeneous-agent New Keynesian framework in which cyclical inequality
and risk shape transmission and policy, and its tractable pricing
benchmark uses the lagged market price as a reference. Maliar and
Naubert (2025) directly compare aggregate-lag and own-lag price
adjustment when central-bank responses are endogenous. The present paper
is narrower in household risk but extends this analytical comparison
along a different dimension: sticky, type-specific wages and labor
supply make profit incidence, earnings reallocation, and distributional
welfare explicit. Its novelty is therefore the closed-form
distributional mapping and its qualified timing comparison, not the
first use of an aggregate price reference or the first analytical
heterogeneous-agent monetary model.

The present paper keeps the closed-form discipline of this literature
but changes the labor side of the model. The issue is not simply to add
sticky wages. Once wages are sticky, heterogeneity also works through
labor supply, wage-setting, and profits. These margins determine whether
monetary policy redistributes income toward hand-to-mouth households or
whether wage rigidity offsets the usual TANK amplification mechanism.

\subsection{TANK Models with Wage
Rigidity}\label{tank-models-with-wage-rigidity}

Several TANK papers introduce wage rigidity through institutional
assumptions that preserve tractability. Colciago (2011) and Ascari et
al. (2017) use union-based wage setting, which effectively imposes
common wages and common labor supply across household types. This is
useful for deriving aggregate dynamics, but it removes the type-specific
wage and hours gaps through which profit redistribution affects
household allocations.

Recent sticky-wage TANK applications make clear why this missing margin
matters. Bilbiie and Känzig (2023) connect profits, inflation, and
demand under wage rigidity, but their tractable mechanism relies on
tighter distributional restrictions than those used here. Ida and Okano
(2024) show that nominal wage stickiness changes fiscal multipliers when
liquidity-constrained households are present, but their main analysis is
numerical rather than a closed-form characterization of monetary
multipliers. Bilbiie et al. (2024) emphasize that the relative
stickiness of wages and prices shapes profit cyclicality. Taken
together, this work shows that wage rigidity matters for TANK dynamics
while leaving open an analytical model with price rigidity, wage
rigidity, type-specific wages, and type-specific labor supply all
present at once.

The type-specific closure is also a substantive boundary of the
comparison. Miyazaki (2026) derives both type-specific wage adjustment
and a reduced-form common-wage closure. Section 5 reports a matched
numerical contrast that suppresses relative wage and hours dispersion
while retaining the aggregate sticky-wage block. This is an
institutional-closure comparison rather than a nested causal parameter:
it indicates how the type-specific earnings-reallocation margin changes
the benchmark responses, but it does not identify an independently
varying structural effect of common wages or common labor.

\subsection{HANK Models with Wage
Rigidity}\label{hank-models-with-wage-rigidity}

The HANK literature incorporates sticky wages with much richer household
heterogeneity. McKay and Wolf (2023) show that distributional
considerations matter for Ramsey policy once wage rigidity and
heterogeneity interact. Auclert et al. (2023) demonstrate that
sticky-wage HANK models can jointly speak to marginal propensities to
consume, marginal propensities to earn, and fiscal multipliers. Related
applications include Hagedorn et al. (2019) and Ferra et al. (2020).
These models are quantitatively rich, but their complexity makes it
difficult to isolate transparent closed-form conditions for when
heterogeneity amplifies or dampens monetary transmission. Earlier work
by Cúrdia and Woodford (2016) also derives welfare-based monetary policy
in a New Keynesian model with household heterogeneity and credit
frictions. Its heterogeneity operates through borrower-saver spreads and
aggregate credit rather than the type-specific sticky-wage and
profit-incidence margins studied here.

Among sticky-wage heterogeneous-agent models, Gerke et al. (2023) are
especially close in spirit to the present paper because they question
the standard assumption of uniform labor supply across households. Their
analysis shows that labor-market structure matters for both inflation
dynamics and optimal policy, but the framework remains computationally
intensive. The present analysis complements that line of work by
separating mechanisms that richer models combine numerically. In
particular, it separates the price-rigidity channel, which moves profits
and real wages in opposite directions, from the wage-rigidity channel,
which can dampen the resulting demand amplification.

\subsection{Contribution}\label{contribution}

The paper makes three contributions. First, it relaxes the common-wage
and common-labor-supply restrictions used in much of the analytical
sticky-wage TANK literature. In the aggregate-lag benchmark, price and
wage rigidities generate type-specific consumption, wage, and hours
gaps, with profits serving as the sufficient statistic for those gaps.
Second, it derives closed-form Phillips and IS curves, together with
impact and infinite-horizon cumulative monetary-policy multipliers, for
a TANK model with price rigidity, wage rigidity, type-specific wages,
and type-specific labor supply. It gives explicit local conditions under
which heterogeneity strengthens monetary transmission in the
aggregate-lag benchmark. On impact, the multipliers rise locally with
the hand-to-mouth share only under the sufficient root condition stated
in Proposition 2. For the infinite-horizon cumulative multipliers, the
corresponding local heterogeneity effect holds under the determinacy
condition in Proposition 3. The benchmark also identifies a sharp
limiting case: under pure wage stickiness, price flexibility shuts down
profit redistribution, so the heterogeneity term in the aggregate IS
relation disappears.

Third, using the conventional own-lag framework in Miyazaki (2026), the
paper conducts a common-primitive comparison with the analytical
aggregate-lag benchmark. Here, common primitives mean identical values
of \((\eta_p,\eta_w,\psi_p,\psi_w)\), not equal reduced-form
Phillips-curve slopes or economic rigidity across timings. Under the
baseline policy rule and household-side wage-subsidy-financing
arrangement, the comparison preserves the monetary-shock sign and the
reported \(\lambda\), \(D_p\), and \(D_w\) orderings over the 27-case
duration grid. The broader policy-rule and financing exercise preserves
positive output and inflation responses and negative profit responses,
on impact and in 21-quarter sums, throughout the reported grid.
Finite-horizon cumulative output, inflation, and absolute profit
responses increase monotonically with the hand-to-mouth share under both
adjustment timings. The corresponding impact ordering is uniform under
own-lag adjustment but conditional under aggregate-lag adjustment, with
small reversals concentrated at stronger output-gap feedback and longer
price-duration targets.

Common lump-sum financing of household wage subsidies changes aggregate
responses only modestly while strengthening the relative-wage-income
offset to direct profit incidence; the firm-side sales-subsidy offset is
not varied. The companion paper explains why type-specific wage
adjustment makes profits cease to be a sufficient statistic and why
own-lag timing makes relative wages inherited. The matched
restricted-closure comparison shows that suppressing relative wage and
hours dispersion has little effect on benchmark impact output and
inflation but makes the consumption gap larger. The comparison here
identifies which qualitative results survive these changes and which
quantitative and impact-ordering results depend on the maintained
closure.

The model sits between analytical TANK models and computational
sticky-wage HANK models. It keeps tractability while making the
distributional mechanism explicit: price rigidity moves profits and
wages in opposite directions, wage rigidity changes the strength of that
redistribution, and household heterogeneity determines how the
redistribution maps into aggregate demand. The next section presents the
model environment that delivers these results.

\section{Model}\label{model}

This section presents the aggregate-lag two-agent New Keynesian (TANK)
benchmark that keeps the distributional channel analytically
transparent. Relative to the standard analytical TANK setup, the model
has four important features: households set type-specific wages, labor
supplies can differ by type, price and wage adjustment costs use lagged
aggregate reference prices and wages, and a subsidy-tax scheme removes
steady-state markups and profits. These choices preserve the wage-profit
redistribution mechanism while delivering a closed-form log-linear
equilibrium. The notation is standard: Latin letters denote variables
and Greek letters denote parameters, except for inflation rates
(\(\pi_t^i\), \(\Pi_t^i\)). Variables without time subscripts denote
steady-state values. Lowercase variables are log deviations from steady
state, except for firm profits, denoted by \(d_{t}:=D_{t}/Y\).

The economy has a labor packer, two household types, two types of firms,
and a consolidated government authority. Variety-level objects are
indexed by \(h\) for household labor varieties and by \(j\) for
intermediate-goods varieties. The Online Appendix uses an additional
superscript \(M\) for pre-aggregation variety-level objects in the
nonlinear derivations. The labor packer turns differentiated household
labor into homogeneous labor inputs sold to intermediate producers.
Labor aggregation follows a constant elasticity of substitution (CES)
specification, where \(\psi_w > 1\) is the elasticity of substitution
across labor types. The resulting labor demand function is given by
\begin{equation}\protect\phantomsection\label{eq-labor-demand}{
N_{t}(h)=N_{t}\left(\frac{W_{t}^{N}(h)}{W_{t}^{N}}\right)^{-\psi_{w}},
}\end{equation} where \(N_{t}(h)\) is labor supplied by household \(h\),
\(N_t\) is aggregate labor input, \(W_{t}^{N}(h)\) is the nominal wage
paid to household \(h\), and \(W_{t}^{N}\) is the aggregate nominal
wage.

Households buy final goods and supply differentiated labor services to
the packer. Their labor varieties give them wage-setting power, subject
to the labor demand curve above. All households have the same
preferences but face quadratic costs when adjusting nominal wages.
Preferences are parameterized by relative risk aversion
(\(\gamma > 0\)), the inverse Frisch elasticity of labor supply
(\(\varphi > 0\)), and the discount factor (\(0 < \beta < 1\)). The wage
adjustment cost is quadratic in the deviation from the past aggregate
nominal wage: \[
\frac{\eta_w}{2}\left(\frac{W_{t}^{N}(h)-W_{t-1}^N}{W^N_{t-1}}\right)^{2}Y_t,
\] where \(\eta_w\) represents the degree of wage adjustment cost. The
lagged aggregate wage is interpreted as a reduced-form external wage
norm rather than as a uniquely microfounded quadratic adjustment
cost.\footnote{External wage norms and wage leadership affect wage
  setting in the staggered-contract model and Austrian bargaining
  evidence of Knell and Stiglbauer (2012). The fair-wage model of Kuang
  and Wang (2017) uses aggregate and past wages among the reference
  points governing worker effort. These mechanisms motivate an
  external-norm interpretation, but neither paper derives the exact
  quadratic cost or aggregate-lag timing used here. A standard
  sticky-wage specification instead defines type-specific wage inflation
  relative to each type's own lagged wage,
  \(\Pi_t^i = \Pi_t^p W_t^i / W_{t-1}^i\) for \(i\in\{H,S\}\). Section
  5.2 compares the two timing specifications at the same primitive
  parameters.} Using this reference is a deliberate analytical
simplification rather than the conventional Rotemberg timing. A standard
type-specific lagged wage introduces wage-dispersion state variables and
forward-looking terms (Miyazaki 2026). Section 5.2 applies that
framework in a numerical comparison with the same primitive parameters.

The household sector consists of Type H (hand-to-mouth) and Type S
(saver) households. Total population is normalized to one. Type H
households make up a fraction \(\lambda \in [0,1)\) of the economy, and
Type S households make up the remaining \(1 - \lambda\). Type H
households face binding liquidity constraints and consume their labor
income each period. Type S households have access to financial markets
and smooth consumption through asset holdings.

Within each type, households share risk perfectly. Thus all households
of type \(i\in\{H,S\}\) choose the same consumption \(C_t^i\), labor
supply \(N_t^i\), and real wage \(W_t^i\) in equilibrium. Aggregate
consumption, hours, and real wages are \[
\begin{aligned}
C_{t}&=\lambda C_{t}^{H}+(1-\lambda)C_{t}^{S},\\
N_{t}&=\left[
\lambda(N_{t}^{H})^{\frac{\psi_{w}-1}{\psi_{w}}}
+(1-\lambda)(N_{t}^{S})^{\frac{\psi_{w}-1}{\psi_{w}}}
\right]^{\frac{\psi_{w}}{\psi_{w}-1}},\\
W_{t}&=\left[
\lambda(W_{t}^{H})^{1-\psi_{w}}
+(1-\lambda)(W_{t}^{S})^{1-\psi_{w}}
\right]^{\frac{1}{1-\psi_{w}}},
\end{aligned}
\] where \(W_{t}=W_t^N/P_t\) is the real wage. The wage markup for each
type measures the wedge between the real wage and the marginal rate of
substitution between labor and consumption. For type \(i\in\{H,S\}\),
this marginal rate of substitution is
\((C_t^i)^\gamma (N_t^i)^\varphi\).

On the production side, final goods firms aggregate differentiated
intermediate goods into a homogeneous final good using a standard CES
technology with substitution elasticity \(\psi_p > 1\).

Intermediate-goods firms produce differentiated varieties using
homogeneous labor and face a quadratic aggregate-reference adjustment
cost when changing prices. They hire labor from the labor packer and
sell their output to final goods firms. Production is linear in labor
and subject to an aggregate productivity shock: \[
\begin{aligned}
Y_t(j)& = \exp(a_t) N_t(j),\\
a_t&=\rho_a a_{t-1}+e_t^a,\;0<\rho_a<1,\\
\end{aligned}
\] The technology shock \(\exp(a_t)\) affects all firms equally and
follows an AR(1) process with innovation \(e^a_t\).

The benchmark price adjustment cost is \[
\frac{\eta_p}{2}\left(\frac{P_{t}(j)-P_{t-1}}{P_{t-1}}\right)^{2}Y_t,
\] where \(\eta_p\) measures how costly price adjustment is. Following
the aggregate-reference analytical benchmarks in Bilbiie (2025) and
Maliar and Naubert (2025), the lagged aggregate price index is the
adjustment reference.\footnote{Bilbiie (2025) uses the lagged market
  price index as the adjustment reference in an analytical benchmark,
  and Maliar and Naubert (2025) explicitly compare aggregate-lag and
  own-lag price adjustment. Micro price evidence also shows that
  repricing responds to deviations from competitors' average prices
  (Campbell and Eden 2014; Gagnon et al. 2013). These studies motivate
  an aggregate-reference interpretation but do not uniquely derive the
  exact quadratic cost used here or justify applying the same timing to
  wages.} This specification is not the canonical own-lag Rotemberg
cost: it removes forward-looking price-setting dynamics and permits a
transparent closed-form solution. Section 5.2 compares it with standard
own-lag price and wage adjustment. In equilibrium, all firms make
identical choices, so firm-specific indices can be dropped. The price
markup is defined as the ratio of the firm's price to marginal cost; in
this environment, marginal cost equals the real wage divided by
productivity.

The government plays two roles: fiscal and monetary. For tractability,
the government uses a subsidy-tax scheme that removes steady-state
markup distortions and normalizes steady-state profits to
zero.\footnote{This subsidy-tax arrangement is introduced only to remove
  steady-state markup distortions and steady-state profits. It is a
  standard analytical device that allows the paper to focus on the
  transitional dynamics generated by nominal rigidities and household
  heterogeneity, rather than on welfare losses associated with
  inefficient steady-state markups.}

Monetary policy follows a Taylor-type rule: \[
\begin{aligned}
R^N_{t} & =\frac{\exp(-m_{t})}{\beta}(\Pi_t^p)^{\phi},\;\phi>1,\\
m_{t} & =\rho_m m_{t-1}+e^m_{t},\;0<\rho_m<1,
\end{aligned}
\] The parameter \(\phi > 1\) ensures that the Taylor principle holds:
real rates rise when inflation rises. Monetary policy shocks \(m_t\)
follow their own AR(1) process.

The following definitions are useful later. Wage inflation for each
household type is
\begin{equation}\protect\phantomsection\label{eq-wage-inflation}{
\Pi_{t}^{i}:=P_t W^i_{t}/W^N_{t-1}=\Pi^p_tW^i_{t}/W_{t-1},
\qquad i\in\{H,S\}.
}\end{equation} Wage inflation is defined relative to the aggregate wage
in the previous period, not each type's own past wage. This choice
differs from conventional sticky-wage specifications and is what
delivers the closed-form analytical solution in the benchmark. Total
nominal wage inflation and the aggregate wage markup are denoted by
\(\Pi_t^w\) and \(\mu_t^w\), respectively.

Market clearing requires production to equal consumption plus the
resources spent on price and wage adjustment: \[
\begin{aligned}
Y_{t}&=C_{t}+\frac{\eta_p}{2}(\Pi^p_{t}-1)^{2}Y_{t}\\
&+\lambda\frac{\eta_w}{2}(\Pi^H_t-1)^2Y_t
+(1-\lambda)\frac{\eta_w}{2}(\Pi^S_t-1)^2Y_t.
\end{aligned}
\] The steady state is normalized so that the main aggregate quantities
equal one. Log-linearizing around this steady state gives a square
system for the endogenous log-linear variables, conditional on the
exogenous states \(a_t\) and \(m_t\). The table below combines aggregate
equations with the Type H block. Type S labor, wage, markup, and
wage-inflation variables are recovered from the aggregation identities,
which avoids redundant type-specific equations.
Table~\ref{tbl-equations} shows the nonredundant linearized equations,
while Table~\ref{tbl-parameters} lists the key parameters and their
benchmark values.

\begin{longtable}[]{@{}
  >{\raggedright\arraybackslash}p{(\linewidth - 2\tabcolsep) * \real{0.5000}}
  >{\raggedright\arraybackslash}p{(\linewidth - 2\tabcolsep) * \real{0.5000}}@{}}
\caption{Aggregate-lag TANK
equations}\label{tbl-equations}\tabularnewline
\toprule\noalign{}
\begin{minipage}[b]{\linewidth}\raggedright
Name
\end{minipage} & \begin{minipage}[b]{\linewidth}\raggedright
Equation
\end{minipage} \\
\midrule\noalign{}
\endfirsthead
\toprule\noalign{}
\begin{minipage}[b]{\linewidth}\raggedright
Name
\end{minipage} & \begin{minipage}[b]{\linewidth}\raggedright
Equation
\end{minipage} \\
\midrule\noalign{}
\endhead
\bottomrule\noalign{}
\endlastfoot
Production function & \(y_{t}=a_{t}+n_{t}\) \\
Resource constraint & \(y_{t}=c_{t}\) \\
Price markup & \(\mu^{p}_{t}=a_{t}-w_{t}\) \\
Price Phillips curve & \(\eta_{p}\pi^{p}_{t}=-\psi_{p}\mu^{p}_{t}\) \\
Wage inflation & \(\pi^{w}_{t}=w_{t}-w_{t-1}+\pi^{p}_{t}\) \\
Wage markup & \(\mu^{w}_{t}=w_{t}-\gamma c_{t}-\varphi n_{t}\) \\
Wage Phillips curve & \(\eta_{w}\pi^{w}_{t}=-\psi_{w}\mu^{w}_{t}\) \\
Profits & \(d_{t}=a_{t}-w_{t}\) \\
Euler equation, S & \(r_{t}=\gamma E_{t}\Delta c^{S}_{t+1}\) \\
Budget constraint, H & \(c^{H}_{t}=w^{H}_{t}+n^{H}_{t}\) \\
Labor demand, H & \(n^{H}_{t}=n_{t}-\psi_{w}(w^{H}_{t}-w_{t})\) \\
Aggregate consumption &
\(c_{t}=\lambda c^{H}_{t}+(1-\lambda)c^{S}_{t}\) \\
Aggregate labor & \(n_{t}=\lambda n^{H}_{t}+(1-\lambda)n^{S}_{t}\) \\
Aggregate wage & \(w_{t}=\lambda w^{H}_{t}+(1-\lambda)w^{S}_{t}\) \\
Aggregate wage markup &
\(\mu^{w}_{t}=\lambda \mu^{H}_{t}+(1-\lambda)\mu^{S}_{t}\) \\
Aggregate wage inflation &
\(\pi^{w}_{t}=\lambda\pi^{H}_{t}+(1-\lambda)\pi^{S}_{t}\) \\
Wage inflation, H & \(\pi^{H}_{t}=\pi^{p}_{t}+w^{H}_{t}-w_{t-1}\) \\
Wage markup, H &
\(\mu^{H}_{t}=w^{H}_{t}-\gamma c^{H}_{t}-\varphi n^{H}_{t}\) \\
Wage Phillips curve, H & \(\eta_{w}\pi^{H}_{t}=-\psi_{w}\mu^{H}_{t}\) \\
Taylor rule & \(r_{t}=\phi\pi^{p}_{t}-E_{t}\pi^{p}_{t+1}-m_{t}\) \\
\end{longtable}

In Table 1, superscripts \(H\) and \(S\) refer to hand-to-mouth and
saver households, while superscripts \(p\) and \(w\) refer to price and
wage inflation or markups. The variable \(d_t\) denotes dividends, or
profits, scaled by steady-state output.

\begin{longtable}[]{@{}
  >{\raggedright\arraybackslash}p{(\linewidth - 4\tabcolsep) * \real{0.6300}}
  >{\raggedright\arraybackslash}p{(\linewidth - 4\tabcolsep) * \real{0.2200}}
  >{\raggedright\arraybackslash}p{(\linewidth - 4\tabcolsep) * \real{0.1500}}@{}}
\caption{Common benchmark
parameters}\label{tbl-parameters}\tabularnewline
\toprule\noalign{}
\begin{minipage}[b]{\linewidth}\raggedright
Name
\end{minipage} & \begin{minipage}[b]{\linewidth}\raggedright
Parameter
\end{minipage} & \begin{minipage}[b]{\linewidth}\raggedright
Value
\end{minipage} \\
\midrule\noalign{}
\endfirsthead
\toprule\noalign{}
\begin{minipage}[b]{\linewidth}\raggedright
Name
\end{minipage} & \begin{minipage}[b]{\linewidth}\raggedright
Parameter
\end{minipage} & \begin{minipage}[b]{\linewidth}\raggedright
Value
\end{minipage} \\
\midrule\noalign{}
\endhead
\bottomrule\noalign{}
\endlastfoot
Discount factor & \(0<\beta<1\) & 0.99 \\
Inverse of Frisch elasticity & \(\varphi>0\) & 2 \\
Inverse of elasticity of intertemporal substitution & \(\gamma>0\) &
2 \\
Persistence of technology shocks & \(0<\rho_{a}<1\) & 0.95 \\
Persistence of monetary shocks & \(0<\rho_{m}<1\) & 0.50 \\
Taylor rule parameter & \(\phi>1\) & 1.50 \\
Price adjustment cost & \(\eta_{p}\geq0\) & 104.854 \\
Elasticity of substitution between intermediate inputs & \(\psi_{p}>1\)
& 9 \\
Wage adjustment cost & \(\eta_{w}\geq0\) & 524.272 \\
Elasticity of substitution between labor inputs & \(\psi_{w}>1\) &
4.5 \\
Ratio of type H consumers & \(0\leq\lambda<1\) & 0.2 \\
\end{longtable}

The model is calibrated at a quarterly frequency. The calibration sets
\(\beta=0.99\), \(\phi=1.5\), and \(\rho_m=0.50\), values that place the
discount factor and policy response in a conventional quarterly range
while avoiding the unusually persistent monetary shock used in the
earlier numerical illustration (Galí 2015). The elasticities
\(\psi_p=9\) and \(\psi_w=4.5\) imply desired gross price and wage
markups of 1.125 and 1.286, respectively. The calibration retains
\(\lambda=0.2\) so that the recalibration isolates nominal-rigidity and
policy parameters rather than selecting the household share to
strengthen the redistribution mechanism. This value is an illustrative
benchmark rather than an empirical estimate; the reported grid
\(\lambda\in\{0,0.2,0.4\}\) and Figure~\ref{fig-parameter-region} show
how the conclusions vary around it.

The adjustment-cost coefficients are disciplined by Calvo-equivalent
duration targets rather than chosen as free round numbers. Let \(D_j\)
denote the target Calvo mean duration and \(\theta_j=1-1/D_j\) the
corresponding non-adjustment probability. For prices, the Phillips-curve
slope is matched using \[
\eta_p=\frac{\psi_p\theta_p}{(1-\theta_p)(1-\beta\theta_p)}.
\] Setting \(D_p=4\) quarters gives \(\eta_p=104.854\); the sensitivity
analysis uses \(D_p\in\{2,4,6\}\), corresponding to
\(\eta_p\in\{17.822,104.854,257.143\}\). For wages, the mapping depends
on the labor-market structure. Using the Erceg--Henderson--Levin-type
correction discussed by Born and Pfeifer (2020) gives \[
\eta_w=\frac{\psi_w\theta_w(1+\psi_w\varphi)}{(1-\theta_w)(1-\beta\theta_w)}.
\] The benchmark \(D_w=4\) gives \(\eta_w=524.272\); \(D_w\in\{3,4,6\}\)
gives \(\eta_w\in\{264.706,524.272,1285.714\}\). The labels \(D_p\) and
\(D_w\) therefore calibrate the primitive cost coefficients
\((\eta_p,\eta_w)\); they are not literal adjustment durations in either
Rotemberg model, nor do they impose equal reduced-form Phillips-curve
slopes or equal economic rigidity across the two timing specifications.
These mappings provide a transparent scale for the numerical exercises,
but they do not microfound the aggregate-lag reference. The citations
above motivate that reference as a direct price-side analytical
benchmark and a reduced-form external wage norm, not as the unique
implication of an underlying contracting problem. Section 5.2 therefore
assesses the timing assumption directly.

The benchmark values are used for the numerical illustrations in Section
5; the analytical results do not depend on this calibration unless
explicitly stated. The restrictions \(\psi_p>1\) and \(\psi_w>1\) are
imposed in both the structural model and all recalibrated numerical
comparisons.

Table 1 is the aggregate-lag log-linear equilibrium system used in the
analytical derivations. It collects the household Euler equation, the
hand-to-mouth budget constraint, labor demand, markup definitions,
Phillips-curve equations, aggregation conditions, and the Taylor rule.

Two relationships carry most of the intuition. Price rigidity makes
monetary policy move profits and real wages in opposite directions.
Because profits accrue to savers while hand-to-mouth households consume
current labor income, this redistribution enters the aggregate Euler
equation through \(d_t\). At first order, firm dividends coincide with
the price markup: \[
d_t=\mu_t^p=a_t-w_t.
\] Price setting also implies \[
\pi_t^p=-(\psi_p/\eta_p)\mu_t^p,
\] and therefore \[
d_t=-(\eta_p/\psi_p)\pi_t^p.
\] This identity links price inflation to profit redistribution and, in
turn, to the heterogeneity term in the TANK IS curve derived in Section
4.

Several familiar cases are nested by parameter restrictions. Setting
\(\lambda=0\) gives a standard Representative Agent New Keynesian (RANK)
model. If, in addition, \(\eta_w=0\) or \(\psi_w\to\infty\), the model
reduces to a RANK economy with price rigidity alone. Alternatively, when
\(\lambda>0\) and either \(\eta_p=0\) or \(\psi_p\to\infty\), the
framework isolates wage stickiness effects within a heterogeneous agent
environment.

\section{Closed-Form Aggregate-Lag
Results}\label{closed-form-aggregate-lag-results}

This section develops three analytical results for the aggregate-lag
benchmark. First, aggregate price dynamics depend on price and wage
rigidities but not directly on the hand-to-mouth share. Second,
household heterogeneity enters the aggregate IS curve through profit
redistribution, which links price inflation to expected consumption
inequality. Third, the impact and infinite-horizon cumulative
monetary-policy multipliers show when this redistribution channel
amplifies policy transmission locally. The section then shows why the
same channel can be compressed into a higher price-inflation weight
under aggregate-lag adjustment; Section 5.2 establishes that this scalar
representation does not survive own-lag timing.

\subsection{TANK Phillips Curve}\label{tank-phillips-curve}

This subsection derives the TANK Phillips curve, which summarizes the
interaction between price rigidity and wage rigidity. Household
heterogeneity does not enter this aggregate price-setting equation
directly; it enters through the distributional consequences of price
movements and through the IS curve below. Combining the firms' and
households' optimality conditions gives
\begin{equation}\protect\phantomsection\label{eq-TPC}{
\pi_t^p
=
\frac{
\kappa_2 x_t+\pi_{t-1}^p-(\psi_p/\eta_p)\Delta a_t
}{\kappa_1},
}\end{equation} where \[
\begin{aligned}
x_t&= y_t - \frac{\varphi+1}{\varphi +\gamma} a_t,\\
\kappa_1&=1+\psi_p/\eta_p+\psi_w/\eta_w,\\
\kappa_2&=(\psi_p/\eta_p)(\psi_w/\eta_w)(\gamma+\varphi).
\end{aligned}
\]

Equation (\ref{eq-TPC}) is the TANK Phillips curve for the economy with
both price and wage rigidity. It follows from the aggregate
price-setting condition under the benchmark's aggregate-reference
adjustment cost. The output gap \(x_t\), defined as the log deviation of
output from its natural level, drives inflation through marginal cost.
Productivity growth enters the price Phillips curve as a negative direct
term. Holding \(x_t\) and \(\pi_{t-1}^p\) fixed, a positive
\(\Delta a_t\) lowers price inflation.

Both nominal rigidities, prices and wages, enter the coefficients.
Define price and wage flexibility as \((\psi_p/\eta_p,\ \psi_w/\eta_w)\)
and their inverses as rigidity \((\eta_p/\psi_p,\ \eta_w/\psi_w)\).
Effective rigidity is high when adjustment cost \(\eta\) is high or
substitution elasticity \(\psi\) is low. Lower \(\psi\) is equivalent to
higher market power (\(\psi/(\psi-1)\)) or less competitive markets.

The structural coefficients \(\kappa_1\) and \(\kappa_2\) each rise with
flexibility and fall with rigidity. Since \(\kappa_1\) and \(\kappa_2\)
are strictly positive, the Phillips-curve slope \((\kappa_2/\kappa_1)\)
remains positive, while the persistence measure \((1/\kappa_1)\) lies in
\((0,1)\). Greater flexibility strengthens the immediate inflation
response to the output gap and reduces inflation persistence; greater
rigidity does the opposite. The inverse Phillips-curve slope is \[
\frac{\kappa_1}{\kappa_2}=\frac{1}{\gamma+\varphi}
\left\{\frac{1}{\psi_p/\eta_p}+\frac{1}{\psi_w/\eta_w}
+\frac{1}{(\psi_p/\eta_p)(\psi_w/\eta_w)}\right\}.
\] Hence, as flexibility declines or rigidity increases,
\(\kappa_1/\kappa_2\) rises, the slope \(\kappa_2/\kappa_1\) falls, and
persistence \(1/\kappa_1\) increases.

The limiting cases isolate the two sources of nominal rigidity. With
wage flexibility (\(\eta_w=0\) or \(\psi_w\to\infty\)), price stickiness
alone gives \[
\pi_t^p=(\psi_p/\eta_p)(\varphi+\gamma)x_t.
\] Conversely, under price flexibility (\(\eta_p=0\) or
\(\psi_p\to\infty\)), wage rigidity gives \[
\pi_t^p=(\psi_w/\eta_w)(\varphi+\gamma)x_t-\Delta a_{t}.
\] Following Bilbiie and Trabandt (2025), when productivity shocks
vanish (\(a_t=0\)), these specifications become observationally
equivalent, which makes wage and price stickiness hard to distinguish
empirically.

The model also implies the following relations for profits, wages, and
wage inflation: \[
\begin{aligned}
d_{t}&=-\frac{\eta_{p}}{\psi_{p}}\pi_{t}^{p},\\
w_{t}&=a_{t}+\frac{\eta_{p}}{\psi_{p}}\pi_{t}^{p},
\end{aligned}
\]
\begin{equation}\protect\phantomsection\label{eq-wage-inflation2}{\pi_{t}^{w}=(1+\eta_{p}/\psi_{p})\pi_{t}^{p}-(\eta_{p}/\psi_{p})\pi_{t-1}^{p}+\Delta a_{t}.}\end{equation}

Under price rigidity, higher inflation lowers profits and raises real
wages, redistributing income across household types. This redistribution
remains present even without productivity shocks (\(a_t=0\)), breaking
the observational equivalence noted above. Only under price flexibility
(\(\eta_p=0\) or \(\psi_p\to\infty\)) do both profits and wages remain
neutral (\(d_t=w_t=0\)) in the absence of productivity disturbances.

Within the aggregate-lag benchmark, the TANK Phillips curve is
independent of the heterogeneity parameter \(\lambda\). Thus, for a
given output gap, aggregate inflation responds in the same way
regardless of the hand-to-mouth share. Heterogeneity matters elsewhere:
price movements still redistribute income, and the size of that
redistribution rises with price rigidity.

\subsection{TANK IS Curve}\label{tank-is-curve}

This subsection derives the heterogeneity-adjusted IS curve and shows
how liquidity constraints change the standard intertemporal
consumption-saving relation. The first step is to isolate the
distributional mechanism, which is the main departure from the
representative-agent model and the main building block of the TANK IS
curve.

\begin{quote}
\textbf{Proposition 1:} The type-specific allocations satisfy \[
\begin{aligned}
c_{t}^{H}   &=c_{t}-\delta_{c}d_{t},&
c_{t}^{S}   &=c_{t}+\frac{\lambda}{1-\lambda}\delta_{c}d_{t},\\
n_{t}^{H}   &=n_{t}+\delta_{n}d_{t},&
n_{t}^{S}   &=n_{t}-\frac{\lambda}{1-\lambda}\delta_{n}d_{t},\\
w_{t}^{H}   &=w_{t}-\delta_{w}d_{t},&
w_{t}^{S}   &=w_{t}+\frac{\lambda}{1-\lambda}\delta_{w}d_{t},
\end{aligned}
\] where \[
\begin{aligned}
\delta_{c}&=\frac{1+(\eta_{w}/\psi_{w})+\varphi\psi_{w}}
{1+(\eta_{w}/\psi_{w})+\varphi\psi_{w}+\gamma(\psi_{w}-1)},\\
\delta_{n}&=\frac{\gamma\psi_{w}}
{1+(\eta_{w}/\psi_{w})+\varphi\psi_{w}+\gamma(\psi_{w}-1)},\\
\delta_{w}&=\frac{\gamma}
{1+(\eta_{w}/\psi_{w})+\varphi\psi_{w}+\gamma(\psi_{w}-1)}.
\end{aligned}
\] Moreover, \[
c_t^S-c_t^H=\frac{\delta_c}{1-\lambda}d_t,\qquad
n_t^H-n_t^S=\frac{\delta_n}{1-\lambda}d_t,\qquad
w_t^S-w_t^H=\frac{\delta_w}{1-\lambda}d_t,
\] and \[
\delta_c+\delta_n-\delta_w=1.
\]
\end{quote}

Proposition 1 shows that profits are the sufficient statistic for the
distributional gaps in consumption, labor, and wages under aggregate-lag
wage adjustment. In this sense, \(d_t\) measures the redistribution
generated by nominal rigidities. When profits fall, the redistribution
across household types appears immediately in type-specific consumption,
labor, and wages.

The coefficients \((\delta_c,\delta_n,\delta_w)\) determine how that
redistribution is allocated across consumption, labor, and wages. The
factor \(1/(1-\lambda)\) scales the cross-sectional gaps because savers
absorb all firm profits. The derivation is delegated to the Online
Appendix.

\begin{quote}
\textbf{Corollary 1:} The heterogeneity term in the aggregate Euler
equation can be written as \[
\lambda(c_t^S-c_t^H)=\frac{\lambda}{1-\lambda}\delta_c d_t=-\delta_p\pi_t^p.
\]
\end{quote}

Corollary 1 shows that the term \(-\delta_p E_t\Delta\pi_{t+1}^p\) in
the TANK IS curve can be read as the expected change in consumption
inequality, not merely as a reduced-form inflation term.

\begin{quote}
\textbf{Corollary 2:} The comparative statics of the distributional
coefficients are simple. The coefficient \(\delta_c\) increases with
wage adjustment costs \(\eta_w\), whereas \(\delta_n\) and \(\delta_w\)
decrease with \(\eta_w\). With respect to labor-market competitiveness
\(\psi_w\), \(\delta_c\) decreases. The sign of
\(\partial\delta_n/\partial\psi_w\) depends on
\(1-\gamma+2\eta_w/\psi_w\), while \(\delta_w\) is non-monotone in
\(\psi_w\).
\end{quote}

These comparative statics show how labor-market structure shapes the
distributional channel. Greater wage rigidity shifts the redistribution
margin toward consumption: \(\delta_c\) rises, whereas \(\delta_n\) and
\(\delta_w\) fall. Greater labor-market competitiveness lowers the
consumption component \(\delta_c\), while the labor and wage components
depend on the remaining parameters. These properties will be used
repeatedly in Propositions 2 and 3 below, and their proof is reported in
Appendix A.

From the Type S Euler equation and the price-setting relation
\(d_t=-(\eta_p/\psi_p)\pi_t^p\), the TANK IS curve is
\begin{equation}\protect\phantomsection\label{eq-TIS}{
r_{t}=r_{t}^{f}+\gamma(\mathbb{E}_{t}\Delta x_{t+1}-\delta_{p}\mathbb{E}_{t}\Delta\pi_{t+1}^{p}),
}\end{equation} where \[
\begin{aligned}
r_{t}^{f}&=\gamma\frac{\varphi+1}{\varphi+\gamma}\mathbb{E}_{t}\Delta a_{t+1},\\
\delta_{p}&=\frac{\lambda}{1-\lambda}\frac{\eta_p}{\psi_p}\delta_c.
\end{aligned}
\] The natural interest rate \(r_t^f\) is the equilibrium real rate that
would prevail under complete price and wage flexibility
(\(\eta_p/\psi_p=\eta_w/\psi_w=0\)).

The aggregate Euler equation includes the term \(\delta_p\), which
scales with heterogeneity \((\lambda)\) and relative price stickiness
\((\eta_p/\psi_p)\). Intuitively, higher expected inflation signals
future markup compression and wage gains. Agents internalize this
redistribution, so the Euler equation loads on inflation expectations
and the demand response changes. This ``who gains/who loses'' loop
underlies the AD slope, policy multipliers, and welfare results.

The equation collapses to the standard RANK IS curve when
\(\delta_p=0\), so the heterogeneity term disappears. Under pure wage
stickiness (\(\eta_p/\psi_p=0\)), \(\delta_p\) also vanishes.\footnote{In
  the aggregate-lag benchmark, what vanishes under pure wage stickiness
  is the heterogeneity term in the aggregate IS relation, namely
  \(\delta_p = 0\), not the coefficient \(\delta_c\) itself. Since the
  type-specific gaps in Proposition 1 are all proportional to \(d_t\),
  they disappear when price flexibility implies \(d_t = 0\). Under
  own-lag wage adjustment, an inherited relative-wage state can remain
  even when \(d_t=0\). Starting from the symmetric steady state,
  however, an aggregate shock does not create a new relative-wage
  response in the pure-wage-sticky limit.} Moreover, price flexibility
implies \(d_t=0\), so the type-specific gaps disappear as well.

Related analytical TANK IS curves can be recovered as special cases
under tighter distributional assumptions.\footnote{In particular, Ascari
  et al. (2017) use the Colciago framework with \(\delta_c=1\), so the
  real wage gap becomes the relevant transmission margin. Bilbiie and
  Känzig (2023) further specialize to \(\delta_c=1\) and \(a_t=0\),
  obtaining an IS relation in which expected consumption inequality can
  be written as
  \(-\frac{\lambda}{1-\lambda}\{\mathbb{E}_{t}\pi_{t+1}^{p}-\mathbb{E}_{t}\pi_{t+1}^{w}\}\).
  Those formulations reproduce the aggregate dynamics here under tighter
  assumptions such as common wages or common labor supply across
  household types, but they imply different welfare weights because the
  underlying distributional structure is more restrictive.}

Using (\ref{eq-TPC}), the TANK IS curve (\ref{eq-TIS}) can be rewritten
as follows. The corrected Phillips curve implies \[
\kappa_1\mathbb{E}_t\Delta\pi_{t+1}^p
=
\Delta\pi_t^p
+\kappa_2\mathbb{E}_t\Delta x_{t+1}
-(\psi_p/\eta_p)\mathbb{E}_t\Delta^2a_{t+1},
\] and \(d_t=-(\eta_p/\psi_p)\pi_t^p\) implies
\(\Delta\pi_t^p=-(\psi_p/\eta_p)\Delta d_t\). Therefore \[
\begin{aligned}
r_t
&=
r_t^f
+
\tilde{\gamma}\mathbb{E}_t\Delta x_{t+1}
+
\frac{\gamma\lambda\delta_c}{(1-\lambda)\kappa_1}
\left(
\Delta d_t+\mathbb{E}_t\Delta^2a_{t+1}
\right),\\
\tilde{\gamma}
&=
\gamma
\left(
1-\frac{\delta_p\kappa_2}{\kappa_1}
\right)\\
&=
\gamma\left(
1-
\frac{\lambda}{1-\lambda}
\frac{(\gamma+\varphi)(\psi_w/\eta_w)\delta_c}
{1+\psi_p/\eta_p+\psi_w/\eta_w}
\right).
\end{aligned}
\] The literature commonly interprets \(\tilde{\gamma}\) as the
effective slope of the IS relationship under heterogeneous agents. As
long as \(\delta_p\kappa_2/\kappa_1<1\), the slope is negative.

With no wage adjustment cost (\(\eta_w=0\)), the coefficient
\(\kappa_1\) approaches infinity, and the IS relation becomes
\[\begin{aligned}
r_{t}&=r_{t}^{f}+\tilde{\gamma}\mathbb{E}_{t}\Delta x_{t+1},\\
\tilde{\gamma}&=\gamma\left(1-\frac{\lambda}{1-\lambda}(\gamma+\varphi)\delta_{c}\right).
\end{aligned}\] Price rigidity (\(\eta_p/\psi_p\)) does not affect the
IS slope \(\tilde{\gamma}\) in this case; the slope depends only on
heterogeneity \((\lambda)\) and labor-market structure \((\psi_w)\).
This specification nests existing results as special cases:
\(\tilde{\gamma}=\gamma\left(1-\frac{\lambda}{1-\lambda}\varphi\right)\)
under perfect labor market competition (\(\psi_{w}\to\infty\)), matching
Bilbiie (2008). As a limiting log-linear comparison, complete labor
market segmentation (\(\psi_{w}\to1\)) yields
\(\tilde{\gamma}=\gamma\left(1-\frac{\lambda}{1-\lambda}(\varphi+\gamma)\right)\),
consistent with Ascari et al. (2017). Since \(\tilde{\gamma}\) decreases
with heterogeneity (\(\lambda\)), the IS curve becomes flatter and may
even turn inverted.

With wage rigidity present (\(\eta_w\neq 0\)) and complete segmentation
interpreted as the limiting case \(\psi_{w}\to1\), the effective slope
becomes \[
\tilde{\gamma}=\gamma\left(1-\frac{\lambda(\gamma+\varphi)}{1-\lambda}\frac{1/\eta_{w}}{1+\psi_{p}/\eta_{p}+1/\eta_{w}}\right),\]
which increases monotonically with wage adjustment cost \(\eta_w\). This
result aligns with Proposition 1 in Ascari et al. (2017): stronger wage
rigidity prevents the IS curve from becoming inverted.

\subsection{TANK IS-PC-MP Analysis}\label{tank-is-pc-mp-analysis}

This subsection studies monetary transmission in the full TANK system,
setting productivity shocks to zero (\(a_t=0\)). The system combines
three equations: the heterogeneity-adjusted IS curve (\ref{eq-TIS}), the
TANK Phillips curve (\ref{eq-TPC}), and the Taylor rule in Table 1.
Solving them jointly shows how household heterogeneity changes standard
New Keynesian dynamics.

With \(a_t=0\), the solution is \[
\begin{aligned}
\pi_{t}^{p}&=\xi_{1}\pi_{t-1}^{p}+\Omega^{p}m_{t},\\
x_{t}&=\frac{\xi_{1}\kappa_{1}-1}{\kappa_{2}}\pi_{t-1}^{p}+\Omega^{x}m_{t},
\end{aligned}
\] The key objects are the impact multipliers: \[
\begin{aligned}
\Omega^{p}&:=\frac{d\pi_t^p}{dm_t}=\frac{1}{(\kappa_{1}/\kappa_{2})\gamma+(1-\gamma\delta_{p})}\frac{1}{\xi_{2}-\rho_{m}},\\
\Omega^{x}&:=\frac{dx_t}{dm_t}=\frac{\kappa_{1}}{\kappa_{2}}\Omega^{p}=\frac{1}{\gamma+(\kappa_{2}/\kappa_{1})(1-\gamma\delta_{p})}\frac{1}{\xi_{2}-\rho_{m}},
\end{aligned}
\] and \(\xi_{1}\) and \(\xi_{2}\) are the roots of the following
characteristic function: \[
f(\xi)  =\left(1-\gamma\delta_{p}+\frac{\gamma\kappa_{1}}{\kappa_{2}}\right)\xi^{2}-\left(\phi-\gamma\delta_{p}+\frac{\gamma\kappa_{1}}{\kappa_{2}}+\frac{\gamma}{\kappa_{2}}\right)\xi+\frac{\gamma}{\kappa_{2}}.
\] Appendix A provides the detailed derivation.

The analysis assumes
\begin{equation}\protect\phantomsection\label{eq-AD-condition}{
1-\gamma\delta_{p}+\gamma\kappa_{1}/\kappa_{2}>0
,
}\end{equation} ensuring that the characteristic polynomial opens
upward. This is the baseline determinacy condition for the dynamic
IS-PC-MP system. It is weaker than the condition used below for a
downward-sloping AD curve and distinct from the root condition that
signs the impact effect of heterogeneity. Under this assumption, since
\(f(0)\geq 0\) and \(f(1)
<0\), the roots satisfy \(0\leq \xi_{1}<1\) and \(\xi_{2}>1\), which
gives a unique saddle-path solution. The assumption
(\ref{eq-AD-condition}) holds in the absence of heterogeneity
(\(\lambda = 0\)). As discussed in the previous subsections, the slope
of the IS curve is negative when \(\delta_p\kappa_2/\kappa_1<1\). The
negative slope condition ensures (\ref{eq-AD-condition}).

The response coefficients \(\Omega^p>0\) and \(\Omega^x>0\) measure the
immediate effect of a one-time monetary shock (\(e_{t}^{m}=1\),
\(e_{s}^{m}=0\) for \(s>t\)). They depend on the characteristic roots,
which in turn depend on structural parameters. The comparative statics
below account for these interactions.

Start with the benchmark case of pure price rigidity. Under wage
flexibility (\(\eta_{w}/\psi_{w}=0\), equivalently
\(\psi_{w}/\eta_{w}\to\infty\)), the Phillips curve simplifies to
\(\pi_{t}^{p}=(\psi_{p}/\eta_{p})(\varphi+\gamma)x_{t}\), and the policy
multipliers are \[
\begin{aligned}
\xi_{2}&=\frac{(\varphi+\gamma)(\phi-\gamma\delta_{p})+\gamma\eta_{p}/\psi_{p}}{(\varphi+\gamma)(1-\gamma\delta_{p})+\gamma\eta_{p}/\psi_{p}},\\
\Omega^p&=\frac{\varphi+\gamma}
{(\varphi+\gamma)(\phi-\rho_{m})+(1-\rho_{m})\{1-(\varphi+\gamma)\delta_{c}\lambda/(1-\lambda)\}\gamma\eta_{p}/\psi_{p}},\\
\Omega^x&=\frac{\eta_p/\psi_p}
{(\varphi+\gamma)(\phi-\rho_{m})+(1-\rho_{m})\{1-(\varphi+\gamma)\delta_{c}\lambda/(1-\lambda)\}\gamma\eta_{p}/\psi_{p}}.
\end{aligned}
\] With a positive denominator and \(\phi>\rho_m\), price rigidity
(\(\eta_p/\psi_p\)) amplifies the output response \(\Omega^x\). It
reduces the inflation response \(\Omega^p\) provided
\(1-(\gamma+\varphi)\delta_c\lambda/(1-\lambda)>0\); when this term is
negative, the inflation effect is reversed. In this pure-price-rigidity
benchmark, greater heterogeneity and lower labor-market competitiveness
can strengthen monetary transmission when the denominator remains
positive. In the general sticky-price and sticky-wage case, the full
comparative statics are characterized in Proposition 2.

Under pure wage rigidity (\(\eta_{p}/\psi_{p}=0\)), \(\delta_p\)
vanishes, collapsing the aggregate-lag framework to its RANK equivalent
where the standard equivalence result applies. Wage stickiness
parameters (\(\eta_w/\psi_w\)) have opposing effects: they dampen the
inflation response \(\Omega^p\) and amplify the output response
\(\Omega^x\).

The general result is as follows.

\begin{quote}
\textbf{Proposition 2:} Let \(\Omega^p\) and \(\Omega^x\) denote the
impact multipliers of a one-time monetary policy shock on price
inflation and the output gap, respectively.

\begin{enumerate}
\def\labelenumi{\arabic{enumi}.}
\item
  Both \(\Omega^p\) and \(\Omega^x\) decrease with the monetary policy
  response coefficient \(\phi\).
\item
  Let \[
  Z:=\phi+\gamma\frac{\kappa_{1}-1}{\kappa_{2}}-\gamma\delta_{p},
  \qquad
  D:=Z^2+4(\phi-1)\frac{\gamma}{\kappa_2}.
  \] Suppose
  \begin{equation}\protect\phantomsection\label{eq-condition0}{
  \rho_m<\frac12\left(1+\frac{Z}{\sqrt D}\right).
  }\end{equation} Then, locally around parameter values satisfying this
  condition, \(\Omega^p\) and \(\Omega^x\) increase with the share of
  hand-to-mouth households \(\lambda\). If, in addition, \[
  1-\gamma\delta_p>0,
  \] then, locally and for interior rigidity ratios \(\eta_j/\psi_j>0\),
  \(\Omega^x\) decreases with goods- and labor-market competitiveness
  parameters \(\psi_j\) and increases with adjustment-cost parameters
  \(\eta_j\), for \(j \in \{p, w\}\); the corresponding boundary cases
  are understood by continuity.
\item
  Without heterogeneity (\(\lambda = 0\)), condition
  (\ref{eq-condition0}) is not required for the rigidity comparative
  statics in part 2. In addition, \(\Omega^p\) increases with \(\psi_j\)
  and decreases with \(\eta_j\) for \(j \in \{p, w\}\).
\end{enumerate}
\end{quote}

Appendix A contains the proof. The economic intuition is easiest to see
through an aggregate demand-aggregate supply (AD-AS) lens. The TANK
Phillips curve (\ref{eq-TPC}) serves as the aggregate-supply curve. With
the stronger assumption \(1-\gamma \delta_p>0\), the AD curve, derived
by combining the TANK IS curve (\ref{eq-TIS}) with the Taylor rule in
Table 1,\footnote{The expression follows by forward iteration of the
  combined TANK IS curve and monetary policy rule. Appendix A provides
  the corresponding algebra.} is \[
\pi_{t}^{p}=\frac{\gamma\mathbb{E}_{t}\sum_{n=0}^\infty\left(\frac{1-\gamma \delta_{p}}{\phi-\gamma \delta_{p}}\right )^{n}\Delta x_{t+1+n}}{\phi-\gamma \delta_{p}}+\frac{m_{t}}{\phi-\rho_m-\gamma \delta_{p}(1-\rho_{m})}.
\] This assumption ensures that the AD curve slopes downward in
\((x_t,\pi_t^p)\) space. With \(\eta_w/\psi_w=0\), the AD curve becomes
\[
\pi_{t}^{p}=\frac{\gamma(\rho_m-1)x_t+m_{t}}{\phi-\rho_m-\gamma \delta_{p}(1-\rho_{m})}.
\]

An expansionary monetary policy shock shifts this AD curve to the right.
The resulting equilibrium depends on the slopes of both curves, which
are shaped by heterogeneity and rigidities.

Structural parameters affect the AD-AS system through different margins.
Higher nominal rigidities, represented by increases in \(\eta_p\) or
\(\eta_w\), flatten the AS curve. This is the standard New Keynesian
force: stickier prices and wages dampen the inflation response to an
aggregate demand shock and require a larger output adjustment. As a
direct AS-side effect, the inflation multiplier \(\Omega^p\) falls while
the output multiplier \(\Omega^x\) rises.

Household heterogeneity, in contrast, operates primarily through the AD
channel. A larger share of hand-to-mouth households (\(\lambda\))
increases the parameter \(\delta_p\), which magnifies the coefficient on
the monetary shock \(m_t\) in the AD curve. Intuitively, the
policy-induced distributional shift from profits to wages has a larger
net effect on aggregate consumption when more households have a high
marginal propensity to consume. This produces a larger rightward shift
of the AD curve and puts upward pressure on both output and inflation.

These channels interact to determine the overall effectiveness of
monetary policy. While the flatter AS curve from nominal rigidities
works to moderate inflation, the amplified AD shift from household
heterogeneity works to increase it. Under the root condition in
Proposition 2, the \(\lambda\) comparative static is signed: a larger
hand-to-mouth share raises both \(\Omega^p\) and \(\Omega^x\) locally.
The ambiguous object is different. For changes in rigidity parameters,
the direct AS effect and the indirect heterogeneity effect can offset
each other, so the total effect on \(\Omega^p\) is not signed in the
heterogeneous case.

The impact heterogeneity result requires the additional condition
(\ref{eq-condition0}), which depends on monetary-shock persistence and
the characteristic roots. This condition may be violated when price and
wage rigidities become sufficiently large, a scenario explored
numerically in the next section.

The analysis now turns to infinite-horizon cumulative monetary
transmission. With \(\pi_{t-1}^p=0\), \[
\begin{aligned}
\Omega_{\infty}^{p} &:=\sum_{n=0}^{\infty}\frac{d\pi_{t+n}^{p}}{dm_{t}}=\Omega^p\sum_{n=0}^{\infty} \sum_{i=0}^{n} \xi_1^{n-i} \rho_m^{i},\\
\Omega_{\infty}^{x} &:=\sum_{n=0}^{\infty}\frac{dx_{t+n}}{dm_{t}}=\frac{\xi_1\kappa_1-1}{\kappa_2}\Omega^{\infty}_{p}+\frac{\Omega^x}{1-\rho_m}.
\end{aligned}
\] These objects measure infinite-horizon cumulative effects of a
monetary shock on inflation and output.

The infinite-horizon cumulative inflation response \(\Omega^p_\infty\)
satisfies the following characterization:

\begin{quote}
\textbf{Proposition 3:} Let \(\Omega^p_\infty\) and \(\Omega^x_\infty\)
denote the infinite-horizon cumulative multipliers of a one-time
monetary policy shock on price inflation and the output gap,
respectively. They are given by: \[\begin{aligned}
\Omega_{\infty}^{p}&=
\frac{1}{(1-\xi_{1})(1-\rho_{m})}\Omega^p,\\
\Omega_{\infty}^{x}&=
\frac{\eta_{p}/\psi_{p}+\eta_{w}/\psi_{w}}{\varphi+\gamma}\Omega_{\infty}^{p},
\end{aligned}\] indicating that:

\begin{enumerate}
\def\labelenumi{\arabic{enumi}.}
\tightlist
\item
  \(\Omega^p_\infty\) and \(\Omega^x_\infty\) decrease with the monetary
  policy response coefficient \(\phi\).
\item
  \(\Omega^p_\infty\) and \(\Omega^x_\infty\) increase locally with the
  share of hand-to-mouth households \(\lambda\), as long as the
  determinacy condition (\ref{eq-AD-condition}) is maintained.
\end{enumerate}
\end{quote}

Appendix A provides the proof. The restrictive condition
(\ref{eq-condition0}) is not needed for this local infinite-horizon
cumulative result. Thus heterogeneity can raise infinite-horizon
cumulative responses even when the impact response is not signed by
Proposition 2.

Under the stronger condition \(1-\gamma\delta_p>0\), higher adjustment
costs and less market competitiveness have the standard direct AS-side
effects: they reduce impact inflation responses and amplify impact
output responses. In the heterogeneous case, the total impact inflation
response remains ambiguous. With respect to adjustment-cost and
competitiveness parameters, however, infinite-horizon cumulative
inflation effects are also theoretically ambiguous because rigidity,
shock persistence, and heterogeneity channels move in different
directions. Output effects face similar ambiguity because
\(\Omega_\infty^x\) combines the cumulative inflation response with the
rigidity factor \[
\frac{\eta_p/\psi_p+\eta_w/\psi_w}{\varphi+\gamma}.
\] At this point the closed-form expressions no longer give clean signs,
so the next section studies these parameter interactions numerically.

\subsection{Welfare Accounting}\label{welfare-accounting}

This subsection first states a CES-consistent welfare objective that
applies to both adjustment-cost timings and then specializes it to the
aggregate-lag benchmark. The same profit-redistribution channel that
affects monetary multipliers also changes the stabilization objective.
Because price inflation moves profits and real wages in opposite
directions, it generates cross-household consumption, hours, and wage
gaps that are socially costly. The welfare result comes from a
second-order logarithmic approximation to household utilities.

Starting from household utility and taking a second-order approximation
around the efficient steady state, period-\(t\) welfare is \[
\begin{aligned}
\mathbb{W}=&\lambda c^{H}_{t}+(1-\lambda)c^{S}_{t}+\frac{1-\gamma}{2}\{\lambda(c^{H}_{t})^{2}+(1-\lambda)(c^{S}_{t})^{2}\}\\&-\lambda n^{H}_{t}-(1-\lambda)n^{S}_{t}-\frac{1+\varphi}{2}\{\lambda(n^{H}_{t})^{2}+(1-\lambda)(n^{S}_{t})^{2}\}.
\end{aligned}
\] This expression contains both the direct utility effects and the
second-order terms that matter for welfare. Let
\(\sigma_t^q:=q_t^S-q_t^H\) for \(q\in\{c,n,w\}\). After CES
aggregation, welfare can be written as
\begin{equation}\protect\phantomsection\label{eq-general-welfare}{
\begin{aligned}
\mathbb W_t=-\frac12\Bigg[&
(\gamma+\varphi)x_t^2+\eta_p(\pi_t^p)^2+\eta_w(\pi_t^w)^2\\
&+\lambda(1-\lambda)\left\{
\gamma(\sigma_t^c)^2
+\left(\varphi+\frac1{\psi_w}\right)(\sigma_t^n)^2
+\eta_w(\pi_t^S-\pi_t^H)^2
\right\}\Bigg]+t.i.p.
\end{aligned}
}\end{equation} where \(t.i.p.\) denotes terms independent of the
endogenous variables. The term \(1/\psi_w\) is the second-order
contribution of the CES labor aggregator to hours dispersion. The
derivation is reported in the Online Appendix.

In the aggregate-lag benchmark, Proposition 1 and the common lagged wage
reference imply that all three distributional gaps are proportional to
profits. Substitution into (\ref{eq-general-welfare}) gives \[
\begin{aligned}
\mathbb{W}_{t}=&-\frac{\gamma+\varphi}{2}x_{t}^{2}-\frac{\eta_{w}}{2}(\pi_{t}^{w})^{2}-\frac{\eta_{p}}{2}(\pi_{t}^{p})^{2}\\&-\frac{1}{2}\frac{\lambda}{1-\lambda}\left\{\gamma\delta_{c}^{2}+\left(\varphi+\frac{1}{\psi_w}\right)\delta_{n}^{2}+\eta_{w}\delta_{w}^{2}\right\}d_{t}^{2}+t.i.p.,
\end{aligned}\]

Using \(\pi_{t}^{p}=-(\psi_{p}/\eta_{p})d_{t}\), this expression
simplifies to \[
\begin{aligned}
\mathbb{W}_{t}&=-\frac{\gamma+\varphi}{2}x_{t}^{2}-\frac{\eta_{w}}{2}(\pi_{t}^{w})^{2}-\frac{\tilde{\eta}_{p}}{2}(\pi_{t}^{p})^{2}+t.i.p.,\\
\tilde{\eta}_{p}&=\eta_{p}+\frac{\lambda}{1-\lambda}\left(\eta_{p}/\psi_{p}\right)^{2}\left\{\gamma\delta_{c}^{2}+\left(\varphi+\frac{1}{\psi_w}\right)\delta_{n}^{2}+\eta_{w}\delta_{w}^{2}\right\}.
\end{aligned}
\] The representative-agent case is nested at \(\lambda=0\), where
\(\tilde{\eta}_p=\eta_p\). Within the aggregate-lag benchmark, household
heterogeneity therefore raises the coefficient on price volatility
relative to the standard RANK objective. At the benchmark primitives,
\(\tilde{\eta}_p\) rises from 104.854 at \(\lambda=0\) to 170.126 at
\(\lambda=0.2\) and 278.913 at \(\lambda=0.4\), increases of 62.3 and
166.0 percent relative to the representative-agent coefficient.

Hand-to-mouth households (\(\lambda>0\)) cannot smooth consumption
across periods, so they are especially exposed to income-redistribution
shocks, reflected in the \(\delta_c d_t\) and \(\delta_n d_t\) terms.
Since price inflation \(\pi_t^p\) maps one-for-one into the wage-profit
distribution in this benchmark, all three dispersion terms can be
compressed into \(\tilde\eta_p\). This algebra explains why the
benchmark inflation penalty exceeds its RANK counterpart; it is not a
general statement that heterogeneous-agent welfare has a single
additional stabilization target. Note that \[
\begin{aligned}
&(\tilde{\eta}_{p}-\eta_{p})(\pi_{t}^{p})^{2}\\
&=\lambda(1-\lambda)\left\{ \gamma(c_{t}^{S}-c_{t}^{H})^{2}+\left(\varphi+\frac{1}{\psi_w}\right)(n_{t}^{S}-n_{t}^{H})^{2}+\eta_{w}(\pi_t^S-\pi_t^H)^2\right\}\\
&=\lambda(1-\lambda)\left\{ \gamma(c_{t}^{S}-c_{t}^{H})^{2}+\left(\varphi+\frac{1}{\psi_w}\right)(n_{t}^{S}-n_{t}^{H})^{2}+\eta_{w}(w_{t}^{S}-w_{t}^{H})^{2}\right\}.
\end{aligned}
\] The aggregate-lag price-volatility coefficient rises with the extent
of cross-household inequality.

Appendix B reports the discretionary and commitment policy problems
implied by this aggregate-lag objective. These results characterize the
benchmark optimal allocation rather than an implementable interest-rate
rule or an own-lag policy problem. Section 5.2 shows that conventional
own-lag adjustment leaves consumption, hours, and relative wage
inflation as separate history-dependent welfare terms, so
\(\tilde\eta_p\) cannot summarize that model.

\section{Quantitative Implications}\label{quantitative-implications}

This section uses local impulse responses for three purposes. First, it
illustrates the comparative statics in Proposition 2 under the benchmark
calibration in Table~\ref{tbl-parameters}. Second, it examines how the
results change over empirically interpretable Calvo-equivalent price-
and wage-duration targets. Third, it checks the aggregate-lag
adjustment-cost benchmark against a conventional own-lag model with
common primitive parameters. To keep the main text focused, the main
analysis concentrates on monetary transmission and moves
technology-shock results and additional robustness exercises to Appendix
C. All vertical axes are percent deviations from the steady state,
except for the interest rate panel, which is in percentage points. The
aggregate-lag figures are deterministic simulations of the closed-form
model, and the own-lag results solve the full type-specific
expectational system numerically. They are calibrated model comparisons
rather than statistical estimates, so no sampling uncertainty is
involved.

The numerical exercises should be read together with the sufficient
conditions in Proposition 2. The benchmark calibration satisfies the
impact root condition for the local \(\lambda\) comparative static. For
the \(\lambda\) comparison with \(D_p=4\), the right-hand side of
condition (\ref{eq-condition0}) equals 0.936, 0.926, and 0.905 for
\(\lambda=0,0.2,0.4\), respectively, all above \(\rho_m=0.50\). With the
longer price-duration target \(D_p=6\), the corresponding values are
0.894, 0.862, and 0.775, so the root condition also holds in all three
cases. If the previous \(\rho_m=0.85\) is retained as a persistence
check, the condition still holds in five of the six cases; only the
joint extreme \(D_p=6\), \(\lambda=0.4\) falls outside it. The
determinacy coefficient is positive and the roots satisfy
\(0\leq\xi_1<1<\xi_2\) in all six cases. At the benchmark
\(\lambda=0.2\), however, \(1-\gamma\delta_p=-4.520<0\), even though the
effective-IS factor \(1-\delta_p\kappa_2/\kappa_1=0.993>0\).
Consequently, the \(\lambda\) comparisons can be interpreted using
Proposition 2, but the stronger sufficient condition used to sign the
rigidity comparative static of \(\Omega^x\) is not satisfied under the
duration-based calibration. The rigidity figures below are therefore
numerical comparisons rather than applications of that particular
sufficient-condition result.

\begin{figure}

\centering{

\includegraphics[width=1\linewidth,height=\textheight,keepaspectratio]{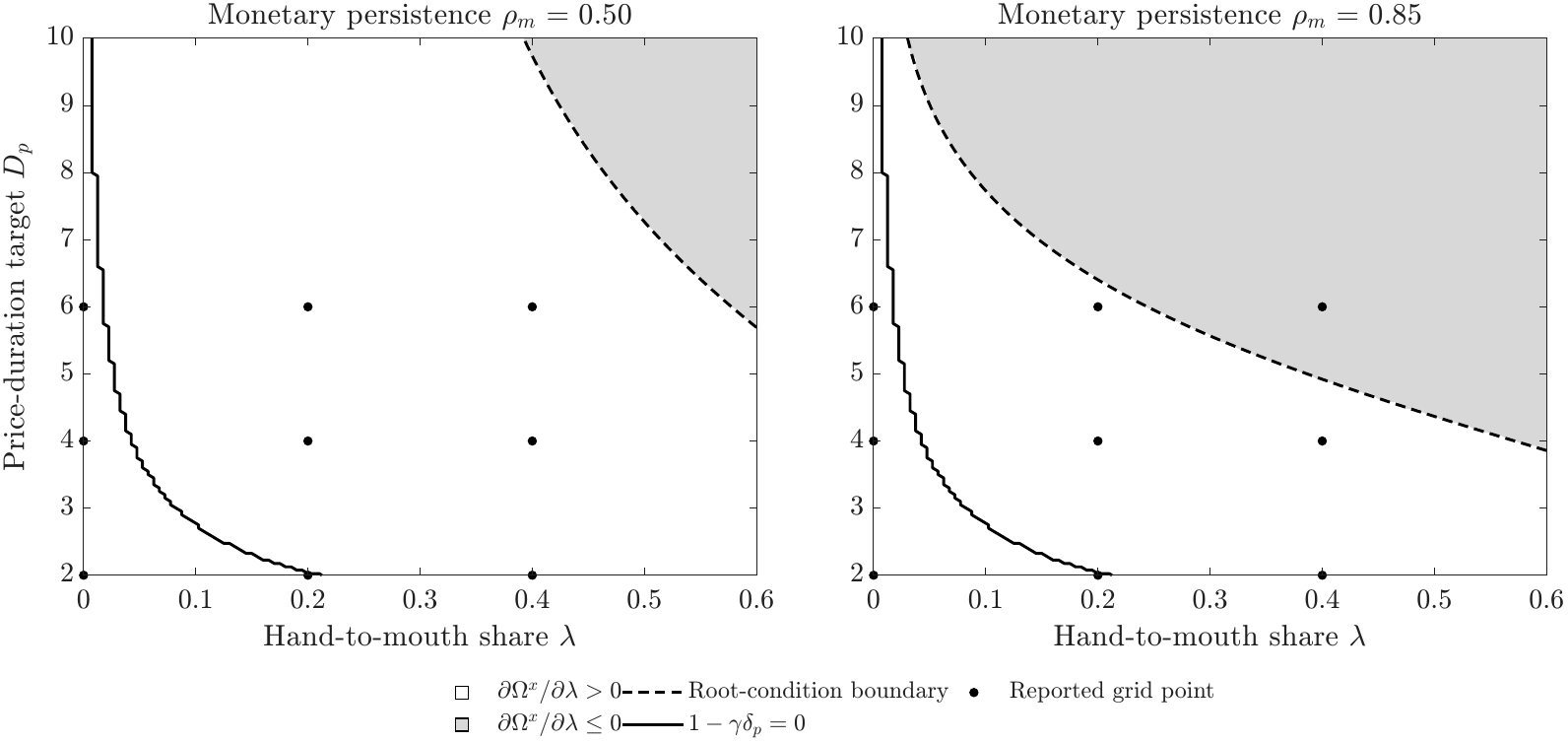}

\par\smallskip
\begin{minipage}{\linewidth}
\footnotesize \textit{Notes:} The figure varies the hand-to-mouth share $\lambda$ and the price-duration target $D_p$, holding $D_w=4$, $\beta=0.99$, $\gamma=\varphi=2$, $\phi=1.5$, $\psi_p=9$, and $\psi_w=4.5$. White denotes $\partial\Omega^x/\partial\lambda>0$ and gray denotes a non-positive derivative. The dashed line is the boundary of condition (\ref{eq-condition0}); the solid line is $1-\gamma\delta_p=0$; black dots mark $\lambda\in\{0,0.2,0.4\}$ and $D_p\in\{2,4,6\}$. Every point in the displayed domain satisfies determinacy. The derivative sign is computed from the analytical root condition and checked against finite differences---centered at interior values and forward at the boundary $\lambda=0$.
\end{minipage}

}

\caption{\label{fig-parameter-region}Parameter regions for impact
amplification}

\end{figure}

Figure~\ref{fig-parameter-region} maps these qualifications beyond the
reported duration grid. The white region satisfies the local root
condition for \(\partial\Omega^x/\partial\lambda>0\), whereas the gray
region does not; the dashed curve traces the analytical root-condition
boundary. All points in the displayed domain are determinate. At the
benchmark persistence \(\rho_m=0.50\), all nine reported
\((\lambda,D_p)\) grid points lie in the amplification region. At
\(\rho_m=0.85\), the boundary moves inward and the reported case
\((\lambda,D_p)=(0.4,6)\) lies outside it; at that point the
root-condition bound is \(0.775\). The solid curve shows that the
stronger condition \(1-\gamma\delta_p>0\) holds on a smaller subset of
the parameter space and is not needed to sign the local \(\lambda\)
effect.

\subsection{Monetary Shocks in the Aggregate-Lag
Benchmark}\label{monetary-shocks-in-the-aggregate-lag-benchmark}

\begin{figure}

\centering{

\includegraphics[width=0.9\linewidth,height=\textheight,keepaspectratio]{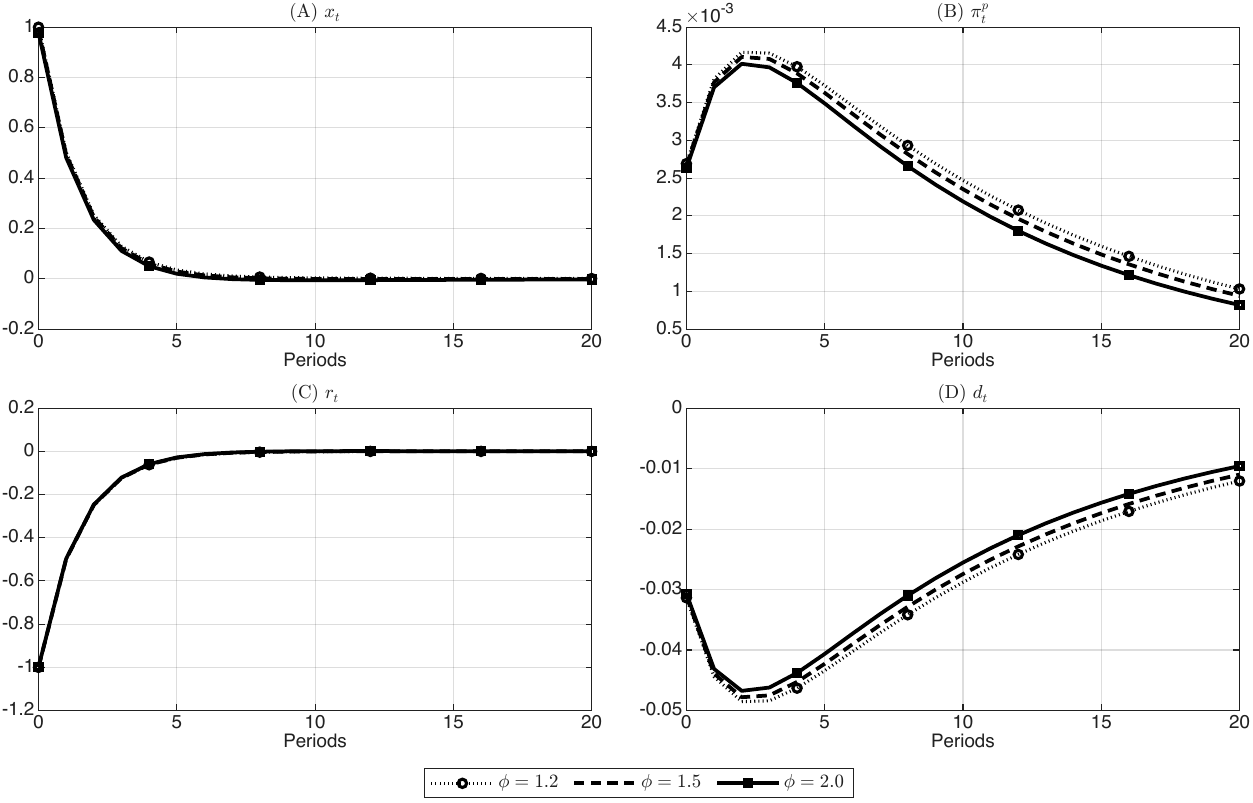}

\par\smallskip
\begin{minipage}{0.90\linewidth}
\footnotesize \textit{Notes:} Panels report $(A)$ $x_t$, $(B)$ $\pi_t^p$, $(C)$ $r_t$, and $(D)$ $d_t$. The dotted, dashed, and solid lines correspond to $\phi=1.2$, $\phi=1.5$, and $\phi=2.0$. Responses are to a one-percent monetary-policy innovation; vertical axes are percent deviations from steady state, except for the real interest rate, which is in percentage points, and $d_t$, which is the profit share scaled by steady-state output.
\end{minipage}

}

\caption{\label{fig-main-phi}Monetary shock with
\(\phi\in\{1.2,1.5,2.0\}\)}

\end{figure}

\begin{figure}

\centering{

\includegraphics[width=0.9\linewidth,height=\textheight,keepaspectratio]{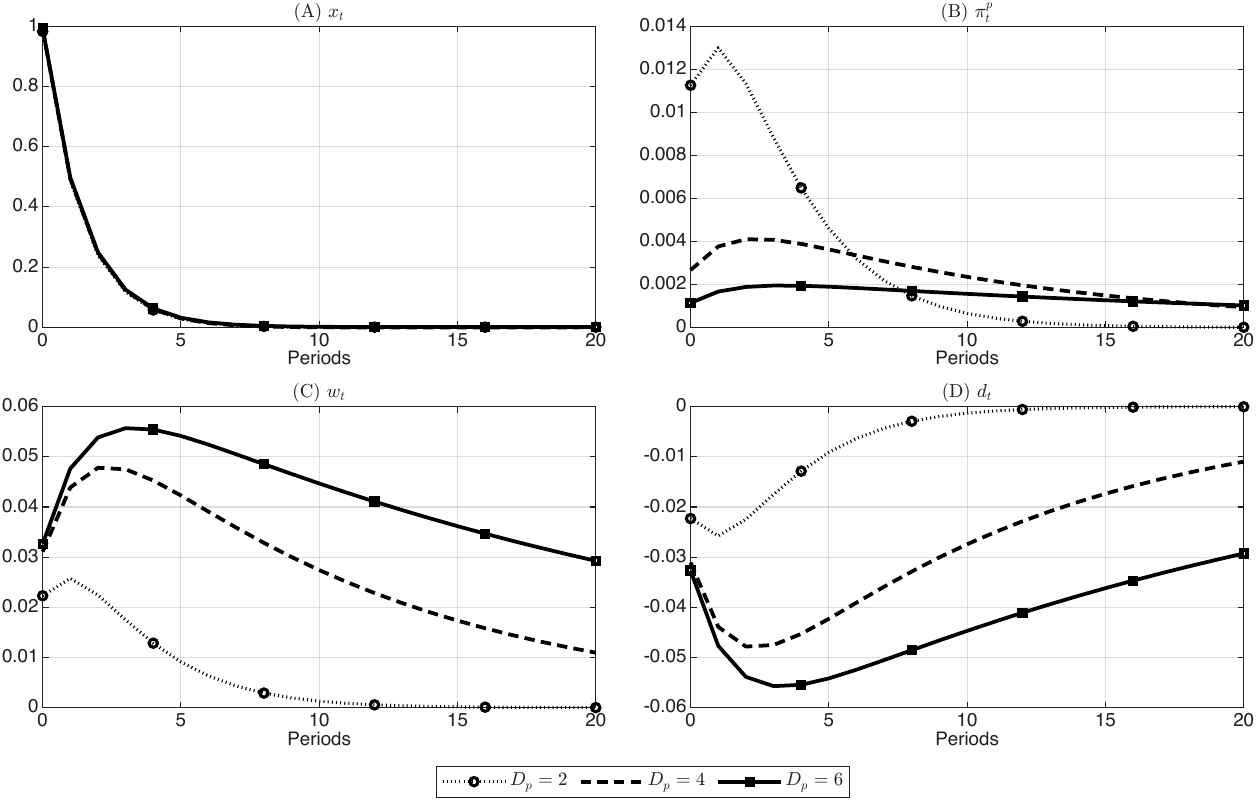}

\par\smallskip
\begin{minipage}{0.90\linewidth}
\footnotesize \textit{Notes:} Panels report $(A)$ $x_t$, $(B)$ $\pi_t^p$, $(C)$ $w_t$, and $(D)$ $d_t$. The dotted, dashed, and solid lines correspond to $D_p=2,4,6$ quarters, or $\eta_p=17.822,104.854,257.143$. Responses are to a one-percent monetary-policy innovation; vertical axes are percent deviations from steady state, and $d_t$ is the profit share scaled by steady-state output.
\end{minipage}

}

\caption{\label{fig-main-etap}Monetary shock with price-duration targets
\(D_p\in\{2,4,6\}\) quarters}

\end{figure}

\begin{figure}

\centering{

\includegraphics[width=0.9\linewidth,height=\textheight,keepaspectratio]{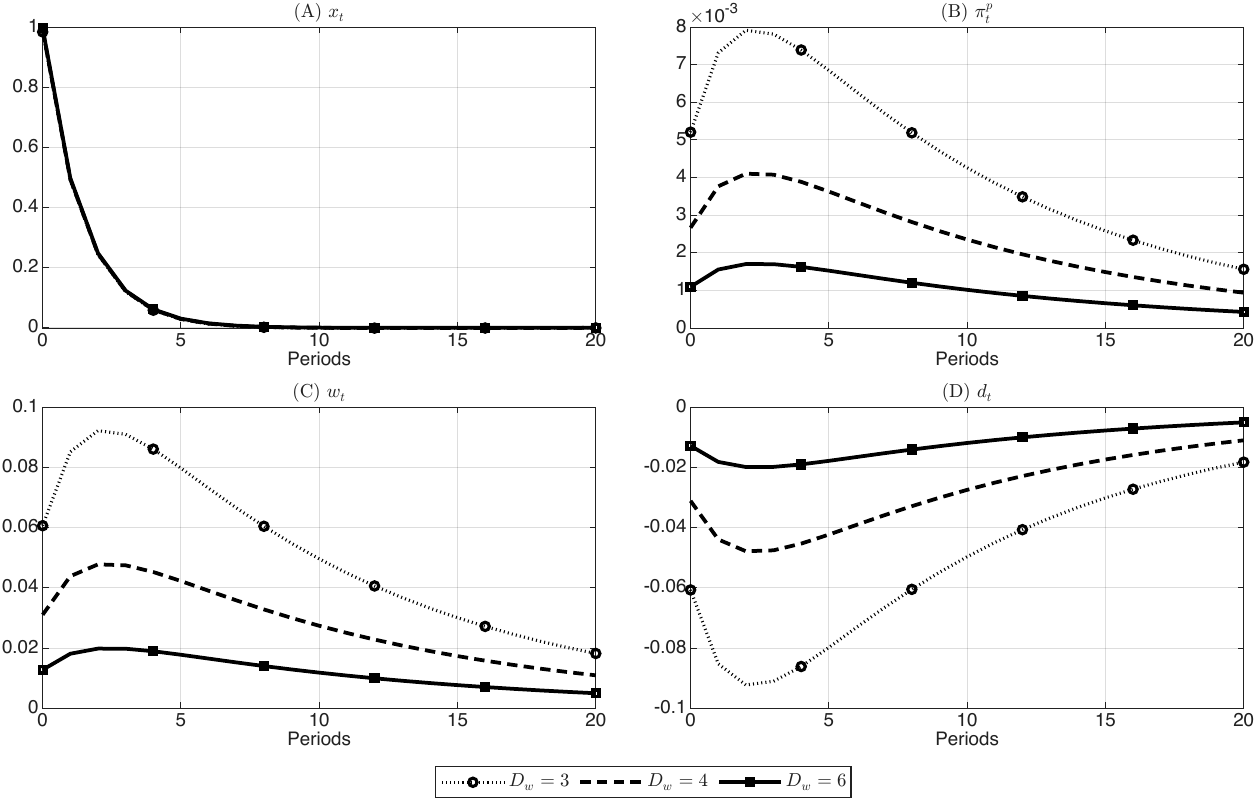}

\par\smallskip
\begin{minipage}{0.90\linewidth}
\footnotesize \textit{Notes:} Panels report $(A)$ $x_t$, $(B)$ $\pi_t^p$, $(C)$ $w_t$, and $(D)$ $d_t$. The dotted, dashed, and solid lines correspond to $D_w=3,4,6$ quarters, or $\eta_w=264.706,524.272,1285.714$. Responses are to a one-percent monetary-policy innovation; vertical axes are percent deviations from steady state, and $d_t$ is the profit share scaled by steady-state output.
\end{minipage}

}

\caption{\label{fig-main-etaw}Monetary shock with wage-duration targets
\(D_w\in\{3,4,6\}\) quarters}

\end{figure}

\begin{figure}

\centering{

\includegraphics[width=0.9\linewidth,height=\textheight,keepaspectratio]{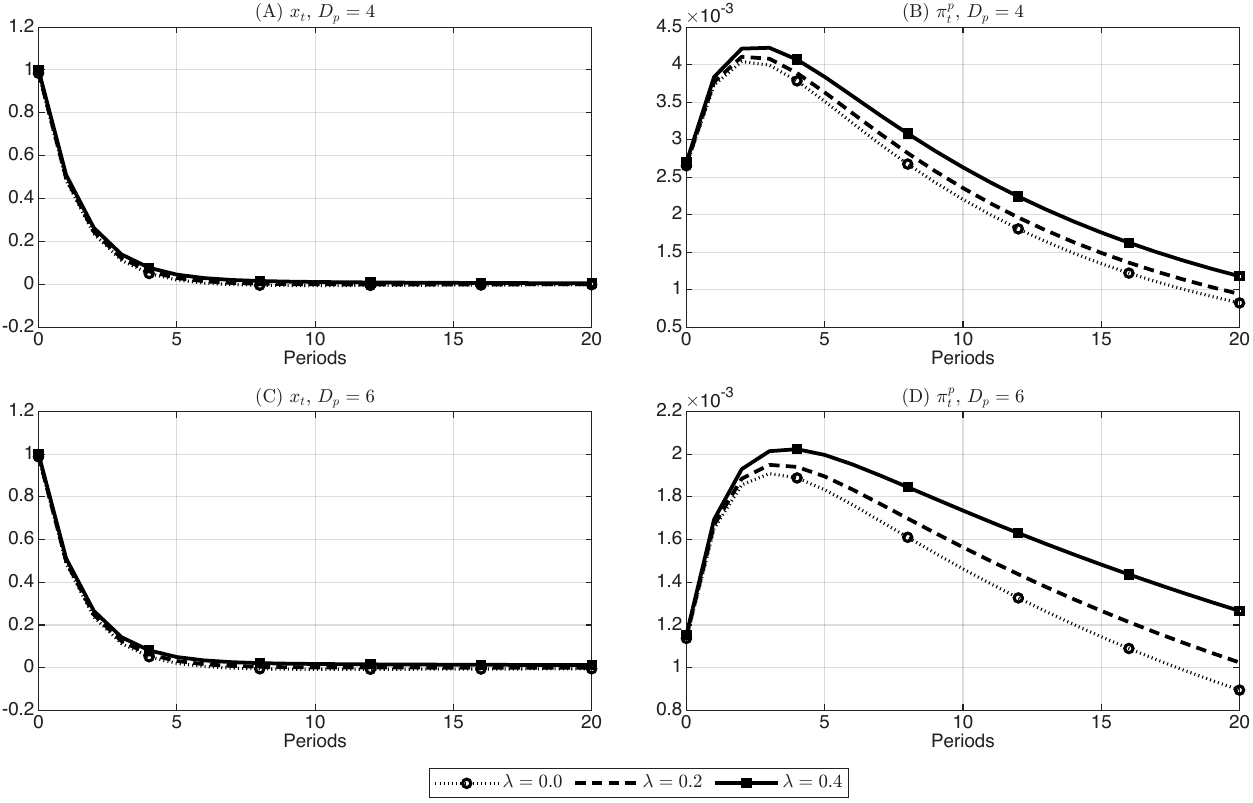}

\par\smallskip
\begin{minipage}{0.90\linewidth}
\footnotesize \textit{Notes:} Panels report $(A)$ $x_t$ with $D_p=4$, $(B)$ $\pi_t^p$ with $D_p=4$, $(C)$ $x_t$ with $D_p=6$, and $(D)$ $\pi_t^p$ with $D_p=6$. The dotted, dashed, and solid lines correspond to $\lambda=0$, $\lambda=0.2$, and $\lambda=0.4$. Responses are to a one-percent monetary-policy innovation; vertical axes are percent deviations from steady state. All six cases satisfy the root and determinacy conditions reported in the text.
\end{minipage}

}

\caption{\label{fig-main-lambda}Monetary shock with
\(\lambda\in\{0,0.2,0.4\}\) and \(D_p\in\{4,6\}\)}

\end{figure}

\begin{figure}

\centering{

\includegraphics[width=0.9\linewidth,height=\textheight,keepaspectratio]{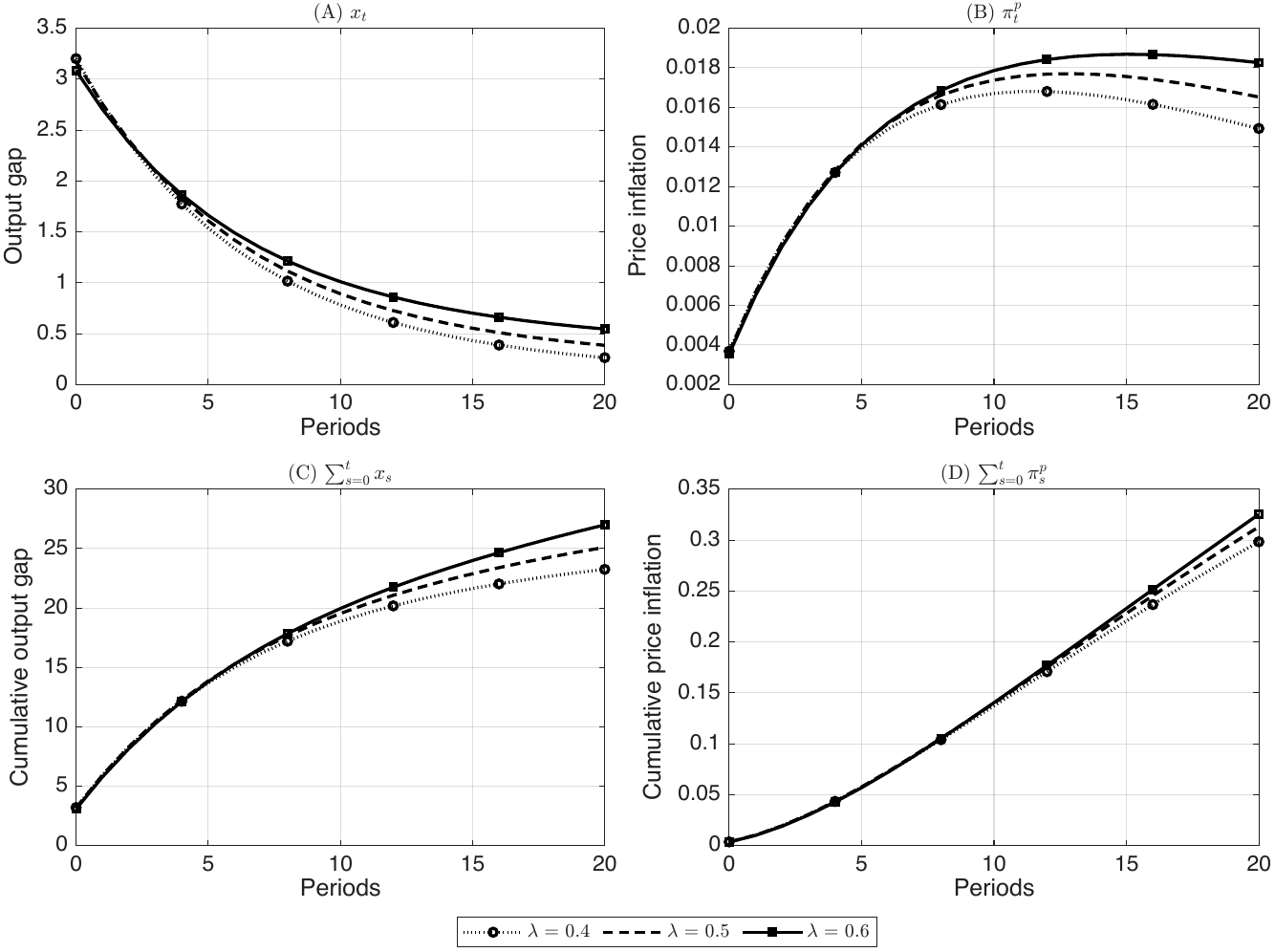}

\par\smallskip
\begin{minipage}{0.90\linewidth}
\footnotesize \textit{Notes:} Panels report $(A)$ the output-gap response, $(B)$ the price-inflation response, $(C)$ the cumulative output-gap response through quarter $t$, and $(D)$ the cumulative price-inflation response through quarter $t$. The dotted, dashed, and solid lines correspond to $\lambda=0.4,0.5,0.6$. The calibration sets $D_p=6$, $D_w=4$, $\rho_m=0.85$, $\beta=0.99$, $\gamma=\varphi=2$, $\phi=1.5$, $\psi_p=9$, and $\psi_w=4.5$. Responses are to a one-percent expansionary monetary-policy innovation; impact responses are percent deviations from steady state and cumulative responses are percent-quarters. All three cases violate the root condition in Proposition 2 but satisfy the determinacy condition in Proposition 3.
\end{minipage}

}

\caption{\label{fig-main-lambda-outside-root}Impact reversal and
cumulative amplification outside the root condition}

\end{figure}

\begin{figure}

\centering{

\includegraphics[width=0.9\linewidth,height=\textheight,keepaspectratio]{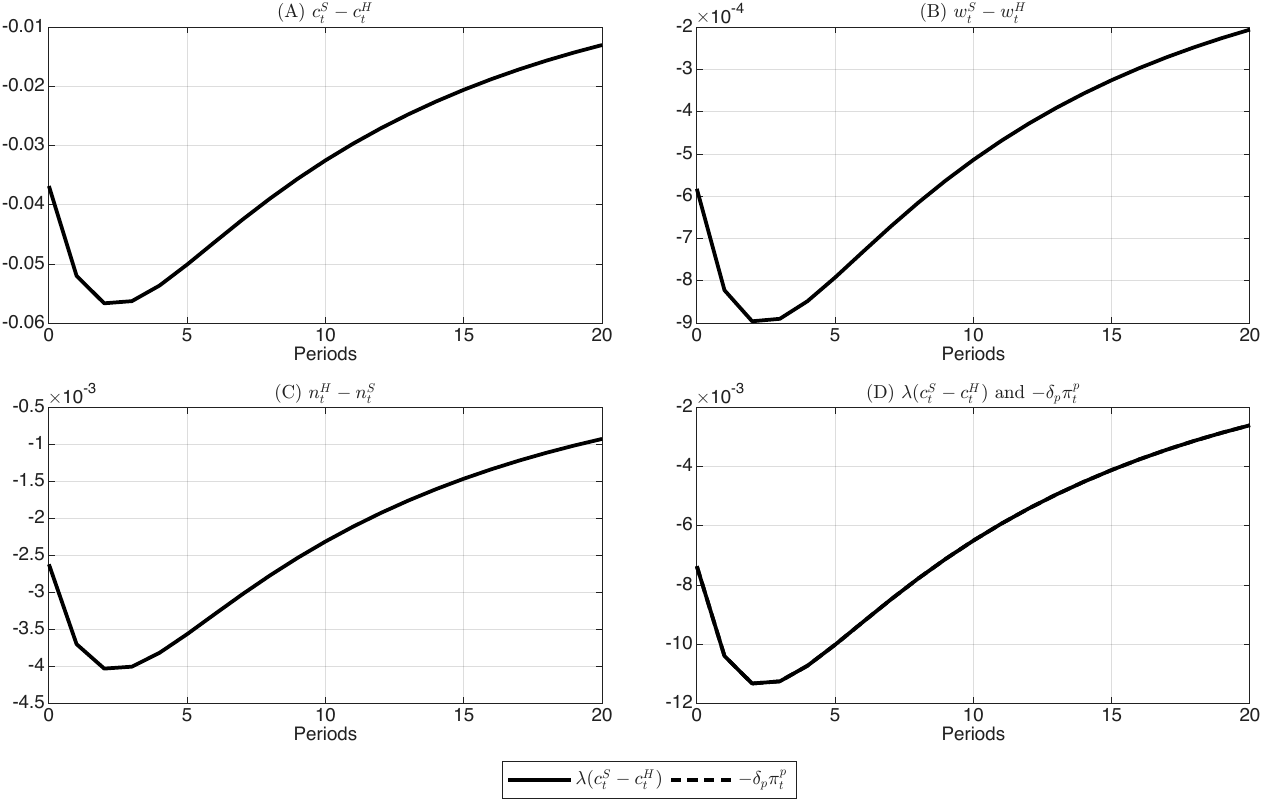}

\par\smallskip
\begin{minipage}{0.90\linewidth}
\footnotesize \textit{Notes:} Panels report $(A)$ $c_t^S-c_t^H$, $(B)$ $w_t^S-w_t^H$, $(C)$ $n_t^H-n_t^S$, and $(D)$ the overlay of $\lambda(c_t^S-c_t^H)$ and $-\delta_p\pi_t^p$ under the aggregate-lag benchmark. Responses are to a one-percent monetary-policy innovation; vertical axes are percent deviations from steady state. The exact overlap in Panel (D) is specific to aggregate-lag adjustment and does not hold in the own-lag model.
\end{minipage}

}

\caption{\label{fig-main-gaps}Aggregate-lag distributional gaps}

\end{figure}

Figures \ref{fig-main-phi}--\ref{fig-main-gaps} summarize the
aggregate-lag comparative statics for a one-percent monetary-policy
innovation. Low parameter values are shown by dotted lines, benchmark
values by dashed lines, and high values by solid lines. Because the
Taylor rule contains \(\exp(-m_t)\), a positive innovation to \(m_t\) is
an expansionary monetary shock. Because natural output does not move
after a monetary shock, it is sufficient to report the output gap
\(x_t\).

Figure~\ref{fig-main-phi} shows the role of the Taylor-rule coefficient.
As Proposition 2 suggests, a stronger policy response dampens both
inflation and the output gap. Over the conventional range
\(\phi\in\{1.2,1.5,2.0\}\), the output and real-rate paths are close,
while the inflation and profit responses display the clearest ordering.

Figure~\ref{fig-main-etap} shows how price rigidity strengthens real
effects and compresses inflation. Moving from a two-quarter to a
six-quarter price-duration target sharply compresses inflation and makes
the wage and profit responses larger and more persistent. The impact
output response changes little, but the slower price adjustment prolongs
the profit-to-wage redistribution behind the heterogeneity term. Because
\(1-\gamma\delta_p>0\) fails at the benchmark, the ordering is reported
as a numerical result rather than attributed to the rigidity
comparative-static clause of Proposition 2.

Figure~\ref{fig-main-etaw} provides the analogous exercise for wage
rigidity. Across the three duration targets, the output response is
nearly unchanged, whereas a longer wage-duration target substantially
reduces inflation, wage, and profit movements. Thus the duration-based
calibration does not support the earlier claim that wage rigidity
sharply raises the output response. Unlike price rigidity, wage rigidity
changes the redistribution channel indirectly through wages and the AS
slope; in this calibration its main quantitative role is to damp the
nominal and profit responses.

Figure~\ref{fig-main-lambda} compares household heterogeneity at the
benchmark price-duration target \(D_p=4\) and the longer target
\(D_p=6\). With \(D_p=4\), raising \(\lambda\) from 0 to 0.4 increases
the impact output response from 0.985 to 1.001 percent and the impact
inflation response from 0.00265 to 0.00270 percent. The impact
differences are therefore monotone but small. The same ordering holds
for \(D_p=6\), and the cumulative separation is more pronounced: the
analytical infinite-horizon cumulative output multiplier rises from
1.718 at \(\lambda=0\) to 2.687 at \(\lambda=0.4\). These comparisons
satisfy the root condition in Proposition 2 and the determinacy
condition in Proposition 3. The recalibration therefore preserves the
profit-redistribution amplification with respect to \(\lambda\), while
showing that the impact magnitude is modest and the more visible
differences occur over the subsequent adjustment path.

The contrast between Propositions 2 and 3 becomes sharper outside the
root condition. To illustrate it,
Figure~\ref{fig-main-lambda-outside-root} retains \(D_p=6\) and
\(D_w=4\), restores the more persistent monetary shock \(\rho_m=0.85\),
and considers \(\lambda\in\{0.4,0.5,0.6\}\). The root-condition bounds
are 0.775, 0.663, and 0.438, respectively, so none of the three cases
satisfies the sufficient condition in Proposition 2. All three
equilibria remain determinate: \(\xi_1\) ranges from 0.969 to 0.985 and
\(\xi_2\) from 1.009 to 1.020.

In this region, the impact ordering reverses. As \(\lambda\) rises from
0.4 to 0.6, the output multiplier falls from 3.201 to 3.083 and the
inflation multiplier from 0.00369 to 0.00355. The responses nevertheless
become more persistent, so the infinite-horizon cumulative output
multiplier rises from 28.62 to 56.93 and the corresponding inflation
multiplier from 0.789 to 1.570. Panels (C) and (D) show that the
cumulative ordering already reverses within the displayed 21-quarter
window. This is not a counterexample to Proposition 2, whose local
amplification result is conditional on the root bound; it is a numerical
case in which that sufficient condition fails while the
determinacy-based result in Proposition 3 continues to hold.

Figure~\ref{fig-main-gaps} links the aggregate-lag numerical results
directly to Proposition 1 and Corollary 1. Within this benchmark,
profits map into type-specific consumption, wage, and labor gaps, and
all three gaps are negative after an expansionary monetary shock because
profits fall and wages rise. At the benchmark, \(\delta_c=0.948\) and
\(\delta_p=2.760\). Panel (D) shows the numerical overlap between
\(\lambda(c_t^S-c_t^H)\) and \(-\delta_p\pi_t^p\); the maximum absolute
discrepancy is \(2.0\times10^{-15}\).

\clearpage

\subsection{Common-Primitive Comparison with the Own-Lag
Framework}\label{common-primitive-comparison-with-the-own-lag-framework}

This subsection uses the conventional own-lag TANK framework developed
in Miyazaki (2026). That paper derives the law of motion for the
inherited relative-wage state and the associated earnings-reallocation
decomposition and compares type-specific wage setting with a
reduced-form common-wage closure. The purpose here is narrower: the
exercise compares own-lag and aggregate-lag adjustment while retaining
the preferences, technology, policy rule, elasticities, and primitive
adjustment-cost coefficients in Table~\ref{tbl-parameters}. It therefore
isolates adjustment-cost timing without rescaling Phillips-curve slopes
by shock persistence, but identical primitive coefficients do not imply
equal reduced-form slopes or equal economic rigidity. Because the
references in Section 3 motivate but do not uniquely derive the exact
quadratic costs---especially on the wage side---this comparison is the
paper's main check on whether the transmission and redistribution
conclusions depend on the aggregate-reference closure. In the own-lag
specification, each firm adjusts relative to its own lagged price and
each household type adjusts relative to its own lagged nominal wage
(Rotemberg 1982; Miyazaki 2026). Both price and wage Phillips curves are
forward looking, and the relative wage is an inherited state.

For completeness, the transfer-free specialization of the canonical
system in Miyazaki (2026) is reported here. Let
\(\sigma_t^w:=w_t^S-w_t^H\), \(\sigma_t^c:=c_t^S-c_t^H\), and
\(\sigma_t^n:=n_t^S-n_t^H\). Also define \(\kappa_p:=\psi_p/\eta_p\),
\(\kappa_w:=\psi_w/\eta_w\), \[
\begin{aligned}
\Theta_w&:=1+\beta+\kappa_p+\kappa_w,\\
\Theta_\sigma&:=1+\beta+\kappa_w\{(1-\gamma)+\psi_w(\gamma+\varphi)\}.
\end{aligned}
\] In the specialization used here, the full type-specific equilibrium
reduces to
\begin{equation}\protect\phantomsection\label{eq-ownlag-canonical}{
\begin{aligned}
r_t&=r_t^f+\gamma E_t\left(\Delta x_{t+1}+\lambda\Delta\sigma_{t+1}^c\right),\\
\pi_t^p&=\beta E_t\pi_{t+1}^p-\kappa_p d_t,\\
\Theta_w w_t&=w_{t-1}+\beta E_tw_{t+1}
 +\kappa_w(\gamma+\varphi)x_t+(\kappa_p+\kappa_w)a_t,\\
\Theta_\sigma\sigma_t^w&=\sigma_{t-1}^w+\beta E_t\sigma_{t+1}^w
 +\frac{\gamma\kappa_w}{1-\lambda}d_t,\\
\sigma_t^c&=(1-\psi_w)\sigma_t^w+\frac{d_t}{1-\lambda},
\qquad d_t=a_t-w_t.
\end{aligned}
}\end{equation} The remaining distributional objects satisfy \[
\sigma_t^n=-\psi_w\sigma_t^w,
\qquad
\pi_t^S-\pi_t^H=\Delta\sigma_t^w,
\qquad
\pi_t^w=\pi_t^p+w_t-w_{t-1}.
\]

As Miyazaki (2026) shows, type-specific adjustments in relative wages
and employment create an earnings-reallocation term
\((1-\psi_w)\sigma_t^w\), while own-lag timing makes the relative wage
an inherited state. For the present comparison, the additional term
means that profits are no longer a sufficient statistic for
cross-household allocations. The price block likewise implies \[
d_t=-\frac{\eta_p}{\psi_p}\left(\pi_t^p-\beta E_t\pi_{t+1}^p\right),
\] so current price inflation alone cannot summarize redistribution.

The common-primitive comparison introduced here varies
\(\lambda\in\{0,0.2,0.4\}\), the price-duration target
\(D_p\in\{2,4,6\}\), and the wage-duration target \(D_w\in\{3,4,6\}\).
For every one of the 27 calibrations, the own-lag and aggregate-lag
models use identical values of \((\eta_p,\eta_w,\psi_p,\psi_w)\). Both
monetary and technology shocks are simulated for 400 quarters; the
finite-horizon cumulative statistics below sum quarters 0 through 20 and
are distinct from the infinite-horizon objects in Proposition 3.
Dynare's first-order solver successfully completes all 27 own-lag cases.
The full type-specific system and the reduced system
(\ref{eq-ownlag-canonical}) produce impulse responses that differ by at
most \(1.1\times10^{-15}\), and the maximum equilibrium-equation
residual is \(6.8\times10^{-16}\).

The monetary-shock comparison is the most relevant test of the paper's
mechanism. The corresponding common-primitive technology-shock
comparison is reported in Appendix C.

Table~\ref{tbl-ownlag-comparison} reports the benchmark-target results.
The output impact is similar across timings: at \(\lambda=0.2\), it is
0.969 percent under own-lag adjustment and 0.991 percent under
aggregate-lag adjustment. The nominal and distributional paths are not
quantitatively invariant. Own-lag inflation is 0.0234 percent on impact,
8.76 times the aggregate-lag response, while the absolute 21-quarter
profit decline is 0.474 times as large. The forward-looking model
therefore produces a sharper initial inflation and profit response but a
less persistent profit redistribution.

\begingroup
\small
\setlength{\tabcolsep}{4pt}

\begin{longtable}[]{@{}
  >{\raggedright\arraybackslash}p{(\linewidth - 14\tabcolsep) * \real{0.1800}}
  >{\raggedleft\arraybackslash}p{(\linewidth - 14\tabcolsep) * \real{0.0600}}
  >{\raggedleft\arraybackslash}p{(\linewidth - 14\tabcolsep) * \real{0.1300}}
  >{\raggedleft\arraybackslash}p{(\linewidth - 14\tabcolsep) * \real{0.1300}}
  >{\raggedleft\arraybackslash}p{(\linewidth - 14\tabcolsep) * \real{0.1200}}
  >{\raggedleft\arraybackslash}p{(\linewidth - 14\tabcolsep) * \real{0.1300}}
  >{\raggedleft\arraybackslash}p{(\linewidth - 14\tabcolsep) * \real{0.1200}}
  >{\raggedleft\arraybackslash}p{(\linewidth - 14\tabcolsep) * \real{0.1300}}@{}}
\caption{Common-primitive monetary-shock
comparison}\label{tbl-ownlag-comparison}\tabularnewline
\toprule\noalign{}
\begin{minipage}[b]{\linewidth}\raggedright
Adjustment timing
\end{minipage} & \begin{minipage}[b]{\linewidth}\raggedleft
\(\lambda\)
\end{minipage} & \begin{minipage}[b]{\linewidth}\raggedleft
Impact \(x_t\)
\end{minipage} & \begin{minipage}[b]{\linewidth}\raggedleft
Impact \(\pi_t^p\)
\end{minipage} & \begin{minipage}[b]{\linewidth}\raggedleft
Impact \(d_t\)
\end{minipage} & \begin{minipage}[b]{\linewidth}\raggedleft
21-quarter \(x_t\)
\end{minipage} & \begin{minipage}[b]{\linewidth}\raggedleft
21-quarter \(d_t\)
\end{minipage} & \begin{minipage}[b]{\linewidth}\raggedleft
\(10^4\mathcal L\)
\end{minipage} \\
\midrule\noalign{}
\endfirsthead
\toprule\noalign{}
\begin{minipage}[b]{\linewidth}\raggedright
Adjustment timing
\end{minipage} & \begin{minipage}[b]{\linewidth}\raggedleft
\(\lambda\)
\end{minipage} & \begin{minipage}[b]{\linewidth}\raggedleft
Impact \(x_t\)
\end{minipage} & \begin{minipage}[b]{\linewidth}\raggedleft
Impact \(\pi_t^p\)
\end{minipage} & \begin{minipage}[b]{\linewidth}\raggedleft
Impact \(d_t\)
\end{minipage} & \begin{minipage}[b]{\linewidth}\raggedleft
21-quarter \(x_t\)
\end{minipage} & \begin{minipage}[b]{\linewidth}\raggedleft
21-quarter \(d_t\)
\end{minipage} & \begin{minipage}[b]{\linewidth}\raggedleft
\(10^4\mathcal L\)
\end{minipage} \\
\midrule\noalign{}
\endhead
\bottomrule\noalign{}
\endlastfoot
Own lag & 0.0 & 0.962 & 0.0226 & -0.0378 & 1.842 & -0.273 & 3.702 \\
Own lag & 0.2 & 0.969 & 0.0234 & -0.0384 & 1.883 & -0.282 & 3.828 \\
Own lag & 0.4 & 0.981 & 0.0247 & -0.0393 & 1.958 & -0.299 & 4.060 \\
Aggregate lag & 0.0 & 0.985 & 0.00265 & -0.0309 & 1.833 & -0.565 &
2.932 \\
Aggregate lag & 0.2 & 0.991 & 0.00267 & -0.0311 & 1.955 & -0.595 &
3.006 \\
Aggregate lag & 0.4 & 1.001 & 0.00270 & -0.0314 & 2.184 & -0.649 &
3.140 \\
\end{longtable}

\endgroup

\emph{Notes:} The duration targets are \(D_p=D_w=4\), corresponding to
the common primitive adjustment costs \(\eta_p=104.854\) and
\(\eta_w=524.272\). Impact responses are percent deviations following a
one-percent expansionary monetary-policy innovation, except \(d_t\),
which is the profit share in percentage points of steady-state output.
Cumulative \(x_t\) is in percent-quarters, while cumulative \(d_t\) is
in percentage-point-of-output quarters; both sum quarters 0--20.
\(\mathcal L\) is the discounted CES-consistent loss over quarters
0--399.

The positive monetary-shock ordering with respect to \(\lambda\)
survives throughout the finite grid. Relative to \(\lambda=0\), all 18
positive-\(\lambda\) cases have larger impact and 21-quarter cumulative
output and inflation responses and larger absolute profit declines. The
output-impact increase ranges from 0.0021 to 0.0299 percentage points,
so the ordering is uniform but often modest. The reported price- and
wage-rigidity orderings each hold across nine comparisons indexed by
\(\lambda\) and the other duration target. A longer price-duration
target raises the impact output response, lowers impact inflation, and
makes the impact profit decline larger. A longer wage-duration target
raises impact output slightly, lowers impact inflation, and sharply
dampens the impact profit decline. Across the full monetary grid, the
own-lag output impact is 94.3--99.4 percent of its aggregate-lag
counterpart, whereas the inflation-impact ratio ranges from 3.48 to
15.87. The absolute 21-quarter cumulative-profit ratio ranges from 0.43
to 0.66. These results preserve the sign of the redistribution mechanism
and the reported \(\lambda\), \(D_p\), and \(D_w\) orderings on the
examined grid, but not the nominal or distributional magnitudes.

Applying the accounting decomposition in Miyazaki (2026) to the present
calibration clarifies the source of the consumption-gap response. At
\(\lambda=0.2\), direct profit incidence lowers \(c_t^S-c_t^H\) by
0.0480 percent on impact, while relative wages and employment offset
0.00595 percentage points, leaving a consumption gap of -0.0420 percent.
The earnings-reallocation term offsets 12.4 percent of direct incidence
on impact and 29.6 percent in the sum over quarters 0--20. This
calculation applies the companion paper's identity to the current
calibration.

The matched closure contrast in Table~\ref{tbl-common-wage-closure}
makes this buffer visible without changing the own-lag price timing or
primitive coefficients. Suppressing relative wage and hours dispersion
leaves the benchmark impact output and inflation responses close to
their type-specific values, but makes the negative consumption gap
larger, especially in the 21-quarter sum.

\begin{longtable}[]{@{}
  >{\raggedright\arraybackslash}p{(\linewidth - 10\tabcolsep) * \real{0.2600}}
  >{\raggedleft\arraybackslash}p{(\linewidth - 10\tabcolsep) * \real{0.1400}}
  >{\raggedleft\arraybackslash}p{(\linewidth - 10\tabcolsep) * \real{0.1400}}
  >{\raggedleft\arraybackslash}p{(\linewidth - 10\tabcolsep) * \real{0.1800}}
  >{\raggedleft\arraybackslash}p{(\linewidth - 10\tabcolsep) * \real{0.1800}}
  >{\raggedleft\arraybackslash}p{(\linewidth - 10\tabcolsep) * \real{0.1000}}@{}}
\caption{Restricted own-lag closure
comparison}\label{tbl-common-wage-closure}\tabularnewline
\toprule\noalign{}
\begin{minipage}[b]{\linewidth}\raggedright
Own-lag closure
\end{minipage} & \begin{minipage}[b]{\linewidth}\raggedleft
Impact \(x_t\)
\end{minipage} & \begin{minipage}[b]{\linewidth}\raggedleft
Impact \(\pi_t^p\)
\end{minipage} & \begin{minipage}[b]{\linewidth}\raggedleft
Impact \(c_t^S-c_t^H\)
\end{minipage} & \begin{minipage}[b]{\linewidth}\raggedleft
21-quarter \(c_t^S-c_t^H\)
\end{minipage} & \begin{minipage}[b]{\linewidth}\raggedleft
\(10^4\mathcal L\)
\end{minipage} \\
\midrule\noalign{}
\endfirsthead
\toprule\noalign{}
\begin{minipage}[b]{\linewidth}\raggedright
Own-lag closure
\end{minipage} & \begin{minipage}[b]{\linewidth}\raggedleft
Impact \(x_t\)
\end{minipage} & \begin{minipage}[b]{\linewidth}\raggedleft
Impact \(\pi_t^p\)
\end{minipage} & \begin{minipage}[b]{\linewidth}\raggedleft
Impact \(c_t^S-c_t^H\)
\end{minipage} & \begin{minipage}[b]{\linewidth}\raggedleft
21-quarter \(c_t^S-c_t^H\)
\end{minipage} & \begin{minipage}[b]{\linewidth}\raggedleft
\(10^4\mathcal L\)
\end{minipage} \\
\midrule\noalign{}
\endhead
\bottomrule\noalign{}
\endlastfoot
Type-specific wages and hours & 0.969 & 0.0234 & -0.0420 & -0.248 &
3.828 \\
Common wage and labor & 0.970 & 0.0237 & -0.0481 & -0.358 & 3.864 \\
\end{longtable}

\emph{Notes:} The calibration sets \(\lambda=0.2\) and \(D_p=D_w=4\) and
applies a one-percent expansionary monetary-policy innovation. Impact
entries are percent deviations; the cumulative consumption gap is in
percent-quarters over quarters 0--20; and \(\mathcal L\) is the
discounted CES-consistent loss over quarters 0--399. The
common-wage/common-labor closure sets relative wage and hours dispersion
to zero, imposes \(c_t^S-c_t^H=d_t/(1-\lambda)\), and retains the
aggregate sticky-wage block. It is an institutional closure contrast,
not a nested causal parameter.

The comparison supports the interpretation of type-specific wages and
hours as a distributional buffer in this calibration. Because the
restriction changes the institutional closure jointly and the compact
grid holds \(D_w=4\), the table does not isolate a structural causal
effect or establish the same quantitative gap for every wage-duration
target.

The welfare calculation is specific to the present timing comparison;
Miyazaki (2026) studies distributional incidence rather than welfare
rankings. For the monetary benchmark, the CES-consistent own-lag loss
rises by 3.40 percent at \(\lambda=0.2\) and 9.68 percent at
\(\lambda=0.4\), relative to \(\lambda=0\). However, the own-lag loss at
\(\lambda=0.2\) is 27.4 percent above the aggregate-lag loss, and the
direct dispersion terms for consumption, hours, and wage inflation
account for only 0.047 percent of total own-lag loss. About 98.6 percent
of the increase from \(\lambda=0\) to 0.2 comes from changes in
aggregate output, price inflation, and aggregate wage inflation rather
than the direct dispersion terms. For comparison, the aggregate-lag
scalar price-inflation coefficient rises by 62.3 percent at
\(\lambda=0.2\) and 166.0 percent at \(\lambda=0.4\), from 104.854 to
170.126 and 278.913, respectively. These coefficient changes are not
realized loss shares: the 0.047 percent figure is the direct-dispersion
share of the own-lag loss along one simulated equilibrium path. Thus the
positive loss ordering with respect to \(\lambda\) survives in the
examined grid, but the loss level, component shares, and scalar
price-inflation-weight interpretation are timing sensitive. These are
losses under the maintained Taylor rule, not welfare rankings of the two
adjustment-cost institutions or solutions to the own-lag optimal-policy
problem.

\subsection{Robustness to the Policy Rule and Household-Side
Wage-Subsidy
Financing}\label{robustness-to-the-policy-rule-and-household-side-wage-subsidy-financing}

The preceding comparison maintains an inflation-only Taylor rule and an
arrangement that rebates each household type's wage subsidy to that
type. This subsection varies the policy rule and household-side
wage-subsidy financing while retaining the common primitive parameters.
The alternative monetary-policy rule is
\begin{equation}\protect\phantomsection\label{eq-taylor-output-robustness}{
r_t=\phi\pi_t^p+\phi_xx_t-E_t\pi_{t+1}^p-m_t,
}\end{equation} where \(\phi_x\in\{0,0.0625,0.125,0.25\}\). The value
\(\phi_x=0.125\) corresponds to an annualized output-gap coefficient of
0.5 in the quarterly rule. The financing alternative replaces the
type-specific wage-subsidy rebates with a common per-capita lump-sum tax
that finances the aggregate wage subsidy. The firm-level offset
associated with the sales subsidy is retained, so the exercise isolates
household-side wage-subsidy financing while preserving the efficient
zero-profit steady state; it is not a general change in fiscal or profit
incidence. The Online Appendix derives the resulting consumption-gap and
relative-wage blocks for both adjustment timings.

Crossing the four policy coefficients, two household-side financing
arrangements, three values of \(\lambda\), three price-duration targets,
and three wage-duration targets gives 216 cases for each adjustment
timing and 432 cases in total. All cases satisfy determinacy. In every
case, an expansionary monetary shock raises the output gap and price
inflation and lowers profits both on impact and in the cumulative sum
over quarters 0--20. For every positive-\(\lambda\) case, direct profit
incidence lowers \(c_t^S-c_t^H\), while relative wages and employment
provide a positive but incomplete offset. Consequently, the consumption
gap remains negative throughout the grid.

\begin{longtable}[]{@{}
  >{\raggedright\arraybackslash}p{(\linewidth - 4\tabcolsep) * \real{0.6800}}
  >{\raggedleft\arraybackslash}p{(\linewidth - 4\tabcolsep) * \real{0.1600}}
  >{\raggedleft\arraybackslash}p{(\linewidth - 4\tabcolsep) * \real{0.1600}}@{}}
\caption{Monetary-transmission robustness
counts}\label{tbl-closure-robustness}\tabularnewline
\toprule\noalign{}
\begin{minipage}[b]{\linewidth}\raggedright
Criterion
\end{minipage} & \begin{minipage}[b]{\linewidth}\raggedleft
Own lag
\end{minipage} & \begin{minipage}[b]{\linewidth}\raggedleft
Aggregate lag
\end{minipage} \\
\midrule\noalign{}
\endfirsthead
\toprule\noalign{}
\begin{minipage}[b]{\linewidth}\raggedright
Criterion
\end{minipage} & \begin{minipage}[b]{\linewidth}\raggedleft
Own lag
\end{minipage} & \begin{minipage}[b]{\linewidth}\raggedleft
Aggregate lag
\end{minipage} \\
\midrule\noalign{}
\endhead
\bottomrule\noalign{}
\endlastfoot
Positive \(x_t\) and \(\pi_t^p\), negative \(d_t\), on impact and in the
sum over quarters 0--20 & 216/216 & 216/216 \\
Negative \(c_t^S-c_t^H\), on impact and in the sum over quarters 0--20,
when \(\lambda>0\) & 144/144 & 144/144 \\
Impact amplification relative to \(\lambda=0\) & 144/144 & 108/144 \\
21-quarter cumulative amplification relative to \(\lambda=0\) & 144/144
& 144/144 \\
\end{longtable}

\emph{Notes:} Each impact or cumulative amplification count compares a
positive hand-to-mouth share, \(\lambda\in\{0.2,0.4\}\), with
\(\lambda=0\), holding adjustment timing, \(\phi_x\), household-side
wage-subsidy financing, \(D_p\), and \(D_w\) fixed. Amplification
requires a larger output response, a larger inflation response, and a
larger absolute profit decline. Cumulative responses sum quarters 0--20.
The first row counts 216 full-model cases per timing; the remaining rows
count the 144 cases per timing with a positive hand-to-mouth share.

The output response in the policy rule attenuates rather than overturns
the mechanism. Relative to \(\phi_x=0\), the standard value
\(\phi_x=0.125\) reduces the absolute impact output response by
9.9--11.2 percent under own-lag adjustment and 10.1--12.6 percent under
aggregate-lag adjustment. The corresponding cumulative output reduction
is 9.1--11.7 percent under own-lag adjustment and 4.6--23.1 percent
under aggregate-lag adjustment. Appendix C reports the ranges for
inflation and profits as well.

The hand-to-mouth-share result is strongest for 21-quarter cumulative
transmission. Under own-lag adjustment, impact and 21-quarter cumulative
output, inflation, and absolute profit responses increase strictly with
\(\lambda\) in all 72 policy-financing-duration configurations. Under
aggregate-lag adjustment, the 21-quarter cumulative responses are also
strictly increasing in all 72 configurations, but the impact responses
are strictly ordered in only 50. Relative to \(\lambda=0\), 108 of the
144 positive-share comparisons amplify the impact responses. The 36
reversals occur at \(\phi_x=0.125\) with \(D_p=6\) and at
\(\phi_x=0.25\) with \(D_p\in\{4,6\}\). They are small: the largest
reduction in the impact output response is 0.0105 percentage points,
equal to 1.32 percent of the corresponding \(\lambda=0\) response. Thus
the safe quantitative conclusion is that the hand-to-mouth share
robustly amplifies impact responses under own-lag adjustment and
21-quarter cumulative responses under both timings; aggregate-lag impact
amplification is conditional.

Changing household-side wage-subsidy financing has a larger effect on
distribution than on aggregate transmission. Across positive-share
cases, the common lump-sum tax changes the macroeconomic response
magnitudes by at most 4.1 percent under own-lag adjustment and 0.9
percent under aggregate-lag adjustment. It reduces the absolute
consumption gap by as much as 10.8 percent under own-lag adjustment and
3.2 percent under aggregate-lag adjustment because the common tax
strengthens the relative-wage-income offset. The sign of the consumption
gap nevertheless remains unchanged. Appendix C gives the full ranges.
This exercise evaluates responses to the expansionary monetary shock
only and does not extend the welfare comparison to the alternative
policy and financing arrangements.

\section{Conclusion}\label{conclusion}

This paper develops a closed-form aggregate-lag TANK benchmark with
price rigidity, wage rigidity, type-specific wages, and type-specific
labor supply. The benchmark isolates a profit-redistribution channel: an
expansionary monetary shock lowers profits and raises real wages,
shifting income toward hand-to-mouth households. Wage rigidity does not
have an unconditional effect on this channel: it shifts the allocation
of a given profit change toward consumption, yet dampens nominal and
profit responses in the benchmark duration calibration. Local impact
amplification with respect to the hand-to-mouth share requires the root
condition in Proposition 2, whereas the local infinite-horizon
cumulative result in Proposition 3 requires determinacy. These
propositions are properties of the tractable aggregate-lag system rather
than general theorems for conventional Rotemberg adjustment.

Applying the own-lag framework in Miyazaki (2026) shows which parts of
this mechanism survive when forward-looking price and wage setting and
inherited relative wages are restored. The comparison holds primitive
cost coefficients fixed but does not equate reduced-form slopes or
economic rigidity across timings. Under the baseline policy rule and
household-side wage-subsidy-financing arrangement, monetary shocks lower
profits throughout the 27-case duration grid, and output, inflation, and
the absolute profit decline increase with the hand-to-mouth share.
Adjustment timing nevertheless changes nominal and distributional
magnitudes substantially. The companion paper's earnings-reallocation
decomposition explains why relative wages and employment partially
offset direct profit incidence and why profits cease to be a sufficient
statistic under own-lag adjustment. A matched institutional-closure
contrast is consistent with this buffer interpretation in the benchmark
calibration, while leaving impact output and inflation close to the
type-specific own-lag values.

The broader policy-rule and household-side financing exercise confirms
the sign of the mechanism within the reported finite grid. Expansionary
monetary shocks raise output and inflation and lower profits on impact
and in sums over quarters 0--20 throughout that grid. In every
positive-share case, the saver-minus-hand-to-mouth consumption gap is
negative both on impact and in the sum over quarters 0--20. These
21-quarter results are numerical finite-grid findings, not Proposition
3's local infinite-horizon comparative static. Finite-horizon cumulative
responses increase monotonically with the hand-to-mouth share under both
timing conventions, and impact responses do so throughout the own-lag
grid. Aggregate-lag impact amplification is more limited: relative to
the representative-agent case, it holds in 108 of 144 positive-share
comparisons, with small reversals under stronger output-gap feedback and
longer price-duration targets.

Common lump-sum financing of household wage subsidies leaves
macroeconomic responses close to baseline while strengthening the
relative-wage-income offset; the firm-side sales-subsidy offset remains
fixed. The robust conclusion is therefore about the direction of profit
redistribution and finite-horizon cumulative monetary transmission
within the examined design, not a timing-invariant impact multiplier.

The welfare comparison remains deliberately narrower. The aggregate-lag
benchmark compresses distributional losses into a higher coefficient on
current price inflation: relative to 104.854 at \(\lambda=0\), the
coefficient is 62.3 percent higher at \(\lambda=0.2\) and 166.0 percent
higher at \(\lambda=0.4\). Own-lag adjustment instead leaves a
multidimensional, history-dependent loss, and direct dispersion accounts
for 0.047 percent of the realized loss at the benchmark \(\lambda=0.2\)
path. The coefficient changes and realized loss share are different
objects and do not establish a general optimal-policy ranking. The
reported welfare magnitudes apply to the baseline policy rule and
financing arrangement and are not extended by the robustness exercise.
Technology shocks likewise show that the positive output ordering with
respect to the hand-to-mouth share is specific to monetary transmission.
Further work could vary the firm-side incidence of the sales subsidy,
profit taxation and dividend ownership, or richer household
heterogeneity.

The central contribution is therefore an analytical benchmark that
exposes the profit-to-consumption, hours, and wage mapping, together
with a common-primitive comparison that marks its limits. Within the
examined grid, restoring conventional own-lag adjustment, adding an
output response to monetary policy, and changing household-side
wage-subsidy financing do not overturn the monetary redistribution
mechanism. They do identify where impact magnitudes, orderings, and
welfare representations remain sensitive to adjustment timing and
closure.

\section*{Declaration of Generative AI and AI-Assisted Technologies in
the Manuscript Preparation
Process}\label{declaration-of-generative-ai-and-ai-assisted-technologies-in-the-manuscript-preparation-process}
\addcontentsline{toc}{section}{Declaration of Generative AI and
AI-Assisted Technologies in the Manuscript Preparation Process}

During the preparation of this work, the author used OpenAI ChatGPT and
OpenAI Codex to support language editing, consistency checks,
LaTeX/Quarto formatting, and code debugging for figure and
submission-file preparation. After using these tools, the author
reviewed and edited the content as needed and takes full responsibility
for the content of the published article.

\section*{References}\label{references}
\addcontentsline{toc}{section}{References}

\protect\phantomsection\label{refs}

\section*{Appendix}\label{appendix}
\addcontentsline{toc}{section}{Appendix}

\subsection*{A Proofs of Proposition 1, Corollary 2, and Propositions
2--3}\label{a-proofs-of-proposition-1-corollary-2-and-propositions-23}
\addcontentsline{toc}{subsection}{A Proofs of Proposition 1, Corollary
2, and Propositions 2--3}

\subsubsection*{Proof of Proposition 1}\label{proof-of-proposition-1}
\addcontentsline{toc}{subsubsection}{Proof of Proposition 1}

Let type differences be denoted by \[
\sigma_t^c:=c_t^S-c_t^H,\qquad
\sigma_t^n:=n_t^S-n_t^H,\qquad
\sigma_t^w:=w_t^S-w_t^H.
\] Since firm profits accrue to savers, the log-linear budget
constraints imply \[
\sigma_t^c=\sigma_t^w+\sigma_t^n+\frac{1}{1-\lambda}d_t.
\] Labor demand gives \[
\sigma_t^n=-\psi_w\sigma_t^w.
\] The type-specific wage Phillips curves imply that the wage-inflation
gap is proportional to the wage-markup gap: \[
\frac{\eta_w}{\psi_w}\sigma_t^w
=-\left\{\sigma_t^w-\gamma\sigma_t^c-\varphi\sigma_t^n\right\}.
\] Substituting the first two equations into the wage-setting condition
gives \[
\left\{
1+\frac{\eta_w}{\psi_w}+\varphi\psi_w+\gamma(\psi_w-1)
\right\}\sigma_t^w
=\frac{\gamma}{1-\lambda}d_t.
\] Therefore \[
\sigma_t^w=\frac{\delta_w}{1-\lambda}d_t,\qquad
\sigma_t^n=-\frac{\delta_n}{1-\lambda}d_t,\qquad
\sigma_t^c=\frac{\delta_c}{1-\lambda}d_t,
\] where \[
\begin{aligned}
\delta_{c}&=\frac{1+(\eta_{w}/\psi_{w})+\varphi\psi_{w}}
{1+(\eta_{w}/\psi_{w})+\varphi\psi_{w}+\gamma(\psi_{w}-1)},\\
\delta_{n}&=\frac{\gamma\psi_{w}}
{1+(\eta_{w}/\psi_{w})+\varphi\psi_{w}+\gamma(\psi_{w}-1)},\\
\delta_{w}&=\frac{\gamma}
{1+(\eta_{w}/\psi_{w})+\varphi\psi_{w}+\gamma(\psi_{w}-1)}.
\end{aligned}
\] Combining these gaps with the aggregation identities \[
c_t=\lambda c_t^H+(1-\lambda)c_t^S,\quad
n_t=\lambda n_t^H+(1-\lambda)n_t^S,\quad
w_t=\lambda w_t^H+(1-\lambda)w_t^S
\] delivers the type-specific allocations stated in Proposition 1. The
identity \(\delta_c+\delta_n-\delta_w=1\) follows directly from the
definitions.

\subsubsection*{Proof of Corollary 2}\label{proof-of-corollary-2}
\addcontentsline{toc}{subsubsection}{Proof of Corollary 2}

This subsection proves Corollary 2. Recall \[
\delta_{c}=\frac{1+(\eta_{w}/\psi_{w})+\varphi\psi_{w}}
{1+(\eta_{w}/\psi_{w})+\varphi\psi_{w}+\gamma(\psi_{w}-1)},
\] \[
\delta_{n}=\frac{\gamma\psi_{w}}
{1+(\eta_{w}/\psi_{w})+\varphi\psi_{w}+\gamma(\psi_{w}-1)},
\qquad
\delta_{w}=\frac{\gamma}
{1+(\eta_{w}/\psi_{w})+\varphi\psi_{w}+\gamma(\psi_{w}-1)}.
\] Let \[
\bar Q_w:=1+\frac{\eta_{w}}{\psi_{w}}+\varphi\psi_{w}+\gamma(\psi_{w}-1).
\] Then \(\bar Q_w>0\), and \[
\delta_c=\frac{1+\eta_w/\psi_w+\varphi\psi_w}{\bar Q_w},
\quad
\delta_n=\frac{\gamma\psi_w}{\bar Q_w},
\quad
\delta_w=\frac{\gamma}{\bar Q_w}.
\]

Differentiating with respect to wage adjustment costs gives \[
\frac{\partial\delta_c}{\partial\eta_w}
=
\frac{\gamma(\psi_w-1)}{\psi_w\bar Q_w^2}\ge0,
\quad
\frac{\partial\delta_n}{\partial\eta_w}
=
-\frac{\gamma}{\bar Q_w^2}<0,
\quad
\frac{\partial\delta_w}{\partial\eta_w}
=
-\frac{\gamma}{\psi_w\bar Q_w^2}<0.
\] Thus greater wage adjustment costs raise \(\delta_c\) and lower
\(\delta_n\) and \(\delta_w\).

Differentiating with respect to labor-market competitiveness gives \[
\frac{\partial\delta_c}{\partial\psi_w}
=
-\frac{
\gamma\left[
1+\varphi+\eta_w(2\psi_w-1)/\psi_w^2
\right]
}{\bar Q_w^2}<0,
\] \[
\frac{\partial\delta_n}{\partial\psi_w}
=
\frac{
\gamma(1-\gamma+2\eta_w/\psi_w)
}{\bar Q_w^2},
\] and \[
\frac{\partial\delta_w}{\partial\psi_w}
=
\frac{
\gamma\{\eta_w/\psi_w^2-(\varphi+\gamma)\}
}{\bar Q_w^2}.
\] Thus \(\delta_c\) decreases with \(\psi_w\), while the signs of the
labor and wage components depend on the remaining parameters. This
proves Corollary 2.

\subsubsection*{Derivation of the AD Curve Used in Section
4.3}\label{derivation-of-the-ad-curve-used-in-section-4.3}
\addcontentsline{toc}{subsubsection}{Derivation of the AD Curve Used in
Section 4.3}

Combining the TANK IS curve and the monetary policy rule gives \[
\gamma E_t\Delta x_{t+1}+m_t
=
\phi\pi_t^p-E_t\pi_{t+1}^p+\gamma\delta_pE_t\Delta\pi_{t+1}^p .
\]

Equivalently, \[
\gamma E_t\Delta x_{t+1}+m_t
=
\left\{
\phi-\gamma\delta_p-(1-\gamma\delta_p)L^{-1}
\right\}\pi_t^p .
\]

Forward iteration under \(1-\gamma\delta_p>0\) and the Taylor-principle
condition yields \[
\pi_t^p
=
\frac{
\gamma E_t\sum_{n=0}^{\infty}
\left(
\frac{1-\gamma\delta_p}{\phi-\gamma\delta_p}
\right)^n
\Delta x_{t+1+n}
}{
\phi-\gamma\delta_p
}
+
\frac{m_t}{
\phi-\rho_m-\gamma\delta_p(1-\rho_m)
}.
\]

This is the AD curve used in Section 4.3.

\subsubsection*{Proof of Proposition 2}\label{proof-of-proposition-2}
\addcontentsline{toc}{subsubsection}{Proof of Proposition 2}

\paragraph*{1. Preliminary: Transformation of the Characteristic
Function}\label{preliminary-transformation-of-the-characteristic-function}
\addcontentsline{toc}{paragraph}{1. Preliminary: Transformation of the
Characteristic Function}

The characteristic function can be written as \[
f(\xi)=
\left(\frac{\gamma\kappa_{1}}{\kappa_{2}}+1-\gamma\delta_p\right)(\xi-1)\left(\xi -\frac{\gamma}{\gamma\kappa_{1}+(1-\gamma\delta_p)\kappa_{2}}\right)-(\phi-1)\xi.
\] Notice that \[
\frac{\gamma}{\gamma\kappa_{1}+(1-\gamma\delta_p)\kappa_{2}}
\] is increasing in \(\delta_p\) and decreasing in \(\kappa_1\). It is
also decreasing in \(\kappa_2\) when \(1-\gamma\delta_p>0\).

The same function can also be written as \[
\begin{aligned}
f(\xi) & =\left(\frac{\gamma\kappa_{1}}{\kappa_{2}}+1-\gamma\delta_p\right)\xi^{2}-\left(\phi-1+\frac{\gamma}{\kappa_{2}}+\frac{\gamma\kappa_{1}}{\kappa_{2}}+1-\gamma\delta_p\right)\xi+\frac{\gamma}{\kappa_{2}}\\
 & =\Xi_{a}\xi^{2}-\Xi_{b}\xi+\Xi_{c},
\end{aligned}
\] where \(\Xi_{a}  =\gamma\kappa_{1}/\kappa_{2}+1-\gamma\delta_p\),
\(\Xi_{b} =\phi-1+\Xi_{a}+\Xi_{c}\), and
\(\Xi_{c}  =\gamma/\kappa_{2}\). Assume \(\Xi_a>0\). Since \(\phi>1\)
and \(\Xi_c>0\), it follows that \(\Xi_b>0\). Let \(\xi_1\) denote the
root smaller than 1 and \(\xi_2\) the root greater than 1.

\paragraph*{\texorpdfstring{2. Effect of
\texorpdfstring{$\phi$}{phi}}{2. Effect of }}\label{effect-of}
\addcontentsline{toc}{paragraph}{2. Effect of
\texorpdfstring{$\phi$}{phi}}

\begin{figure}

\centering{

\includegraphics[width=0.7\linewidth,height=\textheight,keepaspectratio]{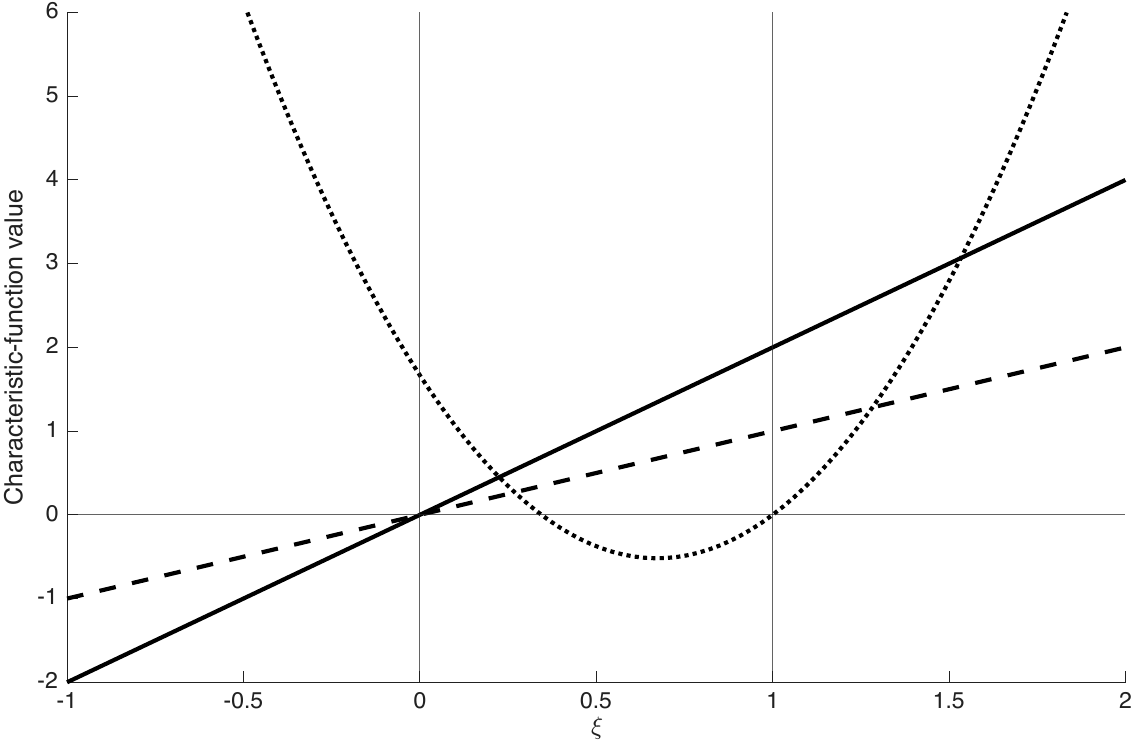}

\par\smallskip
\begin{minipage}{0.7\linewidth}
\footnotesize \textit{Notes:} The dotted curve represents $\left(\frac{\gamma\kappa_{1}}{\kappa_{2}}+1-\gamma\delta_{p}\right)(\xi-1)\left(\xi-\frac{\gamma}{\gamma\kappa_{1}+(1-\gamma\delta_{p})\kappa_{2}}\right)$. The dashed line represents $(\phi-1)\xi$. The solid line represents $(\phi'-1)\xi$ with $\phi'>\phi$. The two intersections of the dotted curve and the dashed line are the initial roots of the characteristic function. The intersections of the dotted curve and the solid line are the roots of the characteristic function with $\phi'>\phi$.
\end{minipage}

}

\caption{\label{fig-fig1}Policy feedback and the characteristic roots}

\end{figure}

The quadratic formula implies that increasing \(\phi\) lowers
\(\xi_{1}\) and raises \(\xi_{2}\). That is, \[
\frac{\partial\xi_1}{\partial\phi}<0,\;\frac{\partial\xi_2}{\partial\phi}>0.
\] This occurs because \(\phi\) directly increases \(\Xi_b\) in the
characteristic equation; Figure~\ref{fig-fig1} gives the corresponding
graphical intuition. As a result, both \(\Omega^{p}\) and \(\Omega^{x}\)
decrease with increasing \(\phi\).

\paragraph*{\texorpdfstring{3. Effect of
\texorpdfstring{$\delta_p$}{delta_p}}{3. Effect of }}\label{effect-of-1}
\addcontentsline{toc}{paragraph}{3. Effect of
\texorpdfstring{$\delta_p$}{delta_p}}

\begin{figure}

\centering{

\includegraphics[width=0.7\linewidth,height=\textheight,keepaspectratio]{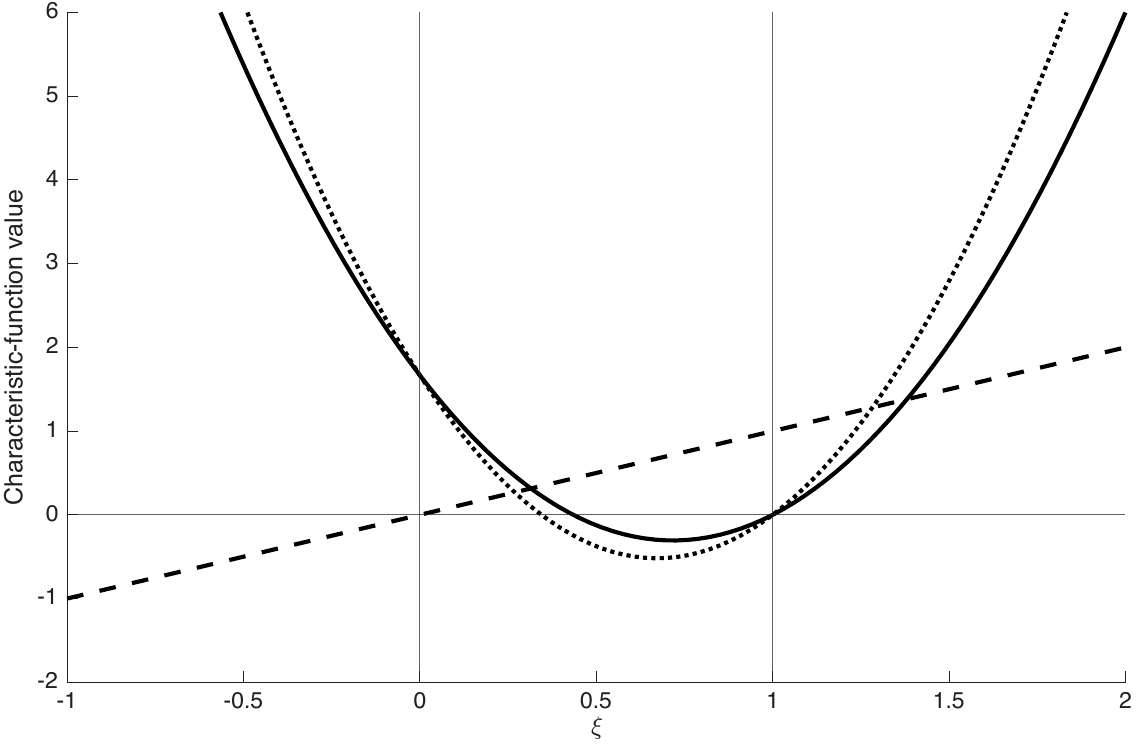}

\par\smallskip
\begin{minipage}{0.7\linewidth}
\footnotesize \textit{Notes:} The dotted curve represents $\left(\frac{\gamma\kappa_{1}}{\kappa_{2}}+1-\gamma\delta_{p}\right)(\xi-1)\left(\xi-\frac{\gamma}{\gamma\kappa_{1}+(1-\gamma\delta_{p})\kappa_{2}}\right)$. The dashed line represents $(\phi-1)\xi$. The solid curve represents $\left(\frac{\gamma\kappa_{1}}{\kappa_{2}}+1-\gamma\delta_{p}'\right)(\xi-1)\left(\xi-\frac{\gamma}{\gamma\kappa_{1}+(1-\gamma\delta_{p}')\kappa_{2}}\right)$ with $\delta_{p}'>\delta_{p}$. The two intersections of the dotted curve and the dashed line are the initial roots of the characteristic function. The intersections of the solid curve and the dashed line are the roots of the characteristic function with $\delta_{p}'>\delta_{p}$.
\end{minipage}

}

\caption{\label{fig-fig2}Profit redistribution and the characteristic
roots}

\end{figure}

Given \(\kappa_1\) and \(\kappa_2\), the quadratic formula implies that
increasing \(\delta_p\) affects both roots as follows: \[
\frac{\partial\xi_1}{\partial\delta_p}>0,\;\frac{\partial\xi_2}{\partial\delta_p}>0.
\] Figure \ref{fig-fig2} gives the same root movement graphically, but
the sign restriction for the multipliers follows from the algebra below.

Since the denominators of \(\Omega^p\) and \(\Omega^x\) are
\(\Xi_a(\xi_{2}-\rho_m)\) and
\((\kappa_{2}/\kappa_1)\Xi_a(\xi_{2}-\rho_m)\), respectively, it is
sufficient to show that increasing \(\delta_p\) lowers
\(\Xi_a(\xi_{2}-\rho_m)\) to prove \[
\frac{\partial\Omega^p}{\partial\delta_p}>0,\;\frac{\partial\Omega^x}{\partial\delta_p}>0
\] given \(\kappa_1\) and \(\kappa_2\).

The larger root satisfies \[
\Xi_{a}\xi_{2} =\frac{\Xi_{b}+\sqrt{\Xi_{b}^{2}-4\Xi_{a}\Xi_{c}}}{2}.
\] Since \[
\begin{aligned}
\Xi_{b}^{2}-4\Xi_{a}\Xi_{c} & =(\phi-1)^{2}+2(\phi-1)(\Xi_{a}+\Xi_{c})+(\Xi_{a}+\Xi_{c})^{2}-4\Xi_{a}\Xi_{c}\\
&=(\phi-1)^{2}+2(\phi-1)(\Xi_{a}+\Xi_{c})+(\Xi_{a}-\Xi_{c})^{2}\\
 & =(\phi-1+\Xi_{a}-\Xi_{c})^{2}+4(\phi-1)\Xi_{c},
\end{aligned}
\] the derivative of \(\Xi_{a}\xi_{2}\) is \[
\begin{aligned}
\frac{\partial\Xi_{a}\xi_{2}}{\partial\delta_{p}} & =\frac{\partial\Xi_{b}/\partial\delta_p}{2}+\frac{\Xi_{b}\partial\Xi_{a}/\partial\delta_p-2\Xi_{c}\partial\Xi_{a}/\partial\delta_p}{2\sqrt{\Xi_{b}^{2}-4\Xi_{a}\Xi_{c}}}\\
 & =-\frac{\gamma}{2}-\frac{\gamma Z}{2\sqrt{D}}.
\end{aligned}
\] where \[
Z
:=
\phi+\gamma\frac{\kappa_1-1}{\kappa_2}
-\gamma\delta_p,
\qquad
D
:=
Z^2+4(\phi-1)\frac{\gamma}{\kappa_2}.
\] Hence \[
\frac{\partial{\Xi_a(\xi_2-\rho_m)}}{\partial\delta_p}
=
\gamma
\left[
\rho_m-\frac12-\frac{Z}{2\sqrt D}
\right].
\] Thus \(\Xi_a(\xi_2-\rho_m)\) decreases with \(\delta_p\) only under
\[
\rho_m
<
\frac12
\left(
1+\frac{Z}{\sqrt D}
\right).
\] Under this additional condition, increasing \(\delta_p\) raises both
\(\Omega^p\) and \(\Omega^x\) given \(\kappa_1\) and \(\kappa_2\).

\paragraph*{\texorpdfstring{4. Effect of
\texorpdfstring{$\lambda$}{lambda}}{4. Effect of }}\label{effect-of-2}
\addcontentsline{toc}{paragraph}{4. Effect of
\texorpdfstring{$\lambda$}{lambda}}

Since \(\lambda\) does not directly affect \(\kappa_1\) or \(\kappa_2\),
it operates only through \(\delta_p\). Given that \(\delta_p\) increases
with \(\lambda\), the local derivatives around parameter values
satisfying condition (\ref{eq-condition0}) obey
\[\frac{d\Omega^p}{d\lambda}>0, \quad \frac{d\Omega^x}{d\lambda}>0.\]

\paragraph*{\texorpdfstring{5. Effects of
\texorpdfstring{$\eta_j$}{eta_j} and \texorpdfstring{$\psi_j$}{psi_j}
(\texorpdfstring{$j=p,w$}{j = p, w})}{5. Effects of  and  ()}}\label{effects-of-and}
\addcontentsline{toc}{paragraph}{5. Effects of
\texorpdfstring{$\eta_j$}{eta_j} and \texorpdfstring{$\psi_j$}{psi_j}
(\texorpdfstring{$j=p,w$}{j = p, w})}

\begin{itemize}
\tightlist
\item
  \emph{Direct Effects (holding} \(\delta_p\) constant)
\end{itemize}

\begin{figure}

\centering{

\includegraphics[width=0.7\linewidth,height=\textheight,keepaspectratio]{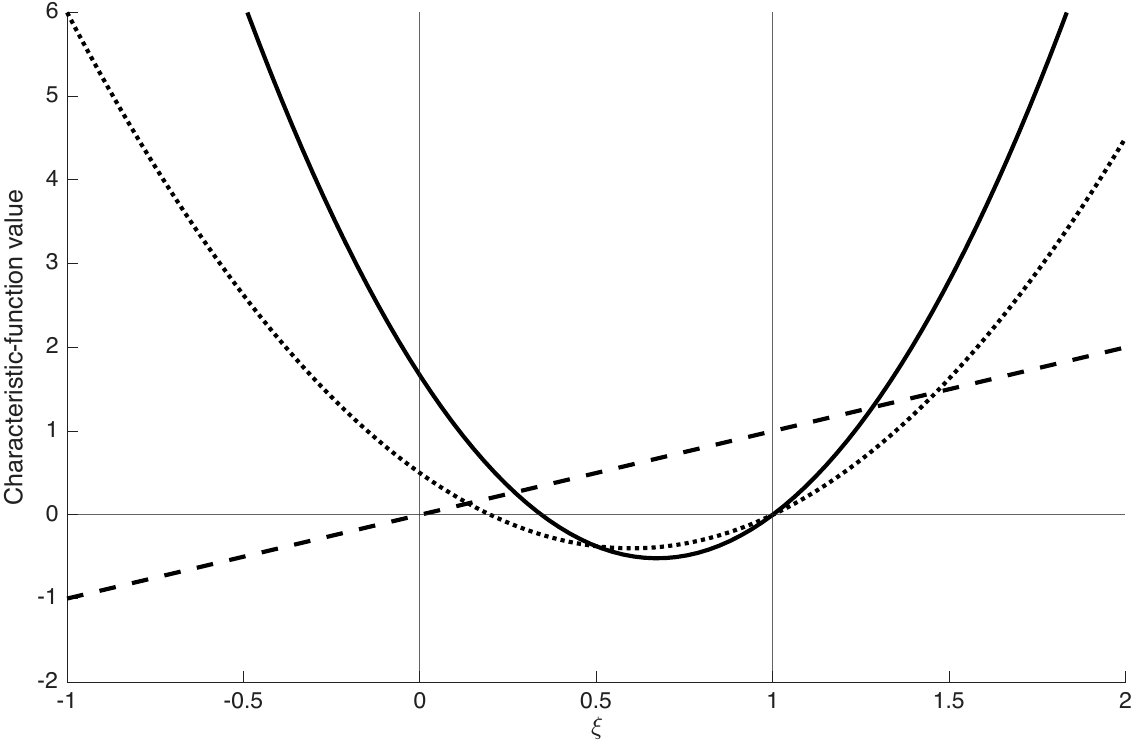}

\par\smallskip
\begin{minipage}{0.7\linewidth}
\footnotesize \textit{Notes:} The dotted curve represents $\left(\frac{\gamma\kappa_{1}}{\kappa_{2}}+1-\gamma\delta_{p}\right)(\xi-1)\left(\xi-\frac{\gamma}{\gamma\kappa_{1}+(1-\gamma\delta_{p})\kappa_{2}}\right)$. The dashed line represents $(\phi-1)\xi$. The solid curve represents $\left(\frac{\gamma\kappa_{1}'}{\kappa_{2}'}+1-\gamma\delta_{p}\right)(\xi-1)\left(\xi-\frac{\gamma}{\gamma\kappa_{1}'+(1-\gamma\delta_{p})\kappa_{2}'}\right)$ with $\kappa_{1}'<\kappa_{1}$, $\kappa_{2}'<\kappa_{2}$, and $\kappa_{2}'/\kappa_{1}'<\kappa_{2}/\kappa_{1}$. The two intersections of the dotted curve and the dashed line are the initial roots of the characteristic function. The intersections of the solid curve and the dashed line are the roots of the characteristic function with $\kappa_{1}'<\kappa_{1}$, $\kappa_{2}'<\kappa_{2}$, and $\kappa_{2}'/\kappa_{1}'<\kappa_{2}/\kappa_{1}$.
\end{minipage}

}

\caption{\label{fig-fig3}Nominal rigidity and the characteristic roots}

\end{figure}

For an algebraic proof, hold \(\delta_p\) fixed and define \[
u_p:=\frac{\eta_p}{\psi_p},\qquad
u_w:=\frac{\eta_w}{\psi_w},\qquad
G:=\gamma+\varphi,\qquad
q:=1-\gamma\delta_p.
\] For interior values \(u_p,u_w>0\), the characteristic coefficients
become \[
\Xi_a=q+\frac{\gamma}{G}(u_p+u_w+u_pu_w),
\qquad
\Xi_c=\frac{\gamma}{G}u_pu_w,
\] and the characteristic function can be written as \[
f(\xi)=(\Xi_a\xi-\Xi_c)(\xi-1)-(\phi-1)\xi.
\] Differentiation with respect to \(u_p\) gives \[
\frac{\partial f(\xi)}{\partial u_p}
=\frac{\gamma}{G}(1-\xi)\{u_w-(1+u_w)\xi\}.
\] At \(r_p:=u_w/(1+u_w)\), \[
f(r_p)
=-\frac{u_w}{(1+u_w)^2}
\left(q+\frac{\gamma}{G}u_w\right)
-(\phi-1)\frac{u_w}{1+u_w}<0
\] when \(q>0\). Because \(f(0)=\Xi_c>0\), the stable root satisfies
\(0<\xi_1<r_p\); because \(\xi_2>1\), it follows that
\(f_{u_p}(\xi_1)>0\) and \(f_{u_p}(\xi_2)>0\). Since
\(f_\xi(\xi_1)<0<f_\xi(\xi_2)\), implicit differentiation yields \[
\frac{\partial\xi_1}{\partial u_p}>0,
\qquad
\frac{\partial\xi_2}{\partial u_p}<0.
\] The same argument with \(p\) and \(w\) interchanged, using
\(r_w:=u_p/(1+u_p)\), gives the corresponding signs for \(u_w\).
Therefore, for \(j\in\{p,w\}\), \[
\frac{\partial\xi_1}{\partial\eta_j}>0,\quad
\frac{\partial\xi_2}{\partial\eta_j}<0,\quad
\frac{\partial\xi_1}{\partial\psi_j}<0,\quad
\frac{\partial\xi_2}{\partial\psi_j}>0.
\] Moreover, \[
\frac{\kappa_2}{\kappa_1}
=\frac{G}{u_p+u_w+u_pu_w},
\] which falls as either \(u_p\) or \(u_w\) rises. Hence, under
\(q=1-\gamma\delta_p>0\), greater rigidity lowers both positive factors
in \[
\{\gamma+(\kappa_{2}/\kappa_{1})(1-\gamma\delta_{p})\}(\xi_{2}-\rho_m),
\] the denominator of \(\Omega^x\). Consequently, \[
\frac{\partial\Omega^x}{\partial\eta_j}>0,\;
\frac{\partial\Omega^x}{\partial\psi_j}<0.
\] Without this additional condition, the sign of the direct effect on
\(\Omega^x\) is not guaranteed. Figure \ref{fig-fig3} gives the
corresponding root movement graphically.

The multiplier sign also depends on the non-root term in the denominator
of \(\Omega^x\), which is why \(1-\gamma\delta_p>0\) is imposed for the
rigidity comparative statics of \(\Omega^x\).

For \(\Omega^p\), the direct rigidity effect works in the standard
direction under \(1-\gamma\delta_p>0\), but the total effect is not
signed in the heterogeneous case because the indirect effect through
\(\delta_p\) can offset it. Accordingly, Proposition 2 does not state a
general rigidity comparative static for \(\Omega^p\) when \(\lambda>0\).
Under the additional condition \(1-\gamma\delta_p>0\), it also follows
that \[
\frac{\partial\Omega^x}{\partial\eta_j}>0,\;
\frac{\partial\Omega^x}{\partial\psi_j}<0.
\]

\begin{itemize}
\tightlist
\item
  \emph{Indirect Effects (through changes in} \(\delta_p\))
\end{itemize}

By Corollary 2 and the definition of \(\delta_p\), \(\delta_p\)
increases with \(\eta_p\) and \(\eta_w\) and decreases with \(\psi_p\)
and \(\psi_w\). Under condition (\ref{eq-condition0}) and
\(1-\gamma\delta_p>0\), the total effects for \(\Omega^x\) are:

For \(\Omega^x\), \[\begin{aligned}
\frac{d\Omega^x}{d\eta_j}&=
\frac{\partial\Omega^x}{\partial\eta_j}
+\frac{\partial\Omega^x}{\partial\delta_p}
\frac{d\delta_p}{d\eta_j}>0,\\
\frac{d\Omega^x}{d\psi_j}&=
\frac{\partial\Omega^x}{\partial\psi_j}
+\frac{\partial\Omega^x}{\partial\delta_p}
\frac{d\delta_p}{d\psi_j}<0.
\end{aligned}
\] The direct rigidity effect on \(\Omega^x\) then has the same sign as
the indirect effect through \(\delta_p\), so the two effects reinforce
each other. Without these restrictions, the total effect of \(\eta_j\)
or \(\psi_j\) on \(\Omega^x\) is not signed in general. For
\(\Omega^p\), on the other hand, the effects are ambiguous.

\paragraph*{\texorpdfstring{6. Special Case:
\texorpdfstring{$\lambda=0$}{lambda = 0}}{6. Special Case: }}\label{special-case}
\addcontentsline{toc}{paragraph}{6. Special Case:
\texorpdfstring{$\lambda=0$}{lambda = 0}}

When \(\lambda=0\), \(\delta_p=0\). Hence the indirect effect through
\(\delta_p\) disappears and \(1-\gamma\delta_p=1>0\). The rigidity
comparative statics for \(\Omega^p\) therefore reduce to the direct
representative-agent case: \[
d\Omega^p/d\eta_j=\partial\Omega^p/\partial\eta_j<0,
\qquad
d\Omega^p/d\psi_j=\partial\Omega^p/\partial\psi_j>0.
\] Thus condition (\ref{eq-condition0}), which is needed for the
heterogeneous impact \(\lambda\) effect, is not needed for this
representative-agent comparison.

\subsubsection*{Proof of Proposition 3}\label{proof-of-proposition-3}
\addcontentsline{toc}{subsubsection}{Proof of Proposition 3}

First, evaluate the double sum \[
S = \sum_{n=0}^{\infty} \sum_{i=0}^{n} \xi_1^{n-i} \rho_m^{i}
\]

Consider the case with \(\xi_1>\rho_m\). The inner sum is \[
\sum_{i=0}^{n} \xi_1^{n-i} \rho_m^{i} = \xi_1^n \sum_{i=0}^{n} \left( \frac{\rho_m}{\xi_1} \right)^i = \xi_1^n \cdot \frac{1 - \left( \frac{\rho_m}{\xi_1} \right)^{n+1}}{1 - \frac{\rho_m}{\xi_1}} = \frac{\xi_1^{n+1}- \rho_m^{n+1}}{\xi_1-\rho_m}
\] The double sum therefore becomes \[
S = \sum_{n=0}^{\infty} \frac{\xi_1^{n+1}-\rho_m^{n+1}}{\xi_1-\rho_m} = \frac{1}{\xi_1 - \rho_m} \left( \sum_{n=0}^{\infty} \xi_1^{n+1} -\sum_{n=0}^{\infty} \rho_m^{n+1}\right)
\]

Convergence of the infinite series requires \(|\xi_1| < 1\) and
\(|\rho_m| < 1\). Under these conditions, \[
\begin{aligned}
\sum_{n=0}^{\infty} \xi_1^{n+1} &= \frac{\xi_1}{1 - \xi_1}\\
\sum_{n=0}^{\infty} \rho_m^{n+1} &= \frac{\rho_m}{1 - \rho_m}
\end{aligned}
\]

Thus \[
\begin{aligned}
S &= \frac{1}{\xi_1-\rho_m } \left( \frac{\xi_1}{1 - \xi_1}-\frac{\rho_m}{1 - \rho_m} \right)\\
&=\frac{1}{(1 - \rho_m)(1 - \xi_1)}
\end{aligned}
\]

When \(\xi_1<\rho_m\), swapping the two parameters in the double sum
gives the same result. When \(\xi_1 = \rho_m = \rho\), the result is
\(1/(1-\rho)^2\).

Using the evaluation of the double sum gives
\begin{equation}\protect\phantomsection\label{eq-omega-infty}{
\begin{aligned}
\Omega^p_\infty
&=\frac{\Omega^p}{(1 - \rho_m)(1 - \xi_1)}\\
&=\frac{1}{1-\rho_m}
\frac{1}{(\kappa_{1}/\kappa_{2})\gamma+(1-\gamma\delta_{p})}\frac{1}{(1-\xi_{1})(\xi_{2}-\rho_{m})}
\end{aligned}
}\end{equation} Some algebra gives \[
\begin{aligned}
(1-\xi_{1})(\xi_{2}-\rho_{m})
&=(1-\xi_{1})(\xi_{2}-1+1-\rho_{m})\\   &=-(1-\xi_{1})(1-\xi_{2})+(1-\xi_{1})(1-\rho_{m})\\
&=-1+(\xi_{1}+\xi_{2})-\xi_{1}\xi_{2}+(1-\xi_{1})(1-\rho_{m}).
\end{aligned}
\] Using the relation between roots and coefficients, \[
\begin{aligned}
\xi_{1}\xi_{2}&=\frac{\gamma/\kappa_{2}}{1-\gamma\delta_{p}+\gamma\kappa_{1}/\kappa_{2}},\\
\xi_{1}+\xi_{2}&=\frac{\phi-\gamma\delta_{p}+\gamma\kappa_{1}/\kappa_{2}+\gamma/\kappa_{2}}{1-\gamma\delta_{p}+\gamma\kappa_{1}/\kappa_{2}},
\end{aligned}
\] it follows that \[
\begin{aligned}
-1+(\xi_{1}+\xi_{2})-\xi_{1}\xi_{2}&=-1+\frac{\phi-\gamma\delta_{p}+\gamma\kappa_{1}/\kappa_{2}}{1-\gamma\delta_{p}+\gamma\kappa_{1}/\kappa_{2}}\\
&=\frac{\phi-1}{1-\gamma\delta_{p}+\gamma\kappa_{1}/\kappa_{2}}.
\end{aligned}
\] Substituting this result into equation (\ref{eq-omega-infty}) gives
\[
\begin{aligned}
\Omega_\infty^p&=\frac{1}{1-\rho_{m}}\frac{1}{(\kappa_{1}/\kappa_{2})\gamma+(1-\gamma\delta_{p})}\frac{1}{(1-\xi_{1})(\xi_{2}-\rho_{m})}\\
&=\frac{1}{1-\rho_{m}}\frac{1}{(\kappa_{1}/\kappa_{2})\gamma+(1-\gamma\delta_{p})}\frac{1}{-1+(\xi_{1}+\xi_{2})-\xi_{1}\xi_{2}+(1-\xi_{1})(1-\rho_{m})}\\
&=\frac{1}{1-\rho_{m}}\frac{1}{\phi-1+\{(\kappa_{1}/\kappa_{2})\gamma+(1-\gamma\delta_{p})\}(1-\xi_{1})(1-\rho_{m})}.
\end{aligned}
\] Since an increase in \(\phi\) lowers \(\xi_1\), \(\Omega_\infty^p\)
decreases with \(\phi\). Given \(\eta_j\) and \(\psi_j\) (\(j=p,w\)), a
local increase in \(\lambda\) that preserves determinacy raises both
\(\xi_1\) and \(\delta_p\), so \(\Omega_\infty^p\) increases. Indeed, an
increase in \(\delta_p\) lowers \[
(\kappa_1/\kappa_2)\gamma+(1-\gamma\delta_p)
\] and raises \(\xi_1\), so both factors in the second term of the
denominator decline. Given \(\delta_p\) and \(1-\gamma\delta_p>0\), an
increase in \(\eta_j\) or a decrease in \(\psi_j\) raises
\(\kappa_1/\kappa_2\) and \(\xi_1\), but \(\Omega_\infty^p\) remains
ambiguous with respect to \(\eta_j\) and \(\psi_j\). Without this
stronger condition, the root response is itself not signed, reinforcing
that ambiguity.

For output, \[
\begin{aligned}
\Omega_{\infty}^{x}&=\frac{\xi_{1}\kappa_{1}-1}{\kappa_{2}}\Omega_{\infty}^{p}+\frac{\Omega^{x}}{1-\rho_{m}}\\
&=\frac{\xi_{1}\kappa_{1}-1}{\kappa_{2}}\frac{\Omega^{p}}{(1-\xi_{1})(1-\rho_{m})}+\frac{(\kappa_{1}/\kappa_{2})\Omega^{p}}{1-\rho_{m}}\\
&=\left\{ \frac{\kappa_{1}}{\kappa_{2}}+\frac{\xi_{1}\kappa_{1}-1}{\kappa_{2}}\frac{1}{1-\xi_{1}}\right\}\frac{\Omega^{p}}{1-\rho_{m}}\\
&=\left\{ \frac{\kappa_{1}/\kappa_{2}-1/\kappa_{2}}{1-\xi_{1}}\right\}\frac{\Omega^{p}}{1-\rho_{m}} \\
&=\frac{\psi_{p}/\eta_{p}+\psi_{w}/\eta_{w}}{(\varphi+\gamma)(\psi_{p}/\eta_{p})(\psi_{w}/\eta_{w})}\Omega_{\infty}^{p}\\
&=\frac{\eta_{w}/\psi_{w}+\eta_{p}/\psi_{p}}{\varphi+\gamma}\Omega_{\infty}^{p}.
\end{aligned}
\] Thus \(\Omega_\infty^x\) decreases with \(\phi\) and increases
locally with \(\lambda\) for changes that preserve determinacy.

\subsection*{B Aggregate-Lag Optimal-Policy
Benchmark}\label{b-aggregate-lag-optimal-policy-benchmark}
\addcontentsline{toc}{subsection}{B Aggregate-Lag Optimal-Policy
Benchmark}

This appendix collects the optimal-policy problems implied by the
aggregate-lag specialization of the welfare loss in Section 4.4. All
targeting conditions below condition on the aggregate-lag adjustment
reference used to obtain the closed-form benchmark. They do not
characterize optimal policy in the conventional own-lag economy, where
relative wages are endogenous states and the welfare objective contains
separate consumption-, hours-, and wage-inflation-dispersion terms. The
exercise characterizes the benchmark optimal allocation rather than a
proposed implementable interest-rate rule.

Consider optimal policy under alternative monetary regimes. Under
discretion, the central bank selects \(x_t\), \(\pi_t^p\), and
\(\pi_t^w\) to minimize the period-by-period loss function \[
\min \left\{ (\gamma+\varphi)x_{t}^{2}+\eta_{w}(\pi_{t}^{w})^{2}+\tilde{\eta}_{p}(\pi_{t}^{p})^{2}\right\},\]
subject to (\ref{eq-TPC}) and (\ref{eq-wage-inflation2}).

The model does not exhibit divine coincidence. In a classic RANK model
with price rigidity alone, complete stabilization is possible through
inflation targeting: as Blanchard and Galí (2007) show, \(\pi_t^p=0\)
automatically delivers \(x_t=0\). Breaking this coincidence typically
requires additional disturbances like cost-push shocks to the Phillips
curve. Here, wage stickiness is enough to break that alignment.
Technology shocks, even without cost-push disturbances, can drive a
wedge between natural output under flexible prices and natural output
under flexible wages. Policymakers therefore face a real trade-off:
stabilizing price inflation can require output deviations from the
flexible-wage natural level, and vice versa. This is why the objective
function penalizes both output gaps and wage inflation.

Let \(\zeta_p:=\eta_p/\psi_p\). The discretionary targeting condition is
\[
(\gamma+\varphi)\frac{\kappa_1}{\kappa_2}x_t
+
\tilde{\eta}_p\pi_t^p
+
\eta_w(1+\zeta_p)\pi_t^w
=
0.
\] Using the wage-inflation identity \[
\pi_t^w
=
(1+\zeta_p)\pi_t^p-\zeta_p\pi_{t-1}^p+\Delta a_t,
\] the discretionary solution is \[
\begin{aligned}
x_t
=&
-\frac{\kappa_2}{\kappa_1(\gamma+\varphi)}
\left[
\{\tilde{\eta}_p+\eta_w(1+\zeta_p)^2\}\pi_t^p
-\eta_w\zeta_p(1+\zeta_p)\pi_{t-1}^p
+\eta_w(1+\zeta_p)\Delta a_t
\right]\\
=&
-\frac{\psi_p\psi_w}{\eta_p\eta_w+\eta_w\psi_p+\eta_p\psi_w}
\left[
\left\{
\tilde{\eta}_p+\eta_w\left(1+\frac{\eta_p}{\psi_p}\right)^2
\right\}\pi_t^p
-\eta_w\frac{\eta_p}{\psi_p}
\left(1+\frac{\eta_p}{\psi_p}\right)
\pi_{t-1}^p
+\eta_w
\left(1+\frac{\eta_p}{\psi_p}\right)
\Delta a_t
\right].
\end{aligned}
\] The monetary authority chooses \(x_t\) to balance price-inflation and
output-gap stabilization within the benchmark constraint set. The
negative relationship is as expected: tighter output gaps reduce
inflationary pressure. Holding the other states and constraints fixed, a
larger \(\lambda\) raises the coefficient on price inflation through
\(\tilde{\eta}_p\). This conditional targeting relation does not
establish that an implementable interest-rate rule should react more
aggressively, and it does not characterize the own-lag policy problem.

Commitment policy follows the timeless perspective in Woodford (2003)
for RANK economies. Under commitment, the central bank chooses the
entire future sequence \(\{x_t,\;\pi_t^p,\;\pi_t^w\}_{t=0}^\infty\) to
minimize the intertemporal loss function \[
\min \sum_{t=0}^\infty\beta^t\left\{ (\gamma+\varphi)x_{t}^{2}+\eta_{w}(\pi_{t}^{w})^{2}+\tilde{\eta}_{p}(\pi_{t}^{p})^{2}\right\},
\] subject to equations (\ref{eq-TPC}) and (\ref{eq-wage-inflation2}).
The commitment targeting condition is \[
\begin{aligned}
(\gamma+\varphi)\frac{\kappa_1}{\kappa_2}x_t
=&
\frac{\beta(\gamma+\varphi)}{\kappa_2}E_t x_{t+1}
-
\tilde{\eta}_p\pi_t^p
-
\eta_w(1+\zeta_p)\pi_t^w\\
&+
\beta\eta_w\zeta_p E_t\pi_{t+1}^w.
\end{aligned}
\] Using \[
E_t\pi_{t+1}^w
=
(1+\zeta_p)E_t\pi_{t+1}^p
-
\zeta_p\pi_t^p
+
E_t\Delta a_{t+1},
\] this condition can be written as the following reduced targeting
representation: \[
\begin{aligned}
x_t
={}&\frac{\beta}{\kappa_1}E_t x_{t+1}\\
&-\frac{\kappa_2}{\kappa_1(\gamma+\varphi)}
\left[
\{\tilde{\eta}_p+\eta_w(1+\zeta_p)^2+\beta\eta_w\zeta_p^2\}\pi_t^p
-\eta_w\zeta_p(1+\zeta_p)\pi_{t-1}^p
+\eta_w(1+\zeta_p)\Delta a_t
\right]\\
&+\frac{\beta\kappa_2}{\kappa_1(\gamma+\varphi)}
\eta_w\zeta_p
\left[
(1+\zeta_p)E_t\pi_{t+1}^p
+
E_t\Delta a_{t+1}
\right].
\end{aligned}
\] The commitment solution differs from the discretionary solution
because expected-future and history-dependent terms, as well as the
discount factor \(\beta\), enter directly. The authority can influence
private-sector expectations, but the sign and magnitude of the resulting
output-gap response remain state dependent.

Under discretion, the benchmark central bank balances price-inflation
and output-gap stabilization costs period by period. Commitment uses
intertemporal linkages to manage expectations, but the same
aggregate-lag trade-off between inflation and output stabilization
remains in both regimes. Neither result extends mechanically to own-lag
adjustment.

\subsection*{C Additional Quantitative Results and
Robustness}\label{c-additional-quantitative-results-and-robustness}
\addcontentsline{toc}{subsection}{C Additional Quantitative Results and
Robustness}

This appendix reports the detailed policy-rule and household-side
wage-subsidy-financing results summarized in Section 5.3, followed by
technology-shock responses and other parameter checks.

\subsubsection*{Policy-Rule and Household-Side Wage-Subsidy-Financing
Robustness}\label{policy-rule-and-household-side-wage-subsidy-financing-robustness}
\addcontentsline{toc}{subsubsection}{Policy-Rule and Household-Side
Wage-Subsidy-Financing Robustness}

Table~\ref{tbl-policy-rule-ranges} compares the standard output-gap
coefficient \(\phi_x=0.125\) with the inflation-only rule. The output
response in the policy rule attenuates all three monetary-transmission
margins. The aggregate-lag cumulative output response is the most
variable because policy feedback interacts with the persistence
generated by the price-duration target.

\begingroup
\small
\setlength{\tabcolsep}{4pt}

\begin{longtable}[]{@{}
  >{\raggedright\arraybackslash}p{(\linewidth - 12\tabcolsep) * \real{0.1800}}
  >{\raggedleft\arraybackslash}p{(\linewidth - 12\tabcolsep) * \real{0.1300}}
  >{\raggedleft\arraybackslash}p{(\linewidth - 12\tabcolsep) * \real{0.1300}}
  >{\raggedleft\arraybackslash}p{(\linewidth - 12\tabcolsep) * \real{0.1300}}
  >{\raggedleft\arraybackslash}p{(\linewidth - 12\tabcolsep) * \real{0.1500}}
  >{\raggedleft\arraybackslash}p{(\linewidth - 12\tabcolsep) * \real{0.1400}}
  >{\raggedleft\arraybackslash}p{(\linewidth - 12\tabcolsep) * \real{0.1400}}@{}}
\caption{Response attenuation under the standard output-gap
coefficient}\label{tbl-policy-rule-ranges}\tabularnewline
\toprule\noalign{}
\begin{minipage}[b]{\linewidth}\raggedright
Adjustment timing
\end{minipage} & \begin{minipage}[b]{\linewidth}\raggedleft
Impact \(x_t\)
\end{minipage} & \begin{minipage}[b]{\linewidth}\raggedleft
Impact \(\pi_t^p\)
\end{minipage} & \begin{minipage}[b]{\linewidth}\raggedleft
Impact \(|d_t|\)
\end{minipage} & \begin{minipage}[b]{\linewidth}\raggedleft
21-quarter \(x_t\)
\end{minipage} & \begin{minipage}[b]{\linewidth}\raggedleft
21-quarter \(\pi_t^p\)
\end{minipage} & \begin{minipage}[b]{\linewidth}\raggedleft
21-quarter \(|d_t|\)
\end{minipage} \\
\midrule\noalign{}
\endfirsthead
\toprule\noalign{}
\begin{minipage}[b]{\linewidth}\raggedright
Adjustment timing
\end{minipage} & \begin{minipage}[b]{\linewidth}\raggedleft
Impact \(x_t\)
\end{minipage} & \begin{minipage}[b]{\linewidth}\raggedleft
Impact \(\pi_t^p\)
\end{minipage} & \begin{minipage}[b]{\linewidth}\raggedleft
Impact \(|d_t|\)
\end{minipage} & \begin{minipage}[b]{\linewidth}\raggedleft
21-quarter \(x_t\)
\end{minipage} & \begin{minipage}[b]{\linewidth}\raggedleft
21-quarter \(\pi_t^p\)
\end{minipage} & \begin{minipage}[b]{\linewidth}\raggedleft
21-quarter \(|d_t|\)
\end{minipage} \\
\midrule\noalign{}
\endhead
\bottomrule\noalign{}
\endlastfoot
Own lag & 9.93--11.21 & 7.86--12.16 & 9.67--11.49 & 9.09--11.73 &
6.32--12.57 & 7.85--12.16 \\
Aggregate lag & 10.10--12.59 & 10.10--12.59 & 10.10--12.59 & 4.57--23.12
& 6.48--19.72 & 6.48--19.72 \\
\end{longtable}

\endgroup

\emph{Notes:} Entries are percentage reductions in absolute response
magnitudes under \(\phi_x=0.125\) relative to \(\phi_x=0\), holding
household-side wage-subsidy financing, \(\lambda\), \(D_p\), and \(D_w\)
fixed. Ranges cover 54 matched comparisons for each adjustment timing.
Cumulative responses sum quarters 0--20.

\clearpage

Table~\ref{tbl-wage-subsidy-financing-ranges} compares common lump-sum
financing of the aggregate household wage-subsidy bill with the
type-specific rebate; the firm-side sales-subsidy offset remains fixed.
The first-order aggregate effects are small, particularly under
aggregate-lag adjustment. The distributional effects are larger because
the common tax increases the share of direct profit incidence offset by
relative wages and employment.

\begin{longtable}[]{@{}
  >{\raggedright\arraybackslash}p{(\linewidth - 4\tabcolsep) * \real{0.3333}}
  >{\raggedleft\arraybackslash}p{(\linewidth - 4\tabcolsep) * \real{0.3333}}
  >{\raggedleft\arraybackslash}p{(\linewidth - 4\tabcolsep) * \real{0.3333}}@{}}
\caption{Effects of common lump-sum
taxation}\label{tbl-wage-subsidy-financing-ranges}\tabularnewline
\toprule\noalign{}
\begin{minipage}[b]{\linewidth}\raggedright
Response or decomposition
\end{minipage} & \begin{minipage}[b]{\linewidth}\raggedleft
Own lag
\end{minipage} & \begin{minipage}[b]{\linewidth}\raggedleft
Aggregate lag
\end{minipage} \\
\midrule\noalign{}
\endfirsthead
\toprule\noalign{}
\begin{minipage}[b]{\linewidth}\raggedright
Response or decomposition
\end{minipage} & \begin{minipage}[b]{\linewidth}\raggedleft
Own lag
\end{minipage} & \begin{minipage}[b]{\linewidth}\raggedleft
Aggregate lag
\end{minipage} \\
\midrule\noalign{}
\endhead
\bottomrule\noalign{}
\endlastfoot
Impact \(x_t\) & \([-0.052,\ 0.091]\) & \([-0.064,\ 0.040]\) \\
Impact \(\pi_t^p\) & \([-2.456,\ -0.012]\) & \([-0.064,\ 0.040]\) \\
Impact \(|d_t|\) & \([-0.492,\ -0.002]\) & \([-0.064,\ 0.040]\) \\
21-quarter \(x_t\) & \([-1.204,\ -0.009]\) & \([-0.850,\ -0.003]\) \\
21-quarter \(\pi_t^p\) & \([-4.069,\ -0.027]\) &
\([-0.621,\ -0.003]\) \\
21-quarter \(|d_t|\) & \([-2.426,\ -0.013]\) & \([-0.621,\ -0.003]\) \\
Impact \(|c_t^S-c_t^H|\) & \([-5.051,\ -1.339]\) &
\([-2.632,\ -0.654]\) \\
21-quarter \(|c_t^S-c_t^H|\) & \([-10.840,\ -4.625]\) &
\([-3.175,\ -0.659]\) \\
Impact offset share & \([1.267,\ 3.739]\) pp & \([0.641,\ 2.333]\) pp \\
21-quarter offset share & \([3.717,\ 5.659]\) pp & \([0.641,\ 2.333]\)
pp \\
\end{longtable}

\emph{Notes:} For the first eight rows, entries are ranges of percentage
changes in absolute response magnitudes under the common lump-sum tax
relative to the type-specific rebate; negative values denote smaller
responses. The last two rows report percentage-point changes in the
share of direct profit incidence offset by relative wages and
employment. Ranges cover \(\lambda\in\{0.2,0.4\}\), all four values of
\(\phi_x\), and all nine combinations of \(D_p\) and \(D_w\), giving 72
matched comparisons per timing. Cumulative responses sum quarters 0--20.

\begin{figure}

\centering{

\includegraphics[width=0.9\linewidth,height=\textheight,keepaspectratio]{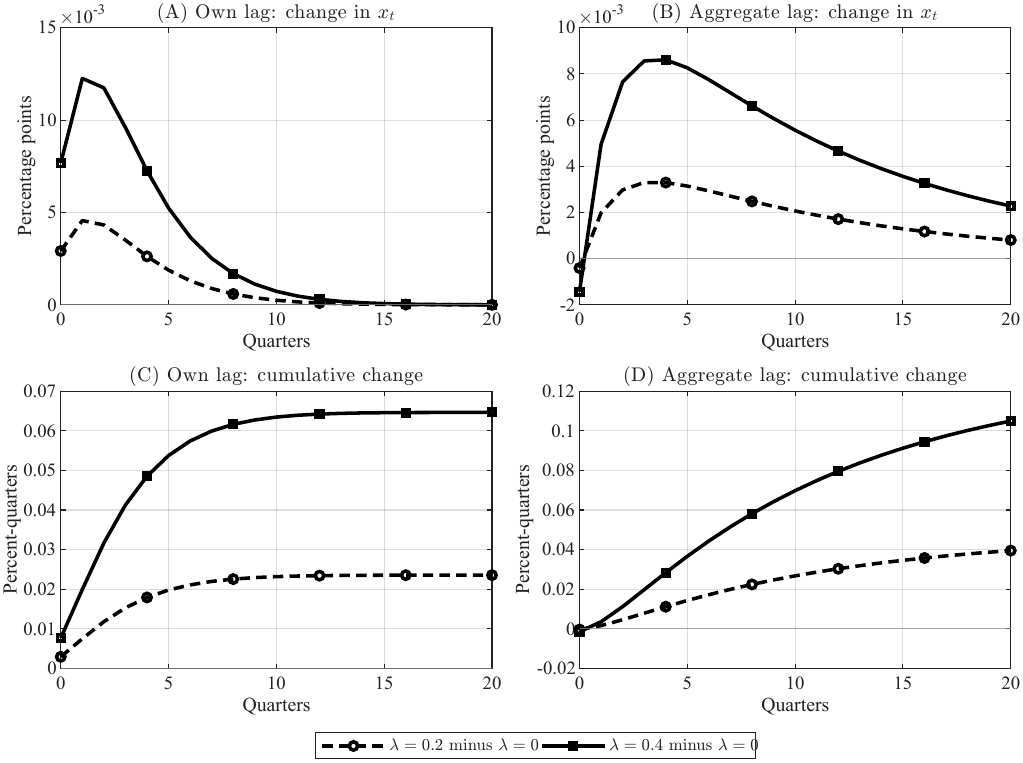}

\par\smallskip
\begin{minipage}{0.90\linewidth}
\footnotesize \textit{Notes:} Panels report differences in the output-gap response relative to $\lambda=0$ for $\lambda=0.2$ and $\lambda=0.4$. Panels (A) and (B) show period responses, while Panels (C) and (D) show cumulative responses through each horizon. The calibration uses $\phi_x=0.25$, $D_p=D_w=4$, and the type-specific wage-subsidy rebate. Responses are to a one-percent monetary-policy innovation. Period differences are in percentage points and cumulative differences are in percent-quarters.
\end{minipage}

}

\caption{\label{fig-app-policy-lambda-edge}Heterogeneity effects under
strong policy feedback}

\end{figure}

The impact qualification is illustrated in
Figure~\ref{fig-app-policy-lambda-edge}. With strong policy feedback,
own-lag output rises with \(\lambda\) on impact, whereas the
aggregate-lag impact response can fall slightly. The aggregate-lag
difference turns positive after impact, and its 21-quarter cumulative
effect is positive.

\clearpage

\subsubsection*{Market-Competitiveness
Checks}\label{market-competitiveness-checks}
\addcontentsline{toc}{subsubsection}{Market-Competitiveness Checks}

The previously reported characteristic-root figures remain unchanged.

\begin{figure}

\centering{

\includegraphics[width=0.9\linewidth,height=\textheight,keepaspectratio]{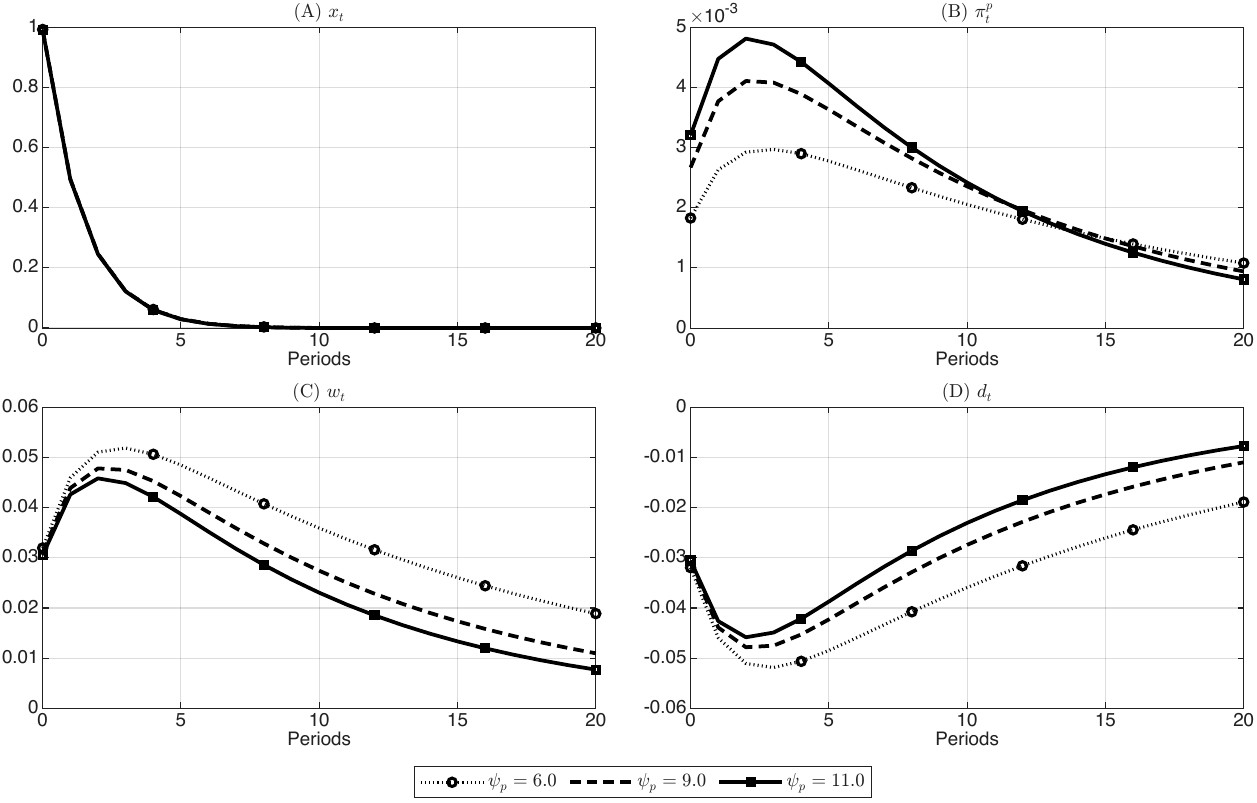}

\par\smallskip
\begin{minipage}{0.90\linewidth}
\footnotesize \textit{Notes:} Panels report $(A)$ $x_t$, $(B)$ $\pi_t^p$, $(C)$ $w_t$, and $(D)$ $d_t$. The dotted, dashed, and solid lines correspond to $\psi_p=6$, $\psi_p=9$, and $\psi_p=11$, with $\eta_p$ held at its benchmark value. Responses are to a one-percent monetary-policy innovation; vertical axes are percent deviations from steady state, and $d_t$ is the profit share scaled by steady-state output.
\end{minipage}

}

\caption{\label{fig-app-psip}Monetary shock with
\(\psi_p\in\{6,9,11\}\)}

\end{figure}

Figure~\ref{fig-app-psip} reports the monetary-shock comparison for
goods-market competitiveness. Higher \(\psi_p\) weakens the effective
price rigidity and therefore raises inflation responses while muting the
redistribution term. The output response is almost invariant over the
narrower admissible range \(\psi_p\in\{6,9,11\}\), while inflation,
wages, and profits display the predicted ordering. The exercise holds
\(\eta_p\) fixed, so it is a structural comparative static in market
competitiveness rather than a constant-duration-target comparison.

\begin{figure}

\centering{

\includegraphics[width=0.9\linewidth,height=\textheight,keepaspectratio]{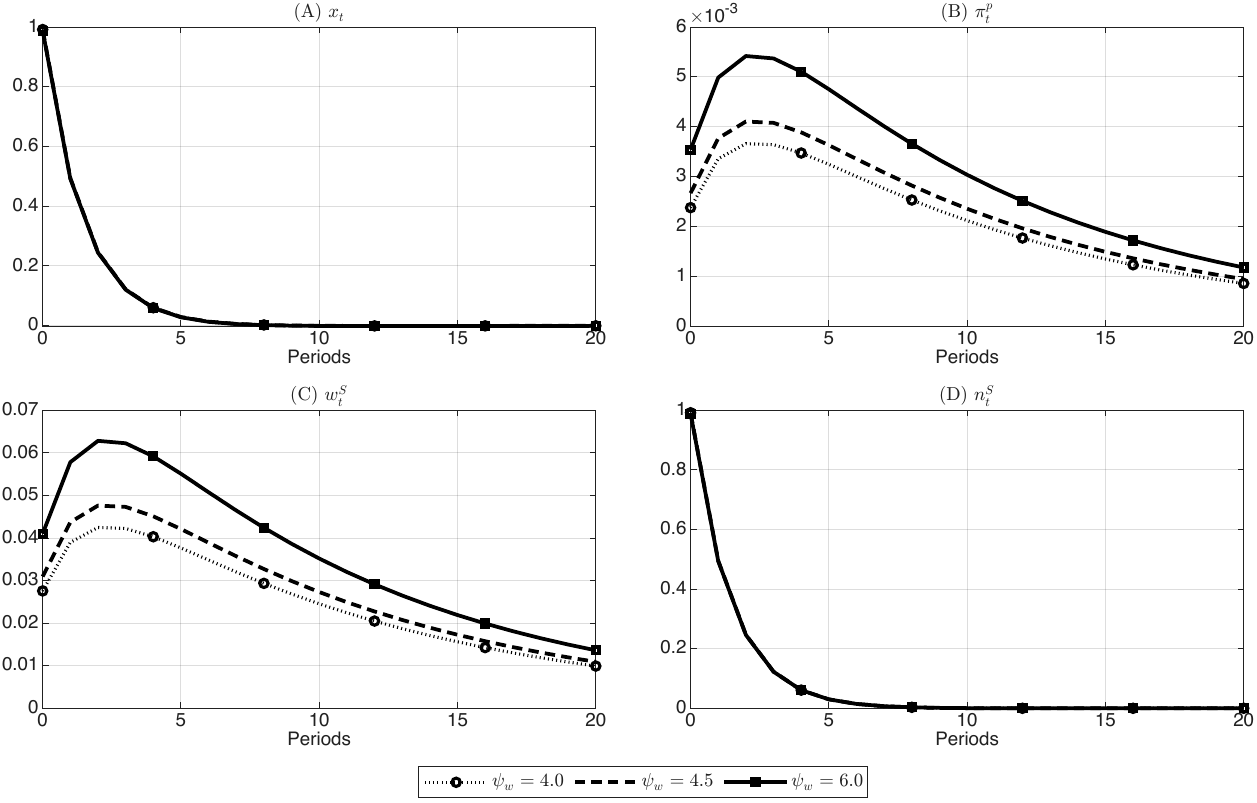}

\par\smallskip
\begin{minipage}{0.90\linewidth}
\footnotesize \textit{Notes:} Panels report $(A)$ $x_t$, $(B)$ $\pi_t^p$, $(C)$ $w_t^S$, and $(D)$ $n_t^S$. The dotted, dashed, and solid lines correspond to $\psi_w=4$, $\psi_w=4.5$, and $\psi_w=6$, with $\eta_w$ held at its benchmark value. Responses are to a one-percent monetary-policy innovation; vertical axes are percent deviations from steady state.
\end{minipage}

}

\caption{\label{fig-app-psiw}Monetary shock with
\(\psi_w\in\{4,4.5,6\}\)}

\end{figure}

Figure~\ref{fig-app-psiw} reports the corresponding comparison for
labor-market competitiveness. The output and saver-hours responses
differ little across \(\psi_w\in\{4,4.5,6\}\), whereas inflation and the
saver wage rise with labor substitutability. The exercise again holds
the corresponding adjustment cost fixed and keeps every case inside the
structural restriction \(\psi_w>1\).

\subsubsection*{Technology-Shock
Results}\label{technology-shock-results}
\addcontentsline{toc}{subsubsection}{Technology-Shock Results}

\begin{figure}

\centering{

\includegraphics[width=0.9\linewidth,height=\textheight,keepaspectratio]{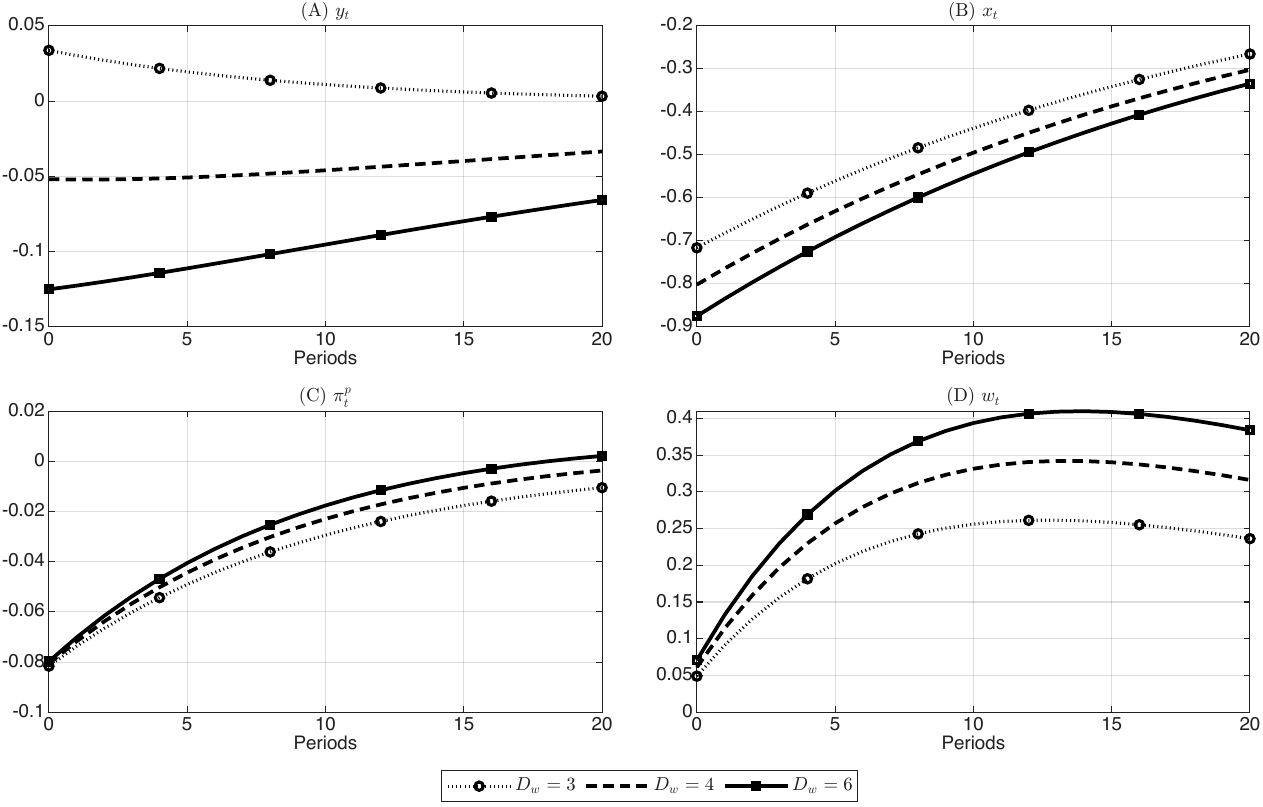}

\par\smallskip
\begin{minipage}{0.90\linewidth}
\footnotesize \textit{Notes:} Panels report $(A)$ $y_t$, $(B)$ $x_t$, $(C)$ $\pi_t^p$, and $(D)$ $w_t$. The dotted, dashed, and solid lines correspond to $D_w=3,4,6$ quarters, or $\eta_w=264.706,524.272,1285.714$. Responses are to a one-percent technology innovation; vertical axes are percent deviations from steady state.
\end{minipage}

}

\caption{\label{fig-app-tech-etaw}Technology shock with wage-duration
targets \(D_w\in\{3,4,6\}\) quarters}

\end{figure}

\begin{figure}

\centering{

\includegraphics[width=0.9\linewidth,height=\textheight,keepaspectratio]{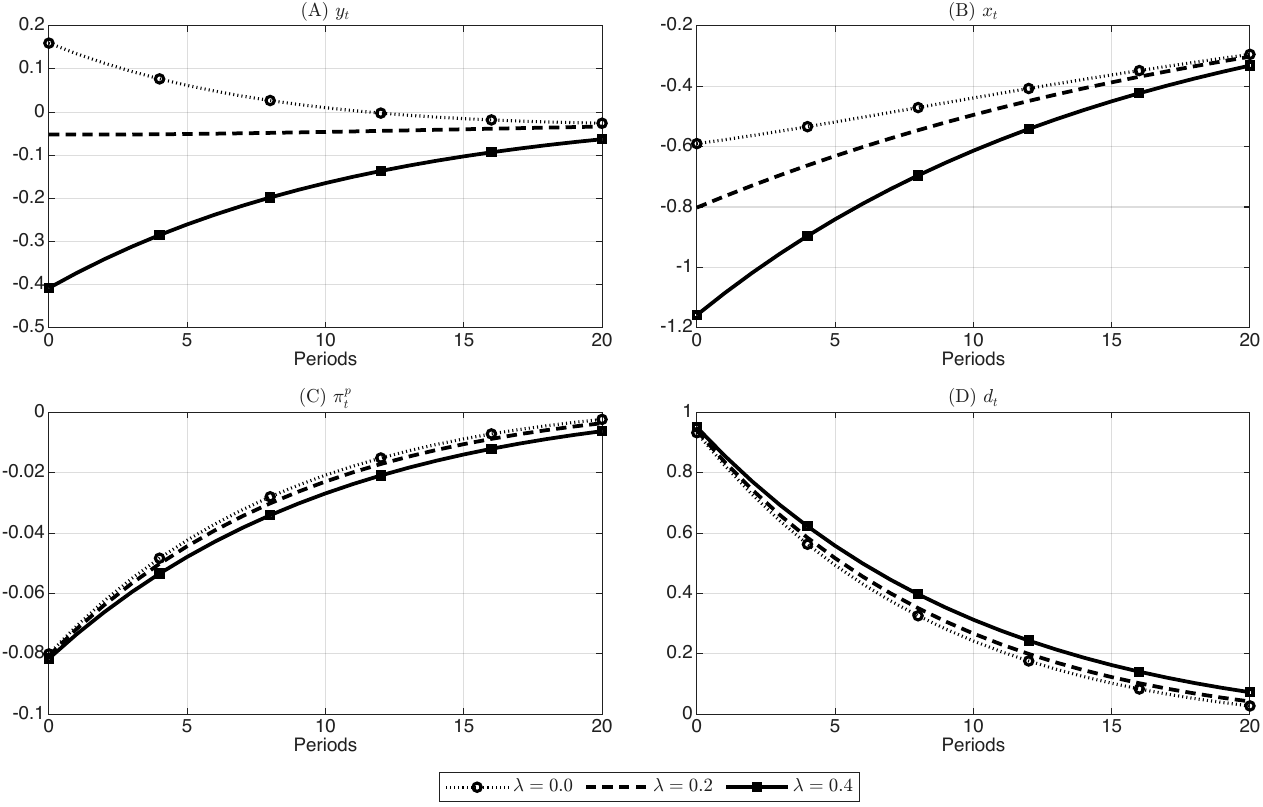}

\par\smallskip
\begin{minipage}{0.90\linewidth}
\footnotesize \textit{Notes:} Panels report $(A)$ $y_t$, $(B)$ $x_t$, $(C)$ $\pi_t^p$, and $(D)$ $d_t$. The dotted, dashed, and solid lines correspond to $\lambda=0$, $\lambda=0.2$, and $\lambda=0.4$. Responses are to a one-percent technology innovation; vertical axes are percent deviations from steady state, and $d_t$ is the profit share scaled by steady-state output.
\end{minipage}

}

\caption{\label{fig-app-tech-lambda-bench}Technology shock with
\(\lambda\in\{0,0.2,0.4\}\) and \(D_p=4\)}

\end{figure}

\begin{figure}

\centering{

\includegraphics[width=0.9\linewidth,height=\textheight,keepaspectratio]{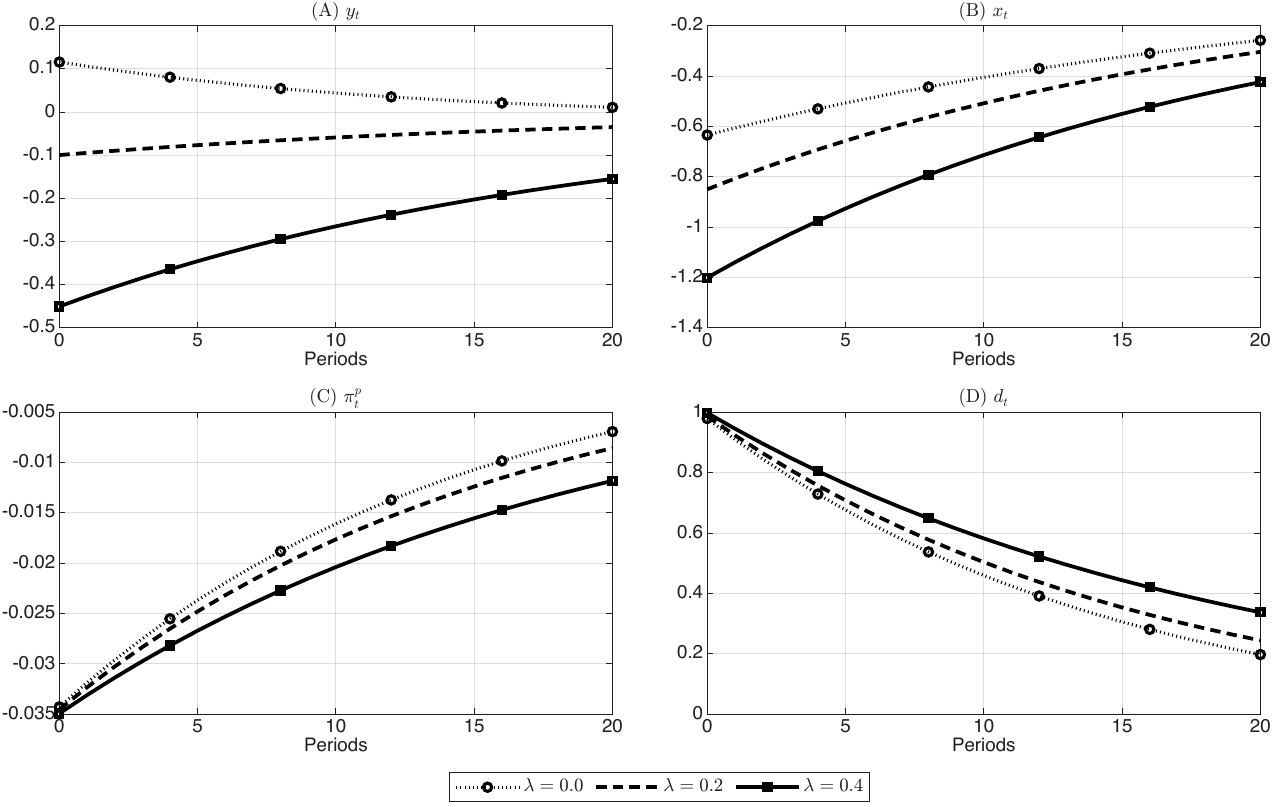}

\par\smallskip
\begin{minipage}{0.90\linewidth}
\footnotesize \textit{Notes:} Panels report $(A)$ $y_t$, $(B)$ $x_t$, $(C)$ $\pi_t^p$, and $(D)$ $d_t$. The dotted, dashed, and solid lines correspond to $\lambda=0$, $\lambda=0.2$, and $\lambda=0.4$. Responses are to a one-percent technology innovation; vertical axes are percent deviations from steady state, and $d_t$ is the profit share scaled by steady-state output.
\end{minipage}

}

\caption{\label{fig-app-tech-lambda-high}Technology shock with
\(\lambda\in\{0,0.2,0.4\}\) and \(D_p=6\)}

\end{figure}

Aggregate-lag technology-shock responses are collected in Figures
\ref{fig-app-tech-etaw}--\ref{fig-app-tech-lambda-high}, followed by a
common-primitive timing comparison in Table~\ref{tbl-ownlag-technology}.
Unlike monetary shocks, technology shocks move output \(y_t\) and the
output gap \(x_t\) differently. The numerical responses show that a
longer wage-duration target lowers both output and the output gap while
making the inflation decline smaller and the real-wage increase larger.

Figure~\ref{fig-app-tech-etaw} reports the aggregate-lag responses for
\(D_w\in\{3,4,6\}\). On impact, the output response moves from 0.03
percent at \(D_w=3\) to approximately \(-0.12\) percent at \(D_w=6\),
and the output gap becomes correspondingly more negative. The inflation
decline is slightly smaller at longer duration targets, while the wage
response is substantially larger and more persistent.

Figure~\ref{fig-app-tech-lambda-bench} reports the aggregate-lag
benchmark technology-shock comparison across \(\lambda\). Under the
benchmark \(D_p=4\), heterogeneity has sizable effects on
technology-shock transmission: a larger \(\lambda\) lowers output and
the output gap, makes deflation somewhat larger, and raises profits.
Figure~\ref{fig-app-tech-lambda-high} reports the corresponding
\(D_p=6\) case. The same aggregate-lag ordering is preserved, with
greater separation across \(\lambda\) in output, the output gap, and
profits.

\clearpage

The common-primitive technology comparison is less stable than the
monetary comparison. At \(\lambda=0.2\), both specifications imply a
negative output-gap response and higher profits, but the magnitudes
differ substantially, as shown in Table~\ref{tbl-ownlag-technology}. At
\(\lambda=0\), the impact output gap is positive under own-lag
adjustment (0.0456 percent) and negative under aggregate-lag adjustment
(-0.590 percent). Within the own-lag grid, increasing \(\lambda\) lowers
the impact and 21-quarter cumulative output responses in all 18
comparisons, the opposite of the monetary-shock ordering. The positive
\(\lambda\) amplification result should therefore be stated specifically
for monetary shocks.

\begin{longtable}[]{@{}
  >{\raggedright\arraybackslash}p{(\linewidth - 12\tabcolsep) * \real{0.1700}}
  >{\raggedleft\arraybackslash}p{(\linewidth - 12\tabcolsep) * \real{0.1400}}
  >{\raggedleft\arraybackslash}p{(\linewidth - 12\tabcolsep) * \real{0.1400}}
  >{\raggedleft\arraybackslash}p{(\linewidth - 12\tabcolsep) * \real{0.1400}}
  >{\raggedleft\arraybackslash}p{(\linewidth - 12\tabcolsep) * \real{0.1400}}
  >{\raggedleft\arraybackslash}p{(\linewidth - 12\tabcolsep) * \real{0.1400}}
  >{\raggedleft\arraybackslash}p{(\linewidth - 12\tabcolsep) * \real{0.1300}}@{}}
\caption{Common-primitive technology-shock
comparison}\label{tbl-ownlag-technology}\tabularnewline
\toprule\noalign{}
\begin{minipage}[b]{\linewidth}\raggedright
Adjustment timing
\end{minipage} & \begin{minipage}[b]{\linewidth}\raggedleft
Impact \(x_t\)
\end{minipage} & \begin{minipage}[b]{\linewidth}\raggedleft
Impact \(\pi_t^p\)
\end{minipage} & \begin{minipage}[b]{\linewidth}\raggedleft
Impact \(d_t\)
\end{minipage} & \begin{minipage}[b]{\linewidth}\raggedleft
21-quarter \(x_t\)
\end{minipage} & \begin{minipage}[b]{\linewidth}\raggedleft
21-quarter \(d_t\)
\end{minipage} & \begin{minipage}[b]{\linewidth}\raggedleft
\(10^4\mathcal L\)
\end{minipage} \\
\midrule\noalign{}
\endfirsthead
\toprule\noalign{}
\begin{minipage}[b]{\linewidth}\raggedright
Adjustment timing
\end{minipage} & \begin{minipage}[b]{\linewidth}\raggedleft
Impact \(x_t\)
\end{minipage} & \begin{minipage}[b]{\linewidth}\raggedleft
Impact \(\pi_t^p\)
\end{minipage} & \begin{minipage}[b]{\linewidth}\raggedleft
Impact \(d_t\)
\end{minipage} & \begin{minipage}[b]{\linewidth}\raggedleft
21-quarter \(x_t\)
\end{minipage} & \begin{minipage}[b]{\linewidth}\raggedleft
21-quarter \(d_t\)
\end{minipage} & \begin{minipage}[b]{\linewidth}\raggedleft
\(10^4\mathcal L\)
\end{minipage} \\
\midrule\noalign{}
\endhead
\bottomrule\noalign{}
\endlastfoot
Own lag & -0.115 & -0.313 & 0.784 & -2.771 & 3.484 & 56.064 \\
Aggregate lag & -0.802 & -0.0806 & 0.939 & -10.835 & 7.288 & 16.459 \\
\end{longtable}

\emph{Notes:} The calibration sets \(\lambda=0.2\) and the duration
targets \(D_p=D_w=4\). Impact responses are percent deviations following
a one-percent technology innovation, except \(d_t\), which is the profit
share in percentage points of steady-state output. Cumulative \(x_t\) is
in percent-quarters, while cumulative \(d_t\) is in
percentage-point-of-output quarters; both sum quarters 0--20.
\(\mathcal L\) is the discounted CES-consistent loss over quarters
0--399.

\end{document}